\documentclass[prd,nofootinbib,preprint,superscriptaddress]{revtex4-1}

\usepackage{amsmath, amssymb, amsthm, graphicx, epsfig, fancyhdr,epsfig, slashed, mathrsfs}

\usepackage{tikzsymbols}
\usepackage{natbib}
\usepackage{float}
\usepackage{graphicx}
\usepackage{amsmath}
\usepackage{amsfonts}
\usepackage{amssymb}
\usepackage[bookmarks=true,bookmarksopen=true,
                    bookmarksnumbered=true,
                    colorlinks=true,linkcolor=red,
                    citecolor=red]{hyperref}
\usepackage{subfigure}
\usepackage{slashed}
 \usepackage{setspace} 
\usepackage{caption}
\newcommand{\lag}{\mathcal L}

\definecolor{lime}{HTML}{A6CE39}
\DeclareRobustCommand{\orcidicon}{
	\begin{tikzpicture}
		\draw[lime, fill=lime] (0,0) 
		circle [radius=0.2] 
		node[white] {{\fontfamily{qag}\selectfont \tiny ID}};
		\draw[white, fill=white] (-0.0625,0.095) 
		circle [radius=0.007];
	\end{tikzpicture}
	\hspace{-2mm}
}

\foreach \x in {A, ..., Z}{\expandafter\xdef\csname orcid\x\endcsname{\noexpand\href{https://orcid.org/\csname orcidauthor\x\endcsname}
		{\noexpand\orcidicon}}
}

\newcommand{\be}{\begin{equation}}
	\newcommand{\ee}{\end{equation}}
\newcommand{\bea}{\begin{eqnarray}}
	\newcommand{\eea}{\end{eqnarray}}

\newcommand{\beq}{\begin{equation}}
	\newcommand{\eeq}{\end{equation}}

\def\nn{\nonumber}

\usepackage[utf8]{inputenc}

\DeclareUnicodeCharacter{2212}{-}
\allowdisplaybreaks
\begin{document}  

\hspace*{\fill} HRI-RECAPP-2023-10

\title{A Scotogenic model with U(1) symmetry and a scalar dark matter}

\author{\small Anjan Kumar Barik} %\orcidA{}
\email{anjanbarik@hri.res.in }
\affiliation{Regional Centre for Accelerator-based Particle Physics,
Harish-Chandra Research Institute,\\ A CI of Homi Bhabha National
Institute, Chhatnag Road, Jhunsi, Prayagraj 211019, India.\vspace{0.2cm}}

\author{\small Najimuddin Khan} %\orcidC{}
\email{nkhan.ph@amu.ac.in }
\affiliation{\small Department of Physics, Aligarh Muslim University, Aligarh -202001, India.\vspace{1.80cm}}

\author{\small Santosh Kumar Rai} %\orcidD{}
\email{skr@hri.res.in }
\affiliation{Regional Centre for Accelerator-based Particle Physics,
Harish-Chandra Research Institute,\\ A CI of Homi Bhabha National
Institute, Chhatnag Road, Jhunsi, Prayagraj 211019, India.\vspace{0.2cm}}

\begin{abstract}
We study a scotogenic model augmented with an additional $U(1)$ gauge and a discrete $Z_{2}$ symmetry. The lightest $Z_{2}$-odd particle in our model becomes the dark matter (DM) candidate while tiny neutrino masses are realized at one loop. We explore the parameter space of the model for which the DM relic density is satisfied, and the correct low-energy neutrino observables are reproduced. The extended gauge symmetry includes beyond Standard Model (SM) particle spectrum consisting of vector-like fermions and scalars. We also highlight possible collider 
signatures of these particles at the LHC.

\end{abstract}
 
\maketitle
\section{Introduction}
The modern era of particle physics has witnessed a remarkable period of success with the development and establishment of the Standard Model (SM). The discovery of a $\sim125$ GeV Higgs at ATLAS~\cite{ATLAS:2012yve} and CMS~\cite{CMS:2012qbp} completed the search for the  
particle spectrum of the SM. Although the SM of particle physics is a highly successful theoretical framework, it has some limitations as it cannot give any prediction of the DM present in the Universe, where it came from, and its nature. Neither can it explain the matter-antimatter asymmetry in the Universe, so pertinent to our very existence. Then there is the matter of gauge hierarchy problem that suggests the existence of some new physics not far above the electroweak scale and the non-zero mass of the neutrinos where SM falls short of expectations. 
These shortcomings motivate us to explore new theoretical ideas and their experimental investigations beyond the SM.
Various astrophysical observations, such as Coma cluster~\cite{Zwicky:1933gu}, the gravitational lensing effects observed in the bullet cluster~\cite{Clowe:2006eq} and 
anomalies in galactic rotation curves~\cite{1939LicOB..19...41B,1980ApJ...238..471R}, have provided compelling evidence for the existence of DM in the 
Universe. These observations indicate that there is additional mass in the Universe that does not interact with 
electromagnetic radiation and cannot be accounted for by the particles described within the SM 
of particle physics. In the SM of particle physics, neutrinos were originally assumed to be massless. 
However, experimental observations of neutrino oscillations have conclusively demonstrated that neutrinos 
have non-zero masses \cite{SNO:2002tuh,Super-Kamiokande:1998kpq}, which has motivated the need for extensions to the SM of particle physics. It is to 
be noted that neutrinos and DM have distinct properties and behaviours.

We study a scotogenic model, which is augmented with an additional $U(1)$ gauge symmetry. 
The success of the Standard Model being a gauge theory motivates us to consider BSM particles to be 
charged under certain gauge groups. The motivation for $U(1)$ gauge symmetry comes from Grand 
Unified Theories (GUT), which contains the SM gauge group and has a rank greater than four. The mathematical 
fact that any theory with  $SU(N)$ gauge symmetry, when spontaneously broken by a scalar in the adjoint 
representation, has its $N-1$ diagonal generators unbroken, therefore the theory has at least $U(1)^{N-1}$ 
symmetry \cite{Langacker:2008yv}. In special cases when more than one real component of a scalar 
multiplet which breaks the $SU(N)$ gauge symmetry becomes equal, there will be a remnant non-abelian 
gauge symmetry, e.g. $SO(10)$ and $E_{6}$ GUT have gauge symmetry of rank $5$ and $6$ respectively 
after symmetry breaking will give SM gauge group along with additional $U(1)$ symmetry.

In the $U(1)$ extension of SM, the new abelian gauge boson can be a messenger between SM and BSM 
particles, even if the SM particles are neutral under the additional symmetry as the interactions can proceed via gauge kinetic mixing (GKM) between the SM $U(1)_Y$ and new $U(1)$ gauge bosons. 
%The $Z'$ in different colliders can lead to the exploration of new BSM sectors. 
The $U(1)$ extended BSM models can solve many phenomenological problems, e.g., $\mu$ problem in the 
MSSM (Minimal supersymmetric SM) by dynamically generating the $\mu$ term, some flavour non-universal 
coupling of new gauge boson to lepton and quark sector can induce flavour violating loop 
decay, e.g., $\mu \to e\gamma$, $b \to s\gamma$ and also can contribute to the anomalous magnetic moment 
of the respective particle.
$U(1)$ extension of SM can also explain neutrino mass, DM, matter-antimatter asymmetry of the universe etc. \cite{Langacker:2008yv, Leike:1998wr, Rizzo:2006nw, Marshak:1979fm, Mohapatra:1980qe, Khalil:2006yi}. 
 
However, introducing additional particles and new symmetries into the SM comes with constraints on their masses and interactions with SM particles, as there is no convincing experimental evidence of their existence. Introduction of a new gauge boson, $(Z')$ would also have to comply with various constraints such as the $Z-Z'$ mixing angle, its direct search limits at colliders, etc. While the precise measurement of the well-established $Z$ decay widths to SM modes must remain within experimental errors \cite{ParticleDataGroup:2020ssz}, $Z'$ search in Drell-Yan processes at LHC~\cite{ATLAS:2017fih, CMS:2019buh, ACCOMANDO:2013zz} put strong limits on its mass. To circumvent these limitations, we adopt a strategy where SM particles remain neutral under the new gauge symmetry, interacting with the $Z'$ solely through the GKM. By maintaining a small GKM, we can effectively avoid these experimental constraints.

In this work, we have extended the SM by a $U(1)$ gauge symmetry with the inclusion of additional $Z_2$ 
symmetry. All the BSM particles are charged under the new gauge symmetry. In the $Z_2$ even sector 
along with SM particles, we have one gauge singlet scalar, which is responsible for breaking the new $U(1)$ gauge 
symmetry spontaneously. In the $Z_2$ odd sector,  we have $SU(2)$ singlet and doublet scalar fields including
vector-like singlet and doublet fermions. We propose a scalar DM by 
making one of the CP even scalars to be lightest among the $Z_2$ odd sector particles.
Subsequently, we investigate the phenomenology of the DM being a $SU(2)$ singlet, doublet, and linear combination of both. 
By studying the DM phenomenology in the above scenarios, we aim to understand the model's viability 
in explaining the observed DM abundance in the Universe and to predict potential signals that can be probed 
experimentally. In addition, we thoroughly investigate relevant experimental and theoretical constraints on the 
model parameters. By considering these constraints, we can identify parameter space regions consistent with all 
available experimental data and gain insight into the model's overall viability and implications for DM phenomenology.

The paper is organized as follows: Section~\ref{sec:2} provides a concise overview of the $U(1)$ extended neutrino DM model. Section~\ref{sec:3} is dedicated to discussing the relevant theoretical and experimental constraints associated with the model. Moving on to Section~\ref{sec:4}, we present the numerical analysis carried out to explore the model parameters 
in the context of neutrino and DM in depth. We discuss the possible collider search channels at the LHC in 
Section~\ref{sec:5}. Finally, Section~\ref{sec:6} summarizes the key findings and draw conclusions based on our study.

%%%%%%%%%%%%%%%%%
\section{Model}
\label{sec:2}
%%%%%%%%%%%%%%%%%
In the proposed model the SM gauge sector is extended by an abelian $U(1)$ gauge symmetry. The model 
also contains new fields that carry charges under both the SM gauge group and the new $U(1)$. A 
discrete $Z_2$ symmetry is also included which is responsible for making the lightest particle odd 
under this symmetry to act as the DM of the Universe. The new particles and their 
corresponding charge assignment are shown in Tab.~\ref{tab1}. The scalar sector of the SM is 
extended to include a new doublet $(\Phi_2)$ and two complex singlet fields $(S_{1}, S_{2})$ while 
the fermion sector has a new vector-like doublet ($\psi$) and singlet fermion ($N$). 
The model provides an explanation for neutrino mass via a radiative seesaw mechanism~\cite{Ma:2006km} while also providing a viable DM candidate.
%%%%%%%%%%%%%%%%%
  	 \begin{table}[h!]
    \begin{center}
	\begin{tabular}{|c|ccccc|}
		\hline
		\hline
		Particles &$\psi=(N_D,E_1^-)^T$ &~~$~~N~~$&~~$~~\Phi_2~~$~~&~~$~~S_{1}~~$~~&~~$~~S_{2}~~$~~\\
		\hline
		$SU(2)_L$& 2 & 1 & 2&  1& 1\\
		\hline
		$U(1)_Y$& -1 & 0 & $1$& 0& 0\\
		\hline
		$U(1)_X$& 1  & 1 & 1  & 2& 1\\
		\hline
		$Z_2$& -1 &-1 &  -1& 1& -1\\
		\hline
		\hline		
	\end{tabular}
	\caption{Additional particle content and their charge assignments under symmetry groups.}\label{tab1}          
    \end{center}
\end{table}
%%%%%%%%%%%%%%%%%
%%%%%%%%%%%%%%%%%

\subsection{Scalars}
\label{sec:scalar}
%%%%%%%%%%%%%%%%%%%%
The model contains a new $Z_2$-odd $SU(2)$ Higgs doublet ($\Phi_2$) and two complex scalar singlets ($S_1, \, S_2$) where $S_1$ is $Z_2$-even while $S_2$ is $Z_2$-odd, in addition to the SM Higgs doublet ($\Phi_1$). The kinetic part of the Lagrangian density for the above-mentioned scalars is given by,
%%%%%%%%%%%%%%%%%
%%%%%%%%%%%%%%%%%
\begin{equation}
\mathcal{L}_{KE}=\sum_{\phi} (D_{\mu}\phi)^{\dagger}(D^{\mu}\phi),
\end{equation}
%%%%%%%%%%%%%%%%%
%%%%%%%%%%%%%%%%%
where $\phi=\Phi_1,\Phi_2,S_1,S_2$ and $D_{\mu}$ represents the covariant derivatives given by,
%%%%%%%%%%%%%%%%%
%%%%%%%%%%%%%%%%%
\begin{eqnarray}
D_{\mu}^{\Phi_1} &=& \partial_\mu - i \frac{g_2}{2}  { \sigma^{a} . W^{a}_\mu}  - i \frac{g_1}{2} { Y} B_\mu \nn\\
D_{\mu}^{\Phi_2} &=& \partial_\mu - i \frac{g_2}{2}  {\sigma^{a} .W^{a}_\mu}  - i \frac{g_1}{2} { Y} B_\mu  - i \frac{g_{x}}{2}  { X} C_\mu \\
D_{\mu}^{S_{{1,2}}} &=& \partial_\mu  - i \frac{g_{x}}{2}  { X} C_\mu \nn
\end{eqnarray}
%%%%%%%%%%%%%%%%%
%%%%%%%%%%%%%%%%%
where $\sigma^a$ are the Pauli matrices and $W^a_{\mu}$ are the $SU(2)_L$ gauge bosons $(a=1,2,3)$. The $B_\mu$ and $C_\mu$ are the $U(1)_Y$ and $U(1)_X$ gauge bosons, respectively. 
The gauge-invariant scalar potential in the Lagrangian is, 
%%%%%%%%%%%%%%%%%
%%%%%%%%%%%%%%%%%
\begin{equation}
%	\mathcal{V} =V_{\Phi_1, \, S_{1}}+V_{\Phi_2, \, S_{2}}+V_{}\,,
    \mathcal{V} =V_1+V_2+V_{12},
	\label{eq:potAll}
\end{equation}
%%%%%%%%%%%%%%%%%
%%%%%%%%%%%%%%%%%
where,
%%%%%%%%%%%%%%%%%
%%%%%%%%%%%%%%%%%
\begin{eqnarray}
V_1 &=& -\mu _{\phi_{1}}^2 \, \Phi _1^{\dagger } \Phi _1  - \mu _{S_{1}}^2 \, S _{1}^{\dagger } S _{1}  + \lambda _1  (\Phi _1^{\dagger } \Phi _1)^2 %\nn\\
%&& 
+ \lambda _{11} \, \Phi _1^{\dagger } \Phi _1 \, S _{1}^{\dagger } S _{1} +\lambda _{s1} (S _{1}^{\dagger } S _{1})^2 \nn \\ 
V_2 &=&\mu _{\phi_{2}}^2  \, \Phi _2^{\dagger } \Phi _2  +\mu _{\phi_{s2}}^2  \, S_{2}^{\dagger } S_{2}  + \lambda _2  (\Phi _2^{\dagger } \Phi _2)^2 %\nn\\
%&& 
+ \lambda _{22}  \, \Phi _2^{\dagger } \Phi _2  \,  S_{2}^{\dagger } S_{2} +\lambda _{s2} (S_{2}^{\dagger } S_{2})^2  \\
V_{12}&=&\lambda _3  \, \Phi _1^{\dagger } \Phi _1  \, \Phi _2^{\dagger } \Phi _2 + \lambda _4  \, 
 \Phi _1^{\dagger } \Phi _2  \, \Phi _2^{\dagger } \Phi _1 + \kappa _{1}  \,  \Phi _1^{\dagger } \Phi _1  \,  S_{2}^{\dagger } S_{2} + \kappa _{2}  \, \Phi _2^{\dagger } \Phi _2  \, S_{1}^{\dagger } S_{1} \nn\\ 
&& +  \, \kappa _{3}  \,  S_{1}^{\dagger } S_{1}  \, S_{2}^{\dagger } S_{2} 
-(\sqrt{2} \mu_{S} \, S_{2}^2 \, S_{1} - \mu_{h} \, \Phi _2^{\dagger } \Phi _1 \, S_{2}+h.c. )~.  \nn
\label{eq:potAll1}
\end{eqnarray}
%%%%%%%%%%%%%%%%%
%%%%%%%%%%%%%%%%%
In the above potential, the $\mu$'s are the bare masses of the scalars while the $\lambda$'s and $\kappa$'s are the quartic couplings. 
Among the two Higgs doublets only the $\Phi_1$ gets a vacuum expectation value (VEV) and is responsible for the electroweak symmetry breaking (EWSB) of $SU(2)_L \times U(1)_Y \to U(1)_{em}$. Similarly, the $Z_{2}$-even SM singlet scalar ($S_1$) gets a VEV and is responsible for the breaking of the $U(1)_X$ gauge symmetry. The scalars can now be written in their explicit component form as,
%%%%%%%%%%%%%%%%%
%%%%%%%%%%%%%%%%%
\begin{equation}
\Phi_1=\begin{pmatrix}
G^+\\ \frac{(\phi_{dr1} + v_1 +i G^0)}{\sqrt{2}}\\
\end{pmatrix}; \quad \quad \quad   
\Phi_2=\begin{pmatrix}
H^+\\ \frac{(\phi_{dr2}+i \phi_{di2})}{\sqrt{2}}\\
\end{pmatrix}. \nn
\end{equation}
%%%%%%%%%%%%%%%%%
 and 
%%%%%%%%%%%%%%%%%
\begin{equation}
S_{1}=\frac{(\phi_{sr1} + v_s + i \phi_{si1})}{\sqrt{2}} ; \quad  \quad \quad       S_{2}=\frac{(\phi_{sr2}+i \phi_{si2})}{\sqrt{2}}. \nn
\end{equation}
%%%%%%%%%%%%%%%%%

The minimization conditions for the potential are,
%%%%%%%%%%%%%%%%%
%%%%%%%%%%%%%%%%%
\begin{eqnarray}
\mu _{\phi_{1}}^2&=& \frac{1}{2} \left(\lambda _{11} v_s^2+2 \lambda _1 v_1^2\right),~~
\mu _{S_{1}}^2= \frac{1}{2} \left(2 \lambda _{s1} v_s^2 +\lambda _{11}  v_1^2 \right)
\label{eq:mini}
\end{eqnarray}
%%%%%%%%%%%%%%%%%
%%%%%%%%%%%%%%%%%
The CP-even components, ($\phi_{dr1}$ and $\phi_{sr1}$) mix after EWSB. Using the minimization conditions, we can write the mass matrix in the ($\phi_{dr1},\phi_{sr1}$) basis as,
%%%%%%%%%%%%%%%%%
%%%%%%%%%%%%%%%%%
\begin{equation}
(\phi_{dr1} ~\phi_{sr1}) \,
%%%
\begin{pmatrix}
\mathcal{M}_{a11} & \mathcal{M}_{a12} \\ \mathcal{M}_{a21} & \mathcal{M}_{a22}\\
\end{pmatrix} \,
%%%
\begin{pmatrix}
\phi_{dr1}  \\ \phi_{sr1} \\
\end{pmatrix},
\label{eq:massM1}
\end{equation}
%%%%%%%%%%%%%%%%%
%%%%%%%%%%%%%%%%%
where, $\mathcal{M}_{a11}= 2 \lambda _1 v_1^2,~~
\mathcal{M}_{a22}= 2 \lambda _{s1}  v_s^2$ and  $\mathcal{M}_{a12}=\mathcal{M}_{a21}=\lambda _{11} v_1 v_s$. The mass eigenstates are obtained by diagonalizing the mass matrix in eqn.~\ref{eq:massM1} with a rotation of the $(\phi_{dr1} ~\phi_{sr1}) $ basis as,
%%%%%%%%%%%%%%%%%
%%%%%%%%%%%%%%%%%
\begin{equation}
\begin{pmatrix}
h  \\ H \\
\end{pmatrix} =
Z_{H}
\begin{pmatrix}
\phi_{dr1}  \\ \phi_{sr1} \\
\end{pmatrix} ,~~~\text{with}~~Z_{H}=%%%%
\begin{pmatrix}
\cos\alpha & -\sin\alpha\\ \sin\alpha &\cos\alpha \\
\end{pmatrix}
%%%
\end{equation}
%%%%%%%%%%%%%%%%%
%%%%%%%%%%%%%%%%%
The mixing angle $\alpha$ between the CP even scalars is given by,
%%%%%%%%%%%%%%%%%
%%%%%%%%%%%%%%%%%
\begin{equation}
 \alpha = \frac{1}{2} \arctan(\frac{2\,\mathcal{M}_{a12}}{\mathcal{M}_{a22}-\mathcal{M}_{a11}})
\end{equation}
%%%%%%%%%%%%%%%%%
%%%%%%%%%%%%%%%%%
The masses of the scalars are,
%%%%%%%%%%%%%%%%%
%%%%%%%%%%%%%%%%%
\begin{eqnarray}
M_h^2 &=& \frac{1}{2} \left[  (\mathcal{M}_{a22}+\mathcal{M}_{a11}) - \sqrt{  (\mathcal{M}_{a22}-\mathcal{M}_{a11})^2 + 4 \mathcal{M}_{a12}^2}  \right] \\
M_H^2 &=& \frac{1}{2} \left[  (\mathcal{M}_{a22}+\mathcal{M}_{a11})  + \sqrt{  (\mathcal{M}_{a22}-\mathcal{M}_{a11})^2 + 4 \mathcal{M}_{a12}^2}  \right]\nn
\label{eq:mass1}
\end{eqnarray}
%%%%%%%%%%%%%%%%%
%%%%%%%%%%%%%%%%%
Here $h$ is considered to be the observed SM-like Higgs boson with mass $m_h \simeq 125$ GeV and $H$ is a heavy scalar. The couplings of the $h$ to the known fermions and gauge bosons get modified depending on the mixing. The CP-odd scalars of $\Phi_1$ and $S_1$ become the Goldstone bosons and get absorbed to give mass to the neutral gauge bosons, while the charged component of $\Phi_1$ is absorbed by the $W^\pm$ boson. 

The charged component $\phi_{d2}^\pm \equiv H^\pm$ of the $Z_2$ odd doublet does not mix with the other scalar fields and its mass is given by,
%%%%%%%%%%%%%%%%%
%%%%%%%%%%%%%%%%%
\begin{equation}
M_{H^\pm}^2= \mu _{\phi _{2}}^2+\frac{1}{2} \kappa _{1} v_s^2  +\frac{1}{2} \lambda _3 v_1^2
\label{eq:mass2}
\end{equation}
%%%%%%%%%%%%%%%%%
%%%%%%%%%%%%%%%%%
The CP-even part  of the scalar fields $\phi_{dr2}$ and $\phi_{sr2}$ from the $Z_2$-odd sector also mix. The masses are given by,
%%%%%%%%%%%%%%%%%
\begin{eqnarray}
M_{H_1}^2 &=& \frac{1}{2} \left[  (\mathcal{M}_{b22}+\mathcal{M}_{b11}) - \sqrt{  (\mathcal{M}_{b22}-\mathcal{M}_{b11})^2 + 4 \mathcal{M}_{b12}^2}  \right] \\
M_{H_2}^2 &=& \frac{1}{2} \left[  (\mathcal{M}_{b22}+\mathcal{M}_{b11})  + \sqrt{  (\mathcal{M}_{b22}-\mathcal{M}_{b11})^2 + 4 \mathcal{M}_{b12}^2}  \right],\nn
\label{eq:mass3}
\end{eqnarray}
%%%%%%%%%%%%%%%%%
%%%%%%%%%%%%%%%%%

with mixing angle $\beta$, where $ \beta = \frac{1}{2} \arctan((2\,\mathcal{M}_{b12})/(\mathcal{M}_{b22}-\mathcal{M}_{b11}))$. Similarly, the masses of the  $Z_2$-odd pseudoscalar fields are,
%%%%%%%%%%%%%%%%%
%%%%%%%%%%%%%%%%%
\begin{eqnarray}
M_{A_1}^2 &=& \frac{1}{2} \left[  (\mathcal{M}_{b22}+4 \, \mu_{S} v_s+\mathcal{M}_{b11}) - \sqrt{  (\mathcal{M}_{b22}+4 \, \mu_{S} v_s-\mathcal{M}_{b11})^2 + 4 \mathcal{M}_{b12}^2}  \right] \\
M_{A_2}^2 &=& \frac{1}{2} \left[  (\mathcal{M}_{b22}+4 \, \mu_{S} v_s+\mathcal{M}_{b11})  + \sqrt{  (\mathcal{M}_{b22}+4 \, \mu_{S} v_s-\mathcal{M}_{b11})^2 + 4 \mathcal{M}_{b12}^2}  \right],\nn
\label{eq:mass3}
\end{eqnarray}
%%%%%%%%%%%%%%%%%
%%%%%%%%%%%%%%%%%
where, 
\begin{eqnarray}
\mathcal{M}_{b11} &=& M_{H^\pm}^2+ \frac{1}{2} \lambda_4 v_1^2 ,~~\mathcal{M}_{b12} =\mathcal{M}_{b21}= \mu_{h} v_1 
\label{eq:massM3}\\
\mathcal{M}_{b22} &=&  \mu _{\phi _{2 s}}^2-2 \mu_{S} v_s+\frac{1}{2} \kappa _{3} v_s^2+\frac{1}{2} \kappa _{2} v_1^2 \nn
\end{eqnarray}
%%%%%%%%%%%%%%%%%
It is to be noted that the mixing angle in the CP-odd sector ($\delta$) is different than the mixing angle in the CP-even sector ($\beta$) due to the presence of an additional term $4 \, \mu_{S} v_s$ in the `22' component of the mass matrix. The other terms are same as in eqn.~\ref{eq:massM3}. The CP odd scalar mixing angle $\delta$ can be related with $\beta$ as,
%%%%%%%%%%%%%%%%%
\begin{equation}
\frac{1}{\tan 2 \delta} = \frac{1}{\tan 2 \beta} + \frac{4 \, \mu_{S} v_s}{ 2 \, \mathcal{M}_{b12} },
\end{equation}
%%%%%%%%%%%%%%%%%
%%%%%%%%%%%%%%%%%
\subsection{Gauge bosons}
The model has the SM gauge bosons and an additional neutral gauge boson because of the new $U(1)_X$ gauge symmetry. As the field strength tensor of an abelian gauge field is gauge invariant, we can write a cross term involving the field strength tensors of the two different $U(1)$ gauge fields \cite{delAguila:1995rb,Chankowski:2006jk}. The kinetic term representing all the gauge fields in the model can be written as, 
%%%%%%%%%%%%%%%%%
%%%%%%%%%%%%%%%%% 
\begin{eqnarray}
	\lag \supset - \frac{1}{4} G^{a,\mu\nu} G_{\mu\nu}^a - \frac{1}{4} W^{b,\mu\nu} W_{\mu\nu}^b -\frac{1}{4} B^{\mu\nu} B_{\mu\nu} - \frac{1}{4} C^{\mu\nu} C_{\mu\nu} + \frac{1}{2}\, \tilde{g}\, B^{\mu\nu} C_{\mu\nu} \, . 
\end{eqnarray}
Where, $G^{a,\mu\nu}$, $W^{b,\mu\nu}$, $B^{\mu\nu}$ and $C_{\mu\nu}$ are field strength tensor for $SU(3)$, $SU(2)$, $U(1)_Y$ and $U(1)_X$ gauge bosons respectively.
%%%%%%%%%%%%%%%%%
%%%%%%%%%%%%%%%%%

We redefine the two abelian gauge fields so that, the kinetic term will be diagonal.
%%%%%%%%%%%%%%%%%
%%%%%%%%%%%%%%%%%
\begin{eqnarray}
	B^\mu &=& B'^\mu + \dfrac{\tilde{g}}{\sqrt{1-\tilde{g}^2}} C'^\mu\, ,~ C^\mu = \dfrac{1}{\sqrt{1-\tilde{g}^2}} C'^\mu   \, . \label{eqn:kinmix}
\end{eqnarray}
%%%%%%%%%%%%%%%%%
%%%%%%%%%%%%%%%%%
After spontaneous symmetry breaking, the gauge bosons corresponding to the broken symmetry become massive. The mass term of the gauge bosons can be obtained from the kinetic part of the $Z_2$-even scalar sector as follows,
\begin{eqnarray}
	\mathcal{L}_{m,\rm gauge} &=&  \frac{1}{4}\begin{vmatrix} \begin{pmatrix}
			g_2 W_\mu^3 - g_1 B_\mu & g_2(W_\mu^1 - iW_\mu^2)\\
			g_2(W_\mu^1 + iW_\mu^2) & -g_2 W_\mu^3 - g_1 B_\mu
		\end{pmatrix} \begin{pmatrix}
			0 \\ \dfrac{v_1}{\sqrt 2} 
		\end{pmatrix}
	\end{vmatrix}^2 + \frac{1}{2} g_{x}^2 v_s^2 C_\mu C^\mu \nonumber\\
	&=& \frac{1}{4}g_2^2 v^2 W^+_\mu W^{-\mu} + \frac{1}{8}v_1^2 \left|\left(g_2W^3_\mu - g_1 B'_\mu - \frac{g_1\tilde{g}}{\sqrt{1-\tilde{g}^2}} C'_\mu\right)\right|^2 + \frac{ g_{x}^2 v_s^2}{2(1-\tilde{g}^2)}C'_\mu C'^\mu \nonumber  \, . 
\end{eqnarray}
%%%%%%%%%%%%%%%%%%%%%%%%%%%%%%%%%%%%%%%%%%%%%
We make the following redefinition of parameters to get a simple expression for the gauge boson mass matrix,
%%%%%%%%%%%%%%%%%%%%%%%%%%%%%%%%%%%%%%%%%%%%%
\begin{eqnarray}
	g'_x = \frac{g_1\tilde{g}}{\sqrt{1-\tilde{g}^2}}  \, ,  \qquad  \frac{g_{x}}{\sqrt{1-\tilde{g}^2}} \to g_x \, . \label{eqn:redef}
\end{eqnarray}
%%%%%%%%%%%%%%%%%%%%%%%%%%%%%%%%%%%%%%%%%%%%%
From the above Lagrangian, the mass matrix for the neutral gauge bosons, in the basis of $\left(B'_\mu\  W^3_\mu\ C'_\mu\right)^T$ is given by,
%%%%%%%%%%%%%%%%%%%%%%%%%%%%%%%%%%%%%%%%%%%%%
\begin{eqnarray}
	\!M^2 = \frac{1}{4}\!\begin{pmatrix}
		g_1^2 v^2   & -g_1 g_2 v^2 &  g_1 g'_x v^2 \\
		-g_1 g_2v^2 & g_2^2 v^2    & -g_2 g'_x v^2  \\
		g_1 g'_x v^2  & ~~-g_2 g'_x v^2  & ~{g'_x}^2 v^2  + 4 g_x^2 v_s^2
	\end{pmatrix}.
\end{eqnarray}
%%%%%%%%%%%%%%%%%%%%%%%%%%%%%%%%%%%%%%%%%%%%%

%%%%%%%%%%%%%%%%%%%%%%%%%%%%%%%%%%%%%%%%%%%%%
  For the diagonalization of the neutral gauge boson mass matrix, we first do an orthogonal transformation in $W^3_\mu$ and $B'_\mu$ plane.
%%%%%%%%%%%%%%%%%%%%%%%%%%%%%%%%%%%%%%%%%%%%%
	\begin{eqnarray}
		\begin{pmatrix} A_\mu \\ X_\mu \\ C'_\mu \end{pmatrix} =
		\begin{pmatrix} \cos\theta_W & \sin\theta_W & 0\\ -\sin\theta_W & \cos\theta_W & 0 \\ 0 & 0 & 1\end{pmatrix}
		\begin{pmatrix} B'_\mu \\ W^3_\mu \\ C'_\mu\end{pmatrix}  \, , 
	\end{eqnarray}
%%%%%%%%%%%%%%%%%%%%%%%%%%%%%%%%%%%%%%%%%%%%%
where $\tan\theta_W = \dfrac{g_1}{g_2}$. The mass term for neutral gauge bosons in the Lagrangian in the above rotated basis is given by,
%%%%%%%%%%%%%%%%%%%%%%%%%%%%%%%%%%%%%%%%%%%%%
	\begin{eqnarray}
		\lag_{m,\text{gauge}} = \frac{1}{8} v_1^2 \left(g_z X_\mu - g'_x C'_\mu\right)^2 + \frac{1}{2} g_x^2 v_s^2 C'_\mu C'^\mu  \, , 
	\end{eqnarray}
%%%%%%%%%%%%%%%%%%%%%%%%%%%%%%%%%%%%%%%%%%%%%
where $g_z = \sqrt{{g_1}^2 + g_2^2}$. From the above Lagrangian, we find that there is no mass term for the $A_{\mu}$ field, which is the photon field and $\theta_W$ is the Weinberg angle.  
	 The mass matrix in $X_\mu$ and $C'_\mu$ basis is given by,
%%%%%%%%%%%%%%%%%%%%%%%%%%%%%%%%%%%%%%%%%%%%%
	\begin{eqnarray}
		\widetilde{M}^2 = \frac{1}{4}\begin{pmatrix} g_z^2 v^2 & -g_z g'_x v^2\\ -g_z g'_x v^2 & ~~{g'_x}^2 v^2  + 4 g_x^2 v_s^2
		\end{pmatrix} \, . 
	\end{eqnarray}
%%%%%%%%%%%%%%%%%%%%%%%%%%%%%%%%%%%%%%%%%%%%%
	The above mass matrix can be diagonalized by the following orthogonal transformation,
%%%%%%%%%%%%%%%%%%%%%%%%%%%%%%%%%%%%%%%%%%%%%
	\begin{eqnarray}
		\begin{pmatrix} Z_\mu \\ Z'_\mu \end{pmatrix} = \begin{pmatrix}  \cos\theta' & \sin\theta' \\ -\sin\theta' & \cos\theta' \end{pmatrix}
		\begin{pmatrix} X_\mu \\ C'_\mu \end{pmatrix}  \, , 
	\end{eqnarray}
%%%%%%%%%%%%%%%%%%%%%%%%%%%%%%%%%%%%%%%%%%%%%
where $$\tan2\theta' =\frac{2g_z g'_x v^2 }{{g'_x}^2 v^2  + 4 g_x^2v_s^2 - g_z^2 v^2}~~~.$$ The final rotation matrix is given by,
%%%%%%%%%%%%%%%%%%%%%%%%%%%%%%%%%%%%%%%%%%%%%
\begin{eqnarray*}
		\!\!\!\!\begin{pmatrix} B_\mu \\ W_\mu^3 \\ C_\mu \end{pmatrix}
		\!&=&\! \begin{pmatrix} \cos\theta_W & ~-\sin\theta_W\cos\theta'  + \dfrac{g_x'}{g_1}\sin\theta' & ~\sin\theta_W \sin\theta' + \dfrac{g_x'}{g_1}\cos\theta'\\ \sin\theta_W & \cos\theta_W \cos\theta' & -\cos\theta_W \sin\theta' \\ 0 &  \sin\theta'\dfrac{\sqrt{g_1^2 + {g_x'}^2}}{g_1}  & \cos\theta'\dfrac{\sqrt{g_1^2 + {g_x'}^2}}{g_1} \end{pmatrix}\!\! \begin{pmatrix} A_\mu \\ Z_\mu \\ Z'_\mu\end{pmatrix}. 
\end{eqnarray*}
%%%%%%%%%%%%%%%%%%%%%%%%%%%%%%%%%%%%%%%%%%%%%
The masses of the physical gauge bosons are,
%%%%%%%%%%%%%%%%%%%%%%%%%%%%%%%%%%%%%%%%%%%%%
	\begin{eqnarray}
		M_{Z,Z'}^2 &=& \frac{1}{8}\Big[g_z^2 v^2 + g'_x v^2 + 4 g_x^2 v_s^2\Big]  \mp  \frac{1}{8}\sqrt{\Big({g'_x}^2 v^2 + g_z^2 v^2 + 4 g_x^2  v_s^2 \Big)^2 - 16\, g_z^2\, g_x^2 v^2\, v_s^2 }  \nonumber \, . 
	\end{eqnarray}
%%%%%%%%%%%%%%%%%%%%%%%%%%%%%%%%%%%%%%%%%%%%%

%%%%%%%%%%%%%%%%%%%%%%%%%%%%%%%%%%%%%%%%%%%%%
%%%%%%%%%%%%%%%%%
\subsection{Fermions}
As the SM fermions are neutral under the new gauge group, the Lagrangian for quarks remains unchanged while the SM leptons can have additional interaction with BSM fermions.  
The kinetic and interaction part for the newly added fermions is given by,
%%%%%%%%%%%%%%%%%
\begin{eqnarray}\label{Eq:lagfermion}
\lag & = & i\, \bar{\psi} \, \slashed{D}^{\psi}\,\psi + i\, \bar{N}_L \, \slashed{D}^{N}\,N_L  + i\, \bar{N}_R \, \slashed{D}^{N}\,N_R + M_L\, \bar{\psi} \,\psi +   M_N\, \bar{N}_L\,N_R  \nonumber \\
&&+ Y_{N_1}\, \bar{\Psi}_L\,\Phi_1\, N_R + Y_{N_2}\, \bar{\psi}_R \,\, \Phi_1\, N_L+ \, Y_{N_L}\, \bar{N}_L \,N^c_L \,S_{1} + \,  Y_{N_R}\, \bar{N}_R\, N^c_R \,S_{1}  \\
&&+ Y_{fd} \,\bar{\psi}_L \, \Phi_{2} e_R + Y_{fx}\,  \bar{\psi}_R\,S_{2} \,\ell_L +Y_{f\nu} \,\bar{\ell}_L \, \widetilde{\Phi}_{2} N_R + h.c.  \nonumber
\end{eqnarray}
%%%%%%%%%%%%%%%%%
When the SM Higgs gets VEV the singlet and doublet neutral BSM fermions mix with each other. Due to the presence of the Majorana mass term, there will be four Majorana fermions.  The mass matrix of neutral BSM fermions, in the basis of $\psi^0 = ({N_D}_L,{N_D}^c_R,N_L,N^c_R)$ is given by,
 
\begin{equation}
\mathcal{M}^n=\begin{pmatrix}
0 & M_L & 0 & \frac{Y_{N_1} v_1}{\sqrt{2}}\\
M_L & 0 & \frac{Y_{N_2} v_1}{\sqrt{2}} & 0 \\
0 &  \frac{Y_{N_2} v_1}{\sqrt{2}} & \sqrt{2}\, Y_{N_L}v_s & M_N \\
\frac{Y_{N_1} v_1}{\sqrt{2}} &0 & M_N & \sqrt{2}\, Y_{N_R} v_s
\end{pmatrix}
\end{equation}

This matrix can be diagonalized by the unitary ($4\times 4$) matrix $\mathcal{U}_N^{ij}$ with $i,j=1,2,3,4$. The new heavy fermionic state $\chi_i$ is given by,
%%%%%%%%%%%%%
%%%%%%%%%%%%%
\begin{equation}
\chi_i= \mathcal{U}_N^{ij} \, \psi^0_j
\label{eq:heavyFermmix}
\end{equation}
%%%%%%%%%%%%%
%%%%%%%%%%%%

For the choices of $Y_{N_1} = Y_{N_2}$ and $Y_{N_L} = Y_{N_R}$, we get the eigenvalues of the heavy neutrinos as,
%%%%%%%%%%%%%%%%%%%%
%%%%%%%%%%%%%%%%%%
\begin{eqnarray}
M_{N_{i=1,2}} &=\frac{ Y_{N_R} \, v_s }{\sqrt{2}} + \frac{({M_L}+ {M_N})}{2} \pm \frac{1}{2}\sqrt{2\, Y_{N_2}^2 v_1^2+(\sqrt{2}\,Y_{N_R}\, v_s-{M_L}+{M_N})^2}\nonumber\\
%%%%
M_{N_{i=3,4}} &=\frac{ Y_{N_R} \, v_s }{\sqrt{2}} - \frac{({M_L}+ {M_N})}{2} \pm \frac{1}{2}\sqrt{2\, Y_{N_2}^2 v_1^2+(\sqrt{2}\,Y_{N_R}\, v_s +{M_L}-{M_N})^2 }\nonumber
\end{eqnarray}
The $\Psi^{\pm}$ being a $Z_{2}$ odd charged fermion does not mix with SM leptons and is a Dirac fermion with mass $M_{L}$.
%%%%%%%%%%%%%%%%%%
%%%%%%%%%%%%%%%%%
\section{Constraints on this model}
\label{sec:3}
%%%%%%%%%%%%%%%%%
%%%%%%%%%%%%%%%%%
The parameter space of our model is subject to various relevant constraints derived from theoretical considerations and observational phenomena. These constraints include absolute vacuum stability, unitarity of the scattering matrix, and the experimental searches for DM. In addition to these constraints, recent measurements of the Higgs invisible decay width and signal strength at the LHC impose further limitations on the parameter space. 

By considering these additional constraints, we aim to refine our understanding of the viable parameter space for our model and further elucidate its implications for the observed phenomena. In the next section, we address these constraints.

\subsection{\bf Stability of the scalar potential}
%%%%%%%%%%%%%%%%%
%%%%%%%%%%%%%%%%%
The scalar potential should be bounded from the below in such a way that, even for large field values, no other negative infinity arises along any field direction. Following the paper~\cite{Chakrabortty:2013mha}, the absolute stability condition for the potential in eqn.~\ref{eq:potAll} are evaluated in terms of the quadratic couplings.
%%%%%%%%%%%%%%%%%
%%%%%%%%%%%%%%%%%
The most stringent conditions are,
\bea
&\lambda _1  \,\geq \,  0,\,\,\,  \lambda _2  \,\geq \,  0,\,\,\,  \lambda _{s1}  \,\geq \,  0, \,\,\, \lambda _{s1}  \,\geq \,  0,\nn\\
%%%%%%%%%%%%%%%%%
& \lambda _3  \,\geq \,   \, -2 \, \sqrt{\lambda _1 \lambda _2}, \,\,\,  \lambda _3+\lambda _4  \,\geq \,   \, -2 \, \sqrt{\lambda _1 \lambda _2} ,\nn\\
%%%%%%%%%%%%%%%%%
& \lambda _{11}  \,\geq \,   \, -2 \, \sqrt{\lambda _1 \lambda _{s2}}, \,\,\,  \lambda _{22}  \,\geq \,   \, -2 \, \sqrt{\lambda _2 \lambda _{s2}} ,\\
%%%%%%%%%%%%%%%%%
& \kappa _{s_2 \phi _1}  \,\geq \,   \, -2 \, \sqrt{\lambda _1 \lambda _{s2}}, \,\,\, \kappa _{s_1 \phi _2}  \,\geq \,   \, -2 \, \sqrt{\lambda _2 \lambda _{s2}} ,\nn\\
%%%%%%%%%%%%%%%%%
&  \kappa _{3}  \,\geq \,   \, -2 \, \sqrt{\lambda _{s2} \lambda _{s2}},\,\,\,  \kappa _{s_2 \phi _1}  \,\geq \,   \, -2 \, \sqrt{\lambda _1 \lambda _{s2}},
\label{eq:stabilityAbs1}
\nn
%%%%%%%%%%%%%%%%%
\eea
We also list the other stability conditions following~\cite{Chakrabortty:2013mha} which are shown in Appendix~\ref{sec:app}.
%%%%%%%%%%%%%%%%%
%%%%%%%%%%%%%%%%%
\subsection{\bf Unitarity Constraints}
%%%%%%%%%%%%%%%%%
%%%%%%%%%%%%%%%%%
The unitarity bound on the extended scalar sectors can be
calculated from the scalar-scalar, gauge boson-gauge boson, and
scalar-gauge boson scattering matrix (S-matrix). The technique was developed in Refs.~\cite{Lee:1977eg,Lee:1977yc}.
Using the Born approximation, the scattering cross-section for any process is given by,
\bea
\sigma = \frac{16 \pi}{s} \sum_{l=1}^\infty (2l+1) |a_l(s)|^2,
\label{eq:uni1}
\eea
%%%%%%%%%%%%%%%%%
where, $s$ is the Mandelstam variable and $a_l(s)$ are the partial wave coefficients corresponding
to specific angular momenta $l$. This leads to the following unitarity constraint: $Re[a_l(s)]<\frac{1}{2}$~\cite{Lee:1977eg,Lee:1977yc}.
At high energy, the dominant contribution to the amplitude $a_l(s)$ of the two-body scattering
processes comes from the diagram involving the quartic couplings. Far from the resonance, 
the contributions from the heavy fields mediated $s, t,u$-channels are negligibly small. The equivalence theorem implies that, for $s \gg 1$ TeV, the amplitude of scattering processes involving
longitudinal gauge bosons can be approximated by the scalar amplitude in which gauge bosons are replaced by their corresponding Goldstone bosons.
So, to test the unitarity of these model parameters, we construct the S-matrix, which consists of only the scalar quartic couplings.
Calculating the unitary bounds in the physical
basis of the scalar fields is also challenging. The S-matrix, expressed in terms of physical fields, can be rotated into an S-matrix for the non-physical fields (before the EWSB) by making a unitary transformation.

One can expand the potential in eqns.~\ref{eq:potAll} and get various quartic couplings in non-physical bases $\phi_{dr1},\, \phi_{dr2},\, \phi_{sr1},\, \phi_{sr2}, \, \phi_{di1}(\equiv G^0),\, \phi_{di2},\, \phi_{si1},\, \phi_{si2},\, \phi_{d1}^\pm(\equiv G^\pm),\, \phi_{d2}^\pm,$. They are given in appendix~\ref{sec:app}.
%%%%%%%%%%%%%%%%%%%%%%%%%%
%%%%%%%%%%%%%%%%%%%%%%%%%%
The full set of all scattering processes can be expressed as a $40 \times 40$ S-matrix. This matrix is composed of three submatrices of dimensions $16 \times 16$, $14 \times 14$, and $10 \times 10$  which have different initial and final states. All the matrices are shown in the appendix~\ref{sec:app}. The eigenvalues of these matrices are,
%%%%%%%%%%%%%%%%%%%%
%%%%%%%%%%%%%%%%%%%%
\begin{eqnarray*}
&& \lambda _{11}, \lambda _{22},  \lambda _3, 2 \lambda _1,   2 \lambda _2,2 \lambda_{s1},   2 \lambda_{s2},  \lambda _3\pm\lambda _4,  \lambda _3+2 \lambda_4
,\kappa _{s_2 \phi _1},      \kappa _{s_1 \phi _2},\kappa _{3},   \lambda_1+\lambda_2 \pm \sqrt{\left(\lambda_1-\lambda_2\right)^2+\lambda_4^2}, \\ 
&&  3 \lambda_1+3 \lambda_2 \pm \sqrt{9 \left(\lambda_1-\lambda_2\right)^2+\left(2 \lambda_3+\lambda_4\right)^2}, 2 \lambda_{s1}+2 \lambda_{s2}\pm \sqrt{\kappa_{s_1 s_2}^2+4 \left(\lambda_{s1}-\lambda_{s2}\right)^2}
\end{eqnarray*}
%%%%%%%%%%%%%%%%%%%%
%%%%%%%%%%%%%%%%%%%%
Throughout the analysis, we consistently apply the bounds imposed by the unitary constraints of scattering processes, which require the eigenvalues of the S-matrix to be less than or equal to $8\pi$.
%%%%%%%%%%%%%%%%%
%%%%%%%%%%%%%%%%%
\subsection{\bf Constraints from electroweak precision experiments}
%%%%%%%%%%%%%%%%%
%%%%%%%%%%%%%%%%%
New physics contributions beyond the electroweak scale have the potential to impact the electroweak precision bounds. These contributions primarily manifest through vacuum polarization corrections, which involve virtual particles in loop diagrams. These parameters were introduced and extensively studied in the works of Peskin and Takeuchi \cite{Peskin:1991sw}, Altarelli et al. \cite{Altarelli:1993bh}, and Baak et al. \cite{Baak:2014ora}. The electroweak precision experiments utilize parameters known as $S$, $T$, and $U$ to impose constraints on the new physics contributions. The $S$ parameter characterizes new physics contributions to the neutral weak current, while the $T$ parameter quantifies the contributions to the difference between the neutral and charged weak currents. On the other hand, the $U$ parameter is solely sensitive to the mass and width of the $W$-boson. The contributions from beyond the BSM physics can arise from the presence of additional scalar and fermion doublets.
One can add the new physics contributions to the SM as,
%%%%%%%%%%%%%%%%%
%%%%%%%%%%%%%%%%%
\begin{equation}
S=S_{SM}+\Delta S_{SD}+\Delta S_{FD}, ~T=T_{SM}+\Delta T_{SD}+\Delta T_{FD},~U=U_{SM}+\Delta U_{SD}+\Delta U_{FD},\nn
\end{equation}
%%%%%%%%%%%%%%%%%
%%%%%%%%%%%%%%%%%
where, $SD$ stands for scalar doublet whereas $FD$ refers fermionic doublet contributions. The contribution from the $Z_2$-odd scalar doublet are given by,
%%%%%%%%%%%%%%%%%
%%%%%%%%%%%%%%%%%
\bea
\Delta S_{SD}&=&\frac{1}{2\pi}\Big[\frac{1}{6}\ln\frac{M_{H_{d}}^2}{M_{H^{\pm}}^2}-\frac{5}{36}+\frac{M_{H_{d}}^2M_{A_{d}}^2}{3(M_{A_{d}}^2-M_{H_{d}}^2)^2}+\frac{M_A^4(M_{A_{d}}^2-3M_{H_{d}}^2)}{6(M_{A_{d}}^2-M_{H_{d}}^2)^3}\ln \frac{M_{A_{d}}^2}{M_{H_{d}}^2} \Big],\\
\Delta T_{SD}&=&\frac{1}{32\pi^2\alpha v^2}\Big[F(M^2_{H^{\pm}},M^2_{H_{d}})+F(M^2_{H^{\pm}},M_{A_{d}}^2)-F(M_{A_{d}}^2,M^2_{H_{d}})\Big].
\eea
%%%%%%%%%%%%%%%%%
%%%%%%%%%%%%%%%%%
and $\Delta U_{SD}$ is neglected in this case due to small mass differences $M_{A_1}-M_{H_1}$ and $M_{H^\pm}-M_{H_1}$ of the inert fields~\cite{Arhrib:2012ia,Barbieri:2006dq}. The loop function
\begin{align*}
F(x,y) =
\begin{cases}
\frac{x+y}{2}-\frac{xy}{x-y}\ln (\frac{x}{y}) & {x \neq y}\\
0 & {x= y}
\end{cases}\,.  
\end{align*}
% $$F=\frac{x+y}{2}-\frac{xy}{x-y}\ln (\frac{x}{y}),$$ 
%for $x\ne y$ otherwise $F= 0$. 
%%%%%%%%%%%%%%%%%
%%%%%%%%%%%%%%%%%
The fermionic contributions to S and T while considering the electric charge neutral BSM singlet and doublet fermions mixing is given by,
\begin{eqnarray}
\pi \Delta S_{FD}  &=& \frac{1}{6}-b_2(M_L,M_L,0)+\sum_{j,k=1}^{4}\left(\left|\mathcal{O}_{1j}\right|^2\left|\mathcal{O}_{1k}\right|^2+\left|\mathcal{O}_{1j}\right|^2\left|\mathcal{O}_{1k}\right|^2\right)b_2(M_{N_j},M_{N_k},0) \nonumber \\
 &&+\sum_{j,k=1}^{4} \Re\left(\mathcal{O}_{1j}\mathcal{O}^*_{1k}\mathcal{O}_{2j}\mathcal{O}^*_{2k}\right)f\left(M_{N_j},M_{N_k}\right)
\end{eqnarray}
%%%%%%%%%%%%%%%%%
%%%%%%%%%%%%%%%%%
\begin{eqnarray}
4\pi s_{w} c_{w}M^2_{Z}\Delta T_{FD}&=&-2\sum_{k=1}^{4}\left(\left|\mathcal{O}_{1k}\right|^2+\left|\mathcal{O}_{2k}\right|^2\right)b_3(M_{N_k},M_L,0)+2\sum_{k=1}^{4}\Re\left(\mathcal{O}_{1k}\mathcal{O}_{2k}\right)b_0(M_L,M_{N_k},0)\nonumber\\
&&-\sum_{k=1}^{4}\Re\left(\mathcal{O}_{1j}\mathcal{O}^*_{1k}\mathcal{O}_{2j}\mathcal{O}^*_{2k}\right)M_{N_j}M_{N_k}b_{0}\left(M_{N_j},M_{N_k},0\right)\nonumber\\
&&+\sum_{j,k=1}^{4}\left(\left|\mathcal{O}_{1j}\right|^2\left|\mathcal{O}_{1k}\right|^2+\left|\mathcal{O}_{1j}\right|^2\left|\mathcal{O}_{1k}\right|^2\right)b_3(M_{N_j},M_{N_k},0),
\end{eqnarray}
%%%%%%%%%%%%%%%%%
%%%%%%%%%%%%%%%%%
where $b_0(M_1,M_2,q^2) = \int_{0}^{1}  \log\left(\frac{\Delta}{\Lambda^2}\right) \,dx, 
~~
b_1(M_1,M_2,q^2) = \int_{0}^{1} x \log\left(\frac{\Delta}{\Lambda^2}\right) \,dx $
%%%%%%%%%%%%%%%%%
%%%%%%%%%%%%%%%%%
$
b_2(M_1,M_2,q^2) =\int_{0}^{1} x(1-x) \log\left(\frac{\Delta}{\Lambda^2}\right) dx,
~
b_3(M_1,M_2,0) = \frac{M^2_2b_1(M_1,M_2,0)+M^2_1b_1(M_1,M_2,0)}{2}
$, $\Delta = M_2^2 x + M_{1}^2 (1-x) -x(1-x)q^2$, and
%%%%%%%%%%%%%%%%%
%%%%%%%%%%%%%%%%%
\begin{eqnarray}
f(M_1,M_2)&=&M_{1}M_{2}\frac{M^4_{2}-M^4_{2}+2M^2_{1}M^2_{1}\log\left(\frac{M^2_{1}}{\Lambda^2}\right)-2M^2_{1}M^2_{1}\log\left(\frac{M^2_{2}}{\Lambda^2}\right)}{2\left(M^2_{1}-M^2_{2}\right)^3}\nn
\end{eqnarray}
%%%%%%%%%%%%%%%%%
%%%%%%%%%%%%%%%%%
In the absence of mixing between the odd neutral fermions, the contributions from the new vectro like fermions $\Delta S_{FD}=0,\Delta T_{FD}=0  $ \cite{Joglekar:2012vc}.
To impose constraints on the new parameters, we utilize the next-to-next-to-leading order (NNLO) global electroweak fit results obtained by the Gfitter group~\cite{Haller:2018nnx}. The results from their study provide significant limitations on the parameter space. The key constraints extracted from the fit~\cite{Haller:2018nnx} at $95\%$ C.L. are, 
\begin{align*}
 \Delta S_{BSM}<0.04\pm0.11, && T_{BSM}<0.09\pm0.14, && \Delta U_{BSM}<-0.02\pm0.11 \,\,\,\, . 
\end{align*}

%%%%%%%%%%%%%%%%%%%%%%%
%%%%%%%%%%%%%%%%%%%%%%%
\subsection{\bf LHC signal strength Constraints}
%%%%%%%%%%%%%%%%%%%%%%%
%%%%%%%%%%%%%%%%%%%%%%%
In the presence of an additional $Z_2$-even scalar singlet ($S_{1}$), the tree-level couplings of a Higgs-like scalar $h$ to SM fermions and gauge bosons undergo modifications due to the mixing. It leads to changes in the coupling strengths.
Furthermore, the Higgs $h$ loop-induced decays are also affected by the extra charged scalar ($H^\pm$) and fermion ($E_1^\pm$). 
The combined effects result in altered signal strength for the Higgs-like scalar $h$ compared to its standard behavior. For the production cross section $\sigma(pp\to h)$ and decay width $h\to \chi $, we have,
%%%%%%%%%%%%%%%%%%%%%%%
%%%%%%%%%%%%%%%%%%%%%%%
\begin{equation}
	\kappa_{p}^2=\frac{\sigma(pp\to h)_{BSM}}{\sigma(pp\to h)_{SM}},\,\,\,\,\,\,\kappa_\chi^2=\frac{\Gamma(h\to \chi)_{BSM}}{\Gamma(h\to \chi)_{SM}},
\end{equation} 
%%%%%%%%%%%%%%%%%%%%%%%
%%%%%%%%%%%%%%%%%%%%%%%
where $\chi$ stands for the final state $b\bar{b},\,\tau^-\tau^+,\,W^-W^+,\,ZZ,\,Z\gamma$ and $\gamma\gamma$. It is possible to impose significant constraints on the scalar mixing from these channels . %even in the absence of loop-induced decay.
The most stringent constraints arise from the $Z\gamma$ and $\gamma\gamma$ channels. The safest bound on the mixing angle is $\cos\alpha >0.995$~\cite{Arroyo-Urena:2019fyd,Arroyo-Urena:2022oft}.
The loop-induced processes, mainly diphoton Higgs signal strength, put the most stringent constraints. The Higgs to diphoton strength, under the narrow width approximation, can be expressed as follows,
%%%%%%%%%%%%%%%%%
%%%%%%%%%%%%%%%%%
\begin{equation}
\mu_{\gamma\gamma}\approx \kappa_{\gamma\gamma}^2=\, \frac{\sigma(gg\rightarrow h)_{BSM}}{\sigma(gg\rightarrow h)_{SM}} \, \frac{Br(h\rightarrow\gamma\gamma)_{BSM}}{Br(h\rightarrow\gamma\gamma)_{SM}}.
\end{equation}
%%%%%%%%%%%%%%%%%
%%%%%%%%%%%%%%%%%
The production of the Higgs boson through gluons is much larger than the fermions. We can write $\frac{\sigma(gg\rightarrow h)_{BSM}}{\sigma(gg\rightarrow h)_{SM}}=\cos^2\alpha\approx 1$, hence,
%%%%%%%%%%%%%%%%%
%%%%%%%%%%%%%%%%%
\begin{equation}
\mu_{\gamma\gamma}=  \frac{\Gamma_{SM}^{Total}}{\Gamma_{BSM}^{Total}}\, \frac{\Gamma(h\rightarrow\gamma\gamma)_{BSM}}{\Gamma(h\rightarrow\gamma\gamma)_{SM}}.
\end{equation}
%%%%%%%%%%%%%%%%%
%%%%%%%%%%%%%%%%%
If the mass of the extra BSM particles is lighter than $\frac{M_h}{2}$, then it will contribute to the invisible decay width of the Higgs boson. We find that such an invisible allowed branching ratio is much less than $10\%$~\cite{CMS:2022qva,ATLAS:2022tnm}. Hence, the ratio $\frac{\Gamma_{SM}^{Total}}{\Gamma_{BSM}^{Total}}$ could vary $0.9-1$. However for new particle's masses $\frac{M_h}{2}>$, one can write $\frac{\Gamma_{SM}^{Total}}{\Gamma_{BSM}^{Total}}\approx \frac{1}{\cos^2\alpha}\approx 1$, hence,
%%%%%%%%%%%%%%%%%
%%%%%%%%%%%%%%%%%
$
\mu_{\gamma\gamma}= \frac{\Gamma(h\rightarrow\gamma\gamma)_{BSM}}{\Gamma(h\rightarrow\gamma\gamma)_{SM}}.
$
%%%%%%%%%%%%%%%%%
%%%%%%%%%%%%%%%%%
At one-loop level, the physical charged Higgs $H^{\pm}$ and fermion $E_1^\pm$ add extra contribution to the decay width $\Gamma(h\rightarrow\gamma\gamma)_{BSM}$ as,
%%%%%%%%%%%%%%%%%%
%%%%%%%%%%%%%%%%%%
\begin{equation}
\Gamma(h\rightarrow \gamma\gamma)_{BSM}=\frac{\alpha^2M_h^3}{256\pi^3v_1^2}\Big|Q^2_{H^{\pm}} \, \frac{v_1 \, g_{hH^+H^-}}{2M^2_{H^{\pm}}} \, F_0(\tau_{H^{\pm}})
%%%%%%
+Q^2_{E_1^\pm} \, g_{hE_1^+E_1^-} \, F_{1/2}(\tau_{E_1^\pm})+C_{SM}\Big|,
\end{equation}
%%%%%%%%%%%%%%%%%%
%%%%%%%%%%%%%%%%%%
where $C_{SM}$ is the SM particle's contributions, $$ C_{SM}=\sum_f N_f^c Q_f^2 y_f F_{1/2}(\tau_f)+y_W F_1(\tau_W) $$ with $\tau_x=\frac{M_h^2}{M_x^2}$. $Q$ stands for the electric charge of particles, and $N_f^c$ denotes the color factor. 
The loop functions $F_{(0,1/2,1)}(\tau)$ can be found in Ref.~\cite{Djouadi:2005gj}. The coupling strengths are, $$g_{hH^+H^-}=-(\cos\alpha \, \lambda_3 \, v_1)+\kappa_{\phi_1 s_2} \, \sin\alpha \, v_s ,\, g_{hE_1^+E_1^-}=0$$ at tree level.
%%%%%%%%%%%%%%%%%
%%%%%%%%%%%%%%%%%
\subsection{$Z'$ searches at LHC}
The presence of $Z'$ in the particle content which mixes with SM gauge boson Z will lead to the constraints on, $Z-Z'$ mixing angle, which should be less than $10^{-3}$, so that well-measured $Z$ decay width to various SM modes, does not modify much \cite{ParticleDataGroup:2020ssz}. Drell-Yan dilepton ($e,\mu$) resonance of $Z'$ constraint from LHC~\cite{ATLAS:2017fih, CMS:2019buh, ACCOMANDO:2013zz} must also be considered. In this model the SM particle contents are neutral under the new gauge symmetry and they only couple to $Z'$ through the GKM. The above collider bounds can be evaded by keeping a small GKM, so that the production cross section of dilepton pairs via $Z'$ resonance at LHC lies below the current experimental limit.
%%%%%%%%%%%%%%%%%
\subsection{\bf Lepton flavour violation constraints}
%%%%%%%%%%%%%%%%%
%%%%%%%%%%%%%%%%%
\begin{figure}[h!]
	\begin{center}
		
		\includegraphics[width=0.47\textwidth,height=0.28\textwidth]{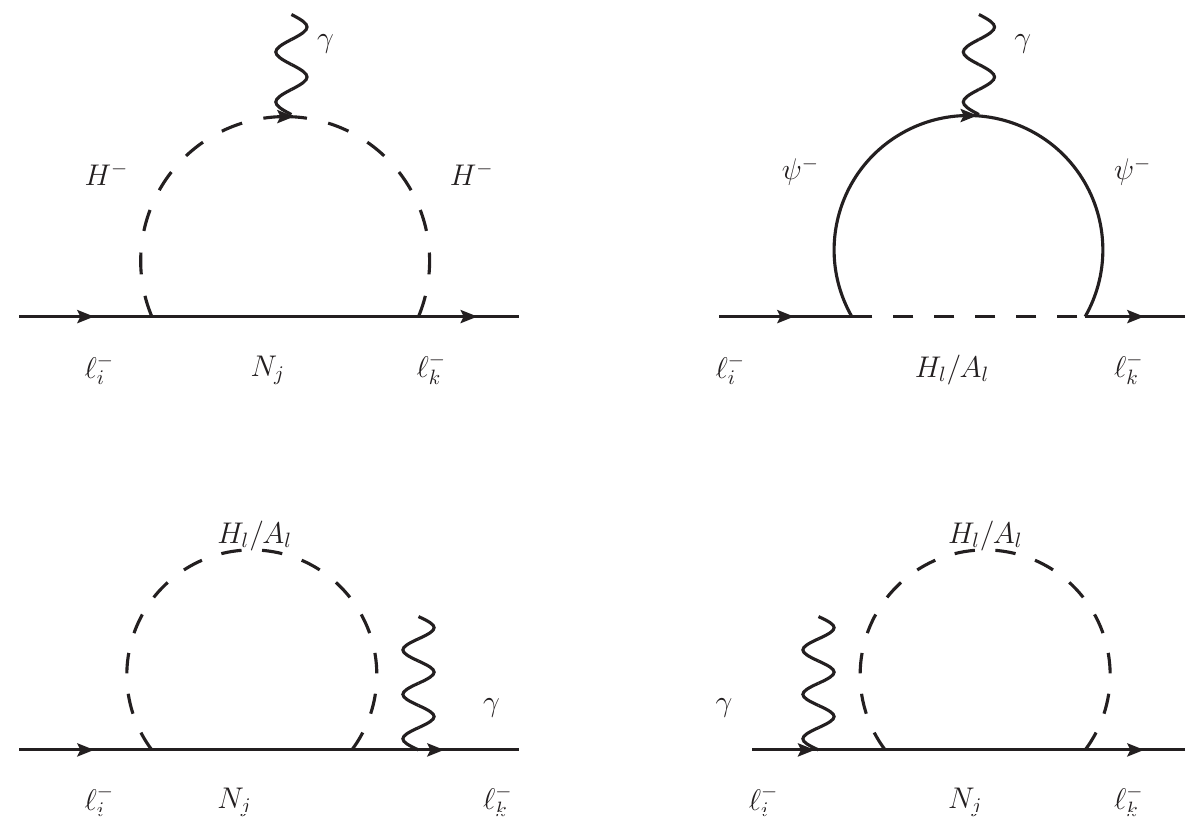}
		\caption{  Feynman diagrams contributing to $l^{-}_{i} \to l^{-}_{k} \gamma \,(i\neq k)$ process at one loop order.}
		\label{fig:lfv}
	\end{center}
\end{figure}
%%%%%%%%%%%%%%%%%
%%%%%%%%%%%%%%%%%
In the SM, lepton flavors (electron, muon, and tau) are conserved, meaning that charged leptons only interact with their corresponding neutrinos and do not change flavors. However, neutrino oscillations data predict the possibility of lepton flavor violation (LFV), where a neutrino of one flavor can transform into another. LFV processes can also occur through the exchange of new particles.
In this model, the presence of dark sector scalars and fermions introduces new interactions that couple to all the SM-charged leptons, potentially leading to lepton flavor-violating processes, e.g., $\mu \to e \gamma$, $\tau \to \mu \gamma$, $\mu \to 3\,e$, etc. The magnetic momentum of SM-charged leptons is also affected by these couplings. The possible processes are shown in Fig.~\ref{fig:lfv}.

The $Y_{X_i,d_i,\nu_i} \sim 0.1$ with $M_{E_i}=\mathcal{O}(1)$ TeV lead to large LFV which is excluded by present data BR$(\mu \rightarrow e\gamma) $ $< 4.2 \times 10^{−13}$~\cite{Baldini:2018nnn}, BR$(\tau \rightarrow e\gamma) $ $< 3.3 \times 10^{−8}$ and BR$(\tau \rightarrow \mu \gamma) $ $< 4.4 \times 10^{−8}$ \cite{BaBar:2009hkt}.
However, by utilizing the first or second-generation Yukawa couplings ($i=1$ or $2$), with values below $10^{-3}$, while keeping other couplings at the order of magnitude $\mathcal{O}(1)$, one can effectively circumvent the constraints imposed by LFV. We also find that large Yukawa couplings ($Y_{X_i,d_i,\nu_i} > 4\pi,~i=1,2,3$ with $M_{E_i}=\mathcal{O}(1)$ TeV) are required to explain the current muon anomalous magnetic moment $\delta a_\mu = (2.49\pm 0.48)\times 10^{-9}$~\cite{Muong-2:2023cdq, Muong-2:2006rrc, Muong-2:2021ojo}. Hence muon anomalous magnetic moment can not be explained within the perturbative regime. Although the measurement of the muon  $g-2$  shows a $5.1 \sigma$ deviation \cite{Muong-2:2023cdq,Venanzoni:2023mbe} from the Standard Model prediction \cite{Aoyama:2020ynm}, there is tension in determining the hadronic vacuum polarization contribution due to discrepancies between lattice QCD calculations \cite{Kurz:2014wya,Davier:2019can,Davier:2017zfy,Keshavarzi:2018mgv,Colangelo:2018mtw,Hoferichter:2019mqg,Keshavarzi:2019abf,Blum:2019ugy,Borsanyi:2020mff,Ce:2022kxy,ExtendedTwistedMass:2022jpw,Chao:2022ycy} and experimental data from $e^- e^+ \to \pi^- \pi^+$ processes \cite{CMD-3:2023alj}. If we consider the hadronic vacuum polarization contribution from the lattice QCD computation the deviation of $(g-2)_\mu$ from SM prediction significantly decreased. Therefore, we will not address the muon $g-2$ puzzle in this work.     
%%%%%%%%%%%%%%%%%%
%%%%%%%%%%%%%%%%%%
\subsection{\bf Constraints from the Neutrino low energy variables}
\begin{figure}[h!]
	\begin{center}
		
			\includegraphics[width=0.47\textwidth,height=0.15\textwidth]{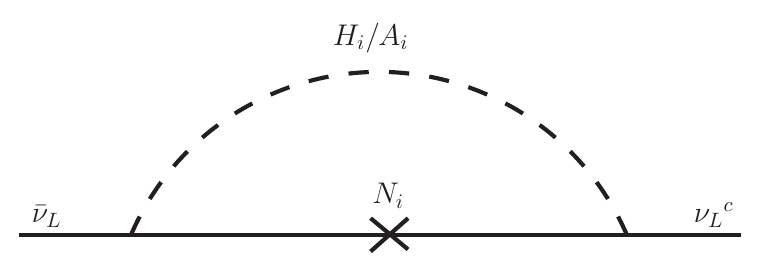}
		\caption{  Neutrino masses diagram at one loop when $H_{j}/A_{j}$ $(j = d,s)$ and $N_{i}$ $(i = 1,2,3,4)$.}
		\label{fig:numass}
	\end{center}
\end{figure}

%%%%%%%%%%%%%%%%%%
%%%%%%%%%%%%%%%%%%
The Yukawa Lagrangian given in eqn.~\ref{Eq:lagfermion} indicates that the mass of SM neutrinos cannot be obtained at the tree level. Instead, the neutrino mass arises through loop diagrams involving the $Z_2$ odd neutral fermions and scalars (see Fig.~\ref{fig:numass}). 
%%%%%%%%%%%%%%%%%%%%%%%
%%%%%%%%%%%%%%%%%%%%%%%
%%%%%%%%%%%%%%%%%%%%%%%
%%%%%%%%%%%%%%%%%%%%%%%
%%%%%%%%%%%%%%%%%%%%%%%
\begin{table}[hb!]
	\begin{tabular}{|c|c|c|c|c|}
		\hline
		 $\Delta m_{21}^2\, (10^{-5}\,\text{eV}^2)$ & $\Delta m_{31}^2\, (10^{-3}\,\text{eV}^2)$ & $\theta_{12}$&$\theta_{13}$& $\theta_{23}$ \\
		\hline
		$6.82-8.04$ & $2.430-2.593$ & $31.27^{\circ}-35.87^{\circ}$ & $8.25^{\circ}-8.98 ^{\circ}$& $39.7^{\circ}-50.9^{\circ}$ \\
		\hline
	\end{tabular}
\caption{ Neutrino oscillation data within $3\sigma$ range.  The CP phase $\delta$ and Majorana phase $\alpha$ are still unconstrained from the experiment~\cite{Esteban:2020cvm, Gonzalez-Garcia:2021dve}.}
\label{tabnu:1}
\end{table}
We are able to generate the neutrino oscillation observables~\cite{Esteban:2020cvm, Gonzalez-Garcia:2021dve}, shown in Table \ref{tabnu:1}. The phases $\alpha$ and $\delta$ are still unconstrained from the experiment. These experimental data restrict the parameters involved in generating neutrino mass and mixing angles. We discuss the detailed parameters in the numerical section.
%%%%%%%%%%%%%%%%%
%%%%%%%%%%%%%%%%%
\subsection{\bf Constraints from dark matter} 
%%%%%%%%%%%%%%%%%%
The combined results from the Wilkinson Microwave Anisotropy Probe (WMAP) satellite, along with other cosmological measurements, have provided constraints on the DM relic density as $\Omega h^2=0.1198\pm 0.0012$~\cite{Aghanim:2018eyx}.
Recent direct-detection experiments, such as Xenon-1T~\cite{Aprile:2018dbl}, LUX-ZEPLIN (LZ)~\cite{LZ:2022ufs}, and PANDAX~\cite{PandaX:2016pdl}, have searched for the presence of weakly interacting massive particles (WIMPs) as potential DM candidates and put stringent bound on DM-nucleon scattering cross-secton. We consider a scalar DM scenario in this model and therefore we shall consider the constraints on the spin-independent (SI) cross-section for direct detection in our analysis.
In this model, the contribution to SI cross-section comes from only the Higgs (light and heavy CP-even) mediated channel. For the DM mass of $\mathcal{O}(10-1000)$\,GeV the stringent bound comes from recent XENON-1T~\cite{Aprile:2018dbl} experiment with $\mathcal{O} (10^{-46} - 10^{-45}) \,{\rm cm}^2$. The null results from these experiments and data from collider experiments that include the invisible Higgs decay (into DM) and the $Z$ boson decay width, put constraints on the couplings 
and mass of the DM.

In this work, we first use {\tt FeynRules}~\cite{Alloul:2013bka} to build the model and then take the help of {\tt micrOMEGAs}~\cite{Belanger:2018mqt} to compute the relic density of the scalar DM.  We also verified these using {\tt SARAH-4.14.3}~\cite{Staub:2013tta,Staub:2015kfa} including {\tt SPheno-4.0.3}~\cite{Porod:2011nf} mass spectrum in {\tt micrOMEGAs}. We present a detailed discussion on DM in the next numerical analysis section. 

%%%%%%%%%%%%%%%%%
\section{Numerical Analysis}
\label{sec:4}
%%%%%%%%%%%%%%%%%
%%%%%%%%%%%%%%%%%
In this section, we present our numerical results for the neutrino sector and DM phenomenology that will provide an insight 
%investigate and understand this model's complex physical phenomena. It enables us to simulate and analyze 
on the characteristics of these elusive particles and their interaction properties.
%%%%%%%%%%%%%%%%%
%%%%%%%%%%%%%%%%%
\subsection{\bf Light neutrino masses and mixing angles}
%%%%%%%%%%%%%%%%%
%%%%%%%%%%%%%%%%%
The nature of the neutrino mass term generated at one loop (see Fig.~\ref{fig:numass}) from the  Lagrangian (eqn.~\ref{Eq:lagfermion}) is Majorana, and its magnitude is determined by the Yukawa couplings, namely $Y_{fx}$ , $Y_{f\nu}$ and the masses of $Z_2$ odd charge neutral scalars and fermions. 
%Additionally, the masses and mixing of the particles involved in the loop diagram (see Fig.~\ref{fig:numass}) contribute to the overall neutrino mass. 
The detailed expression for neutrino mass matrix at one loop level is given by \cite{Ma:2006km},
%%%%%%%%%%%%%%%%%%%%%%%%%%
%%%%%%%%%%%%%%%%%%%%%%%
\begin{table}[b]
	%\centering
	\begin{center}\scalebox{1.0}{
			\begin{tabular}{|c|c|c|}
				\hline &  {\tt BP1} &  {\tt BP2}   \\ \hline
				%\hline
				$M_{N} = \frac{1}{2}\, Y_{N_1}\,v_s$ (GeV) & 500      & 500                   \\
				$M_{L}$ (GeV) & 1000      & 1000                  \\
				$M_{H_{1}}$ (GeV)                & 460.0012  & 479             \\ 
				$M_{H_{2}}$ (GeV) & 400     &  400                 \\
				$M_{A_{1}}$ (GeV)  & 460    &  480         \\    
				$M_{A_{2}}$ (GeV) &  400.0012 &  580      \\  
				$\cos{\beta}$ & 0.710 & 0.910    \\\hline
				$Y_{f\nu_{i}}$ &$ \begin{matrix}
				2.87061\times 10^{-4} \\
				(3.22832 - 0.457563\, i)\times 10^{-3} \\
				1.8424\times 10^{-3}
				\end{matrix}$ &  $ \begin{matrix}
				9.5653 \times 10^{-7} \\
				(1.277+0.0884 i) \times 10^{-5} \\
				3.751 \times 10^{-6}
				\end{matrix}$ \\ \hline
				$Y_{fx_{i}}$ & $ \begin{matrix}
				3.366\times 10^{-4} \\
				(3.0277 - 0.462412\, i)\times 10^{-3} \\
				(1.61062 - 0.0321793\, i)\times 10^{-3}
				\end{matrix}$ & $ \begin{matrix}
				1.8174 \times 10^{-5} \\
				(-4.5597+0.919 i) \times 10^{-6} \\
				(-1.909+1.533 i) \times 10^{-6}
				\end{matrix}$ \\\hline
				$\Delta m_{21}^2\, (10^{-5}\, \text{eV}^2)$ &7.62246  & 7.4539  \\ 
				$\Delta m_{31}^2\, (10^{-3}\, \text{eV}^2)$ &2.44989  & 2.5468  \\ 
				$\theta_{12} $ & $33.4605^{\circ}$  & $33.4605^{\circ}$  \\ 
				$\theta_{13}$ &$8.6192 ^{\circ}$  & $8.6192 ^{\circ}$  \\ 
				$\theta_{23}$ & $42.1304^{\circ}$ & $42.1304^{\circ}$  \\ 
				$\alpha$ & $80^{\circ}$  & $75^{\circ}$  \\ 
				$\delta$ & $200^{\circ}$ & $200^{\circ}$  \\\hline 
				\hline
		\end{tabular}}
	\end{center}
	\caption{ Two {\tt BP}s satisfying the neutrino oscillation data. In {\tt BP1}, mass difference between CP even and CP odd scalars are small hence Yukawa of $\mathcal{O}(10^{-4} - 10^{-3})$ are required to satisfy neutrino observables, while for {\tt BP2}, mass gap between CP even and CP odd scalars are larger therefore, smaller Yukawa coupling need to satisfy neutrino oscilation data.   }
	\label{tab:nubps}
\end{table}
\begin{equation}
  (\mathcal{M}_\nu)_{ij} = \sum _k \sum _l \frac{M^2_{H_l} M_k \, h_{{ik}}^l
  	{h_{{jk}}^l} \, \ln\left(\frac{M^2_{H_l}}{M^2_k}\right)}{16 \pi ^2
  	\left(M^2_{H_l}-M^2_k\right)} - \sum _k \sum _l \frac{M^2_{A_l} M_k \, a_{{ik}}^l
  	{a_{{jk}}^l} \, \ln\left(\frac{M^2_{A_l}}{M^2_k}\right)}{16 \pi ^2
  	\left(M^2_{A_l}-M^2_k\right)}
  	\label{eq:nuMass}
  \end{equation} 
%%%%%%%%%%%%%%%%%%%%%%%
where $h_{{ik}}^l$ and $a_{{ik}}^l$ are coupling of $Z_2$ odd CP-even and odd scalar fields with BSM neutral fermions and neutrinos respectively, are given by, 
%%%%%%%%%%%%%%%%
\begin{equation}
h_{{ik}}^l = \frac{1}{\sqrt{2}}(\mathcal{U}_{N_{k2}}Z_{H_{l2}}Y_{fx_i} - \mathcal{U}_{N_{k4}}Z_{H_{l1}}Y_{f\nu_i}), \, \, \, \,
a_{{ik}}^l = \frac{1}{\sqrt{2}}(\mathcal{U}_{N_{k2}}Z_{A_{l2}}Y_{fx_i} - \mathcal{U}_{N_{k4}}Z_{A_{l1}}Y_{f\nu_i}). \nn
\end{equation}
%%%%%%%%%%%%%%%%%
The scalar mixing parameters $Z_{H}$ and $Z_{A}$ are given in sec.~\ref{sec:scalar}, while the fermionic mixing matrix $\mathcal{U}_{N}$ is shown in eqn.~\ref{eq:heavyFermmix}. 
Note that to achieve a non-zero neutrino mass at the one-loop level, it is crucial to have a mass 
difference between the scalar ($H_l$) and pseudoscalar ($A_l$) particles. This mass difference is 
generated by the trilinear scalar term $\mu_S$ shown in eqn.~\ref{eq:potAll1}. 
To get the neutrino mass eigenvalues, we have to diagonalize the above mass matrix using the well-established PMNS matrix as $ m_{diag}=U_{\rm PMNS}^\dagger M_\nu U_{\rm PMNS}$. It is also essential to ensure that the choice of Yukawa couplings and other parameters involved in light neutrino masses are consistent with the current neutrino oscillation data. 
The neutrino masses depend on various parameters from the scalar as well as the fermionic sectors. A small change in one of the sectors can lead to altering the neutrino masses and mixing angles. We present two benchmark points in Table \ref{tab:nubps}, that give the correct neutrino mass squared differences and mixing angles.

In {\tt BP1}, the mass difference between the CP-even and CP-odd sectors of the $Z_2$-odd scalar particle is extremely small ($\mathcal{O}({\rm MeV})$), which helps us get the observed neutrino data for $\mathcal{O}(0.1-10^{-3})$ Yukawa couplings. We will see that such choices of nearly degenerate scalar masses leads to a considerable increase in co-annihilation processes in the DM sector, ultimately resulting in a reduced relic density of DM. We also find that even with
a significantly larger mass gap between the scalar particles, as shown in {\tt BP2},
one can still generate all the correct neutrino mass differences and mixing angles, provided the relevant Yukawa couplings are chosen much smaller.
%%%%%%%%%%%%%%%%%
\subsection{Dark Matter Analysis }
\label{dm1}
%%%%%%%%%%%%%%%%%
In our analysis, we assume that $H_1$ is the lightest neutral component among the $Z_2$-odd particles which is a potential DM candidate and can give observed relic density of the Universe.
Depending on the mixing angle $\beta$ in the $Z_2$ odd scalar sector, the DM candidate $H_1$ can belong to the doublet, singlet 
or be a linear combination of both. 
The parameter space of the model determines the correct DM relic density, which depends on the annihilation, co-annihilation, and decay processes. 
%%%%%%%%%%%%%%%%%
%%%%%%%%%%%%%%%%%
\begin{figure}[h!]
	\begin{center}
%		\subfigure[]{
			\includegraphics[scale=0.5]{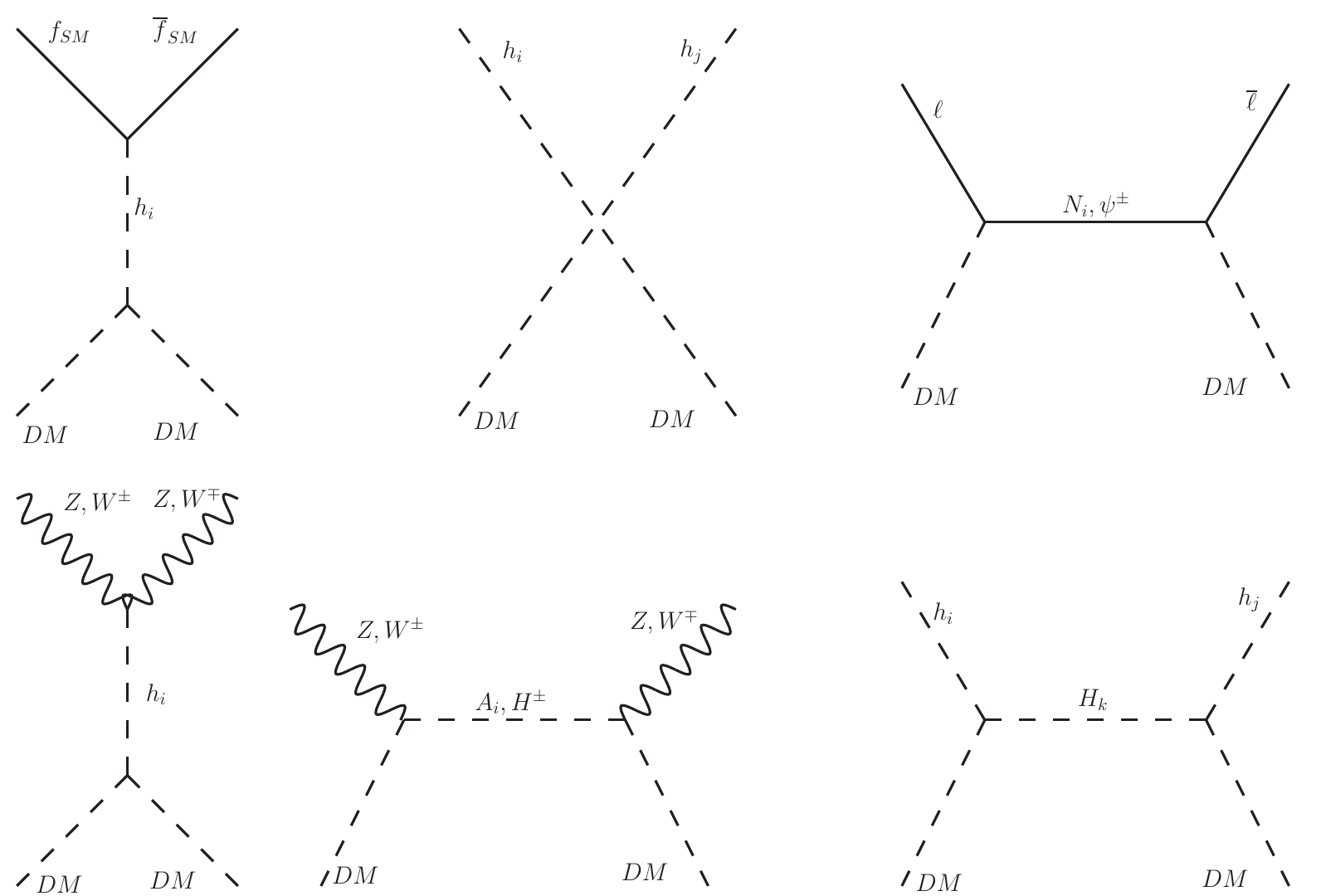}
%			}
		\caption{  The DM annihilation diagrams give the relic density. The vertical axis represents time.}
		\label{fig:DarkAn}
	\end{center}
\end{figure}
%%%%%%%%%%%%%%%%%
%%%%%%%%%%%%%%%%%
\begin{figure}[h!]
	\begin{center}
%		\subfigure[]{
			\includegraphics[scale=0.5]{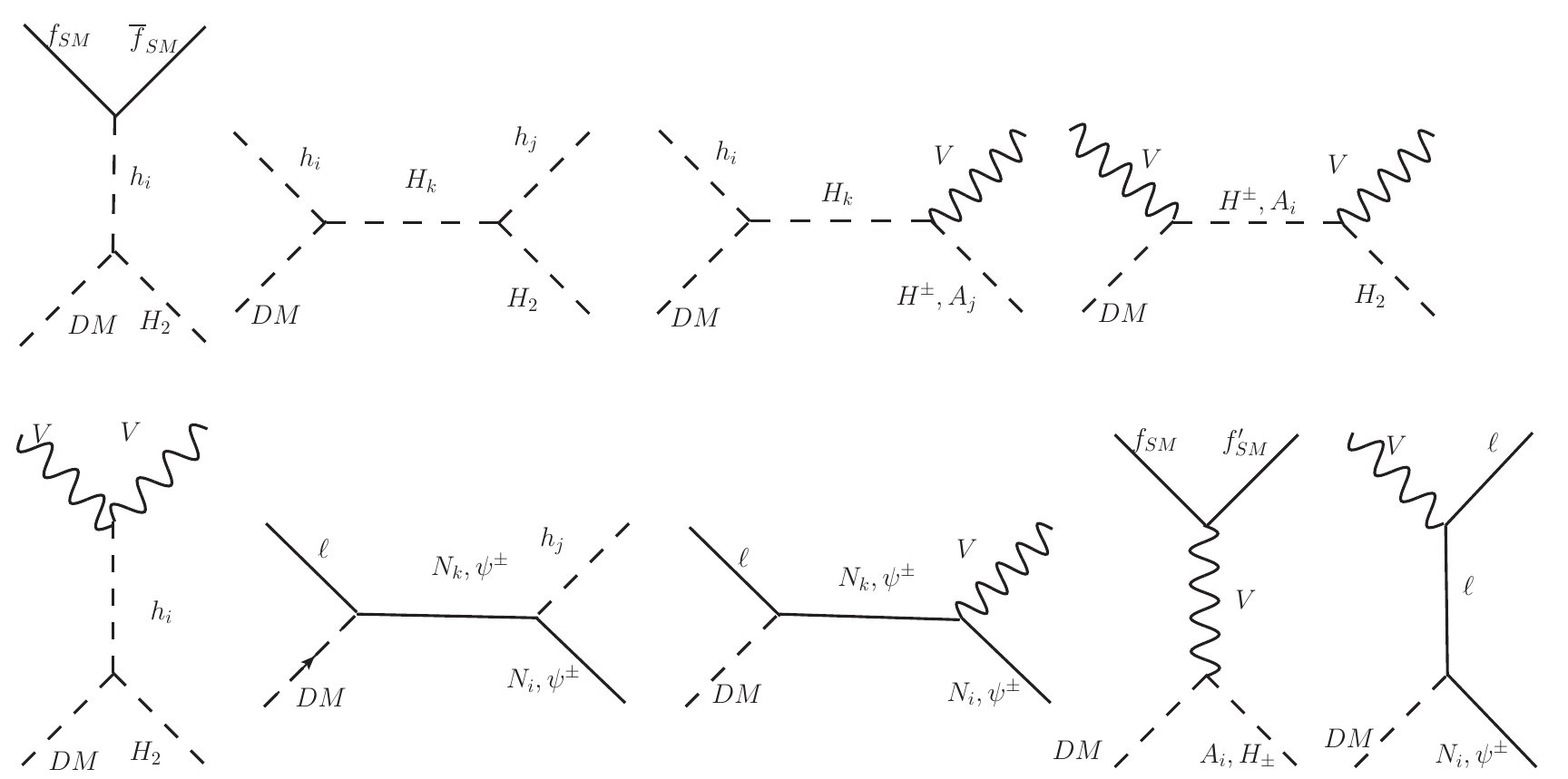}
%			}
		\caption{  The co-annihilation and annihilation diagrams of the DM and the other $Z_2$-odd fermion fields. The vertical axis represents time.}
		\label{fig:DarkCoan}
	\end{center}
\end{figure}
%%%%%%%%%%%%%%%%%
The condition for a decay or a scattering process to be in  thermal equilibrium is ${\Gamma \over H(T) } \geq 1,$ where $\Gamma$ is the rate of the relevant process. For a decay process, $\Gamma$ is the decay width of the decaying particle for that process, while for the annihilation process $\Gamma=n_{eq} <\sigma v>$~\cite{Hall:2009bx} where $n_{eq}$ stands for the equilibrium number density~\cite{Bauer:2017qwy} and  $H(T)$ is the Hubble parameter~\cite{Hall:2009bx}.
We now discuss the DM phenomenology for the freeze-out mechanism, considering all the constraints discussed earlier.
The DM  annihilation occurred in the early Universe when it was in thermal equilibrium with other particles. As the Universe's temperature drops below the DM mass, its production from the bath particles gets suppressed. The DM freezes out when the effective annihilation rate of DM becomes less than the expansion rate of the Universe, resulting in a density determined by the specific parameter ranges. 
In this model, the observed DM relic density can be obtained through annihilation processes shown in Fig. ~\ref{fig:DarkAn} and ~\ref{fig:DarkCoan} which depend upon the Higgs portal couplings and the Yukawa couplings $Y_{fd}$, $Y_{f\nu}$ and $Y_{fx}$.  We will study the scenarios, where the DM scalar belongs dominantly to the $SU(2)$ doublet and when it is dominantly an SM singlet. 
We will also consider the scenario where the DM has a reasonable doublet-singlet admixture. Finally, we study the role of the new gauge boson $Z^{\prime}$ in achieving the correct DM relic density.

In the pure doublet scenario ($\cos\beta=1$), if we remove the new fermion and gauge particle effects, the DM behaves like the DM of an inert Higgs doublet model (IHDM) 
with the absence of the $\lambda_5 \, (\Phi _1^{\dagger } \Phi _2)^2 + h.c.$ term~(see Ref.~\cite{Khan:2015ipa}), which means that $M_{H_1}=M_{A_1}$. This would mean that we never achieve observed relic density due to the large co-annihilation contribution mediated by the SM $Z$ boson. In order to control this large co-annihilation, we need a mass splitting between the $Z_2$-odd CP-even and CP-odd scalar that suppresses the process. In this model such a splitting will be induced by the $(\sqrt{2} \mu_{S} \, S_{2}^2 \, S_{1} - \mu_{h} \, \Phi _2^{\dagger } \Phi _1 \, S_{2}+h.c. )$ terms in the scalar potential (eqn.\ref{eq:potAll1}).
The larger the mass splitting, the more the co-annihilation rate is suppressed. Thus, this mass splitting can be adjusted to achieve the desired co-annihilation rate and match the observed DM relic density. In scenarios where the mass differences between the DM and the other dark sector particles are large, the co-annihilation processes become inefficient for producing the correct DM relic density. In such cases, we rely on DM pair annihilation through the scalar mediators ($h$ and $H$), interacting via scalar portal couplings $g_{h H_1 H_1}$ and $g_{H H_1 H_1}$, respectively. However, to satisfy relic density we would need very large Higgs coupling strength with the DM,
%%%%
$$g_{h H_1 H_1} \sim \cos^2\beta (\lambda_3+\lambda_4)\, v_1-\frac{2 \cos^2 \beta \, \sin^2 \beta \, (M_{H_2}^2-M_{H_1}^2)}{v_1}+\kappa _{1} \sin^2 \beta \, v_1 \,\,\,\, .$$ 
%%%
In such a scenario, most of these points will be excluded by either the direct detection (DD) experiments or Higgs invisible decay width data that depend on the very same large couplings.  Near, $M_{\rm{DM}}\approx \frac{M_h}{2}$, DM can resonantly annihilate to SM Higgs which then decays to SM particles. In the resonance region, a significantly smaller Higgs portal coupling will be sufficient to generate a large annihilation rate that leads to the correct DM relic density. As the couplings are now suppressed, this would help evade the DD constraint and Higgs invisible decay bound.

Note that the co-annihilation channels become crucial for determining the relic density 
of DM in the regions below and above $M_{\rm{DM}}\approx \frac{M_h}{2}$. 
The above mentioned co-annihilation channel is most dominant when the DM annihilates with the CP-odd scalar through the SM $Z$ boson, particularly when their combined mass is near the $Z$ resonance. We find that $M_{H_1}+M_{A_1} - M_{Z} \simeq 25$ GeV also results in dominant co-annihilation. Table~\ref{tabDM:1} presents a couple of representative benchmark points where $M_{DM}=45, 53$ GeV, that highlight the contribution of this co-annihilation process in achieving the correct DM relic density. We have set the mixing angle $\cos\beta$ to $0.9$ that leads to a scalar DM whose dominant composition comes from the $Z_2$-odd scalar doublet. Since the DM belongs to the $SU(2)_L$ scalar doublet, we avoid mass regions with $M_{H_1}+M_{A_1} < M_{Z}$ to prevent any alteration to the well-measured $Z$ boson decay width and branching ratio to SM particles. It is worth pointing out that in the above mentioned benchmark points, with $M_{\rm{A_1}}=54.7$ GeV and $61.7$ GeV, the two $A_1$ particles also annihilate 
via off-shell SM Higgs to $b\bar{b}$, effectively reducing the total number density of the $Z_2$ odd sector and contributing to the DM relic density. 

Similar to the SM $Z$ resonance, one can also exploit the resonant region due to 
the SM Higgs that leads to large rate for DM pair annihilation. We show two benchmark points near the Higgs resonance region in Tab. \ref{tabDM:1anni} with $M_{H_1}=58.82$ and $62.77$ GeV. We choose smaller Higgs portal couplings ($g_{h H_1 H_1}$) to get annihilation rate $<\sigma v>\sim 2\times 10^{-26}~{\rm cm^3/s}$ which gives exact relic density. These two points also comply with the Higgs invisible decay branching fraction constraint and bounds from DD experiments. Another important observation is that DM, being doublet-like, efficiently annihilates into gauge bosons such as $W W^*$ and $Z Z^*$ via the contact vertex and through the $Z_2$ odd sector particles in the t-channel. These DM annihilation channels significantly contribute to the DM relic density in these two representative points. 
%%%%%%%%%%%%%%%%%  
\begin{table*}[h!]
\begin{center}\scalebox{0.8}{
	\begin{tabular}{|p{2.0cm}|p{1.25cm}|p{1.2cm}|p{1.2cm}|p{1.2cm}|p{1.2cm}|p{1.4cm}|p{1.2cm}|p{1.2cm}|p{2.0cm}|p{4.4cm}|}
		\hline
		\hline
		Sl. No. &Mixing $\cos\beta$& $M_{DM}$ (GeV) & $M_{H_2}$ (GeV) & $M_{A_1}$ (GeV) & $M_{H^\pm}$ (GeV) & $Y_{fd,\nu}^{1,3}$ &~~$Y_{fx}^{1,3}$&$\Omega_{DM}h^2$&DD Cross (${\rm cm^3/s}$)&~~~~Contributions \\
		\hline
		\hline
		&&&&&&&&&&$A_1 \,A_1\rightarrow b \overline{b}~~15\%$\\
			D-type&~0.9&~45.0&~86.9&~54.71&~130.0&$10^{-4}$&$10^{-4}$&0.1205&$5.5 \times 10^{-48}$& $H_1 \,A_1\rightarrow q \overline{q}\quad 54 \%$\\
		 BP-1&&&&&&&&&&$H_1 \,A_1\rightarrow l \overline{l}~~9\%$\\
		&&&&&&&&&&$H_1 \,A_1\rightarrow \nu_l \overline{\nu}_l~~21\%$\\
		\hline
		&&&&&&&&&&$A_1 \,A_1\rightarrow b \overline{b}~~18\%$\\
			D-type&~0.9&~53.0&~92.50&~61.777&~133.5&$10^{-4}$&$10^{-4}$&0.1199&$5.4 \times 10^{-48}$& $H_1 \,A_1\rightarrow q \overline{q}\quad 50 \%$\\
		 BP-2&&&&&&&&&&$H_1 \,A_1\rightarrow l \overline{l}~~9\%$\\
		&&&&&&&&&&$H_1 \,A_1\rightarrow \nu_l \overline{\nu}_l~~21\%$\\
		&&&&&&&&&&$H_1 \,H_1\rightarrow W  W^{*}~~1\%$\\
		\hline
		&&&&&&&&&&$H_1 \,H_1\rightarrow W  W^{*}~~80\%$\\
			D-type&~0.9&~72.5&~128.14&~84.206&~152.5&$10^{-4}$&$10^{-4}$&0.1198&$5.4 \times 10^{-48}$& $H_1 \,H_1\rightarrow Z  Z^{*}~~5\%$\\
		 BP-3&&&&&&&&&&$H_1 \,A_1\rightarrow q \overline{q}\quad 10\%$\\
		&&&&&&&&&&$H_1 \,A_1\rightarrow \nu_l \overline{\nu}_l~~5\%$\\
		\hline
		\hline
	\end{tabular}}
\end{center}
	\caption{ Dominated (Co)annihilation processes. $H_1$ is mostly coming from CP-even component of $Z_2$-odd scalar doublet. The fermionic $t,u$-channel contributions almost zero due to small Yukawa couplings. }
	\label{tabDM:1}
\end{table*}
%%%%%%%%%%%%%%%%%
%%%%%%%%%%%%%%%%% 
    
%%%%%%%%%%%%%%%%%  
\begin{table*}[h!]
\begin{center}\scalebox{0.8}{
	\begin{tabular}{|p{2.0cm}|p{1.25cm}|p{1.2cm}|p{1.2cm}|p{1.2cm}|p{1.2cm}|p{1.4cm}|p{1.2cm}|p{1.2cm}|p{2.0cm}|p{4.4cm}|}
		\hline
		\hline
		Sl. No. &Mixing $\cos\beta$& $M_{DM}$ (GeV) & $M_{H_2}$ (GeV) & $M_{A_1}$ (GeV) & $M_{H^\pm}$ (GeV) & $Y_{fd,\nu}^{1,3}$ &~~$Y_{fx}^{1,3}$&$\Omega_{DM}h^2$&DD Cross (${\rm cm^3/s}$)&~~~~Contributions \\
		\hline
	    \hline
		&&&&&&&&&&$H_1 \,H_1\rightarrow b \overline{b}~~72\%$\\
			D-type&~0.9&~58.82&~144.44&~77.419&~138.82&$10^{-4}$&$10^{-4}$&0.1196&$3.3 \times 10^{-48}$& $H_1 \,H_1\rightarrow q \overline{q}\quad 5 \%$\\
		 BP-4&&&&&&&&&&$H_1 \,H_1\rightarrow l \overline{l}~~3\%$\\
		&&&&&&&&&&$H_1 \,H_1\rightarrow W W^* ~~17\%$\\
		&&&&&&&&&&$H_1 \,H_1\rightarrow ZZ^*~~2\%$\\
		\hline
		&&&&&&&&&&$H_1 \,H_1\rightarrow b \overline{b}~~59\%$\\
			D-type&~0.9&~62.77&~148.41&~80.995&~142.77&$10^{-4}$&$10^{-4}$&0.1196&$2.8 \times 10^{-48}$& $H_1 \,H_1\rightarrow q \overline{q}\quad 4 \%$\\
		 BP-5&&&&&&&&&&$H_1 \,H_1\rightarrow l \overline{l}~~3\%$\\
		&&&&&&&&&&$H_1 \,H_1\rightarrow W W^* ~~29\%$\\
		&&&&&&&&&&$H_1 \,H_1\rightarrow ZZ^*~~3\%$\\
		\hline
		\hline
	\end{tabular}}
\end{center}
	\caption{ Dominated annihilation processes. $H_1$ is mostly coming from CP-even component of $Z_2$-odd scalar doublet. The fermionic $t,u$-channel contributions almost zero.}
	\label{tabDM:1anni}
\end{table*}
%%%%%%%%%%%%%%%%%
%%%%%%%%%%%%%%%%%

In this model, similar to the Inert Doublet Model~\cite{Khan:2015ipa}, for doublet-type DM in region $ 72.5\, \rm{GeV} < M_{\rm{DM}} < 400\, \rm{GeV}$  the processes $H_1 \, H_1 \rightarrow VV$ ($V=W$ and $Z$) become dominant and annihilate through standard gauge couplings. A large negative Higgs portal coupling is required to reduce the effective annihilation rate. However, direct detection data rules out such a large Higgs portal coupling, making it impossible to achieve the observed relic density in this region.

%%%%%%%%%%%%%%%%%  
\begin{table*}[h!]
\begin{center}\scalebox{0.8}{
	\begin{tabular}{|p{2.0cm}|p{1.3cm}|p{1.2cm}|p{1.4cm}|p{1.4cm}|p{1.4cm}|p{1.4cm}|p{1.2cm}|p{1.2cm}|p{2.0cm}|p{4.4cm}|}
		\hline
		\hline
		Sl. No. &Mixing $\cos\beta$& $M_{DM}$ (GeV) & $M_{H_2}$ (GeV) & $M_{A_1}$ (GeV) & $M_{H^\pm}$ (GeV) & $Y_{fd,\nu}^{1,3}$ &~~$Y_{fx}^{1,3}$&$\Omega_{DM}h^2$&DD Cross (${\rm cm^3/s}$)&~~~~Contributions \\
		\hline
	    \hline
		&&&&&&&&&&$H^\pm \,H^\pm \rightarrow W W~~18\%$\\
			D-type&~0.9&~420.0&~420.01&~420.002&~420.1&$10^{-4}$&$10^{-4}$&0.1137&$2.2 \times 10^{-49}$& $A_1 \,A_1 \rightarrow W W~~9\%$\\
		 BP-6&&&&&&&&&&$H^\pm \,A_1\rightarrow \gamma W~~9\%$\\
		&&&&&&&&&&$A_1 \,A_1\rightarrow ZZ~~8\%$\\
		&&&&&&&&&&$H^\pm \,H^\pm \rightarrow \gamma \gamma~~8\%$\\
		&&&&&&&&&&$H^\pm \,H_1\rightarrow \gamma W~~7\%$\\
		&&&&&&&&&&$H^\pm \,H^\pm\rightarrow \gamma Z~~6\%$\\
		&&&&&&&&&&$H_1 \,H_1\rightarrow W W~~6\%$\\
		&&&&&&&&&&$H_1 \,H_1\rightarrow Z Z~~6\%$\\
		&&&&&&&&&&$H_1 \,H_2\rightarrow WW~~4\%$\\
		&&&&&&&&&&$H_1 \,H_2\rightarrow ZZ~~3\%$\\
		\hline
		&&&&&&&&&&$H^\pm \,H^\pm \rightarrow W W~~13\%$\\
			D-type&~0.9&~500.0&~500.01&~500.002&~500.1&$10^{-4}$&$10^{-4}$&0.1161&$4.0 \times 10^{-49}$& $A_1 \,A_1 \rightarrow W W~~7\%$\\
		 BP-7&&&&&&&&&&$H^\pm \,A_1\rightarrow \gamma W~~6\%$\\
		&&&&&&&&&&$A_1 \,A_1\rightarrow ZZ~~6\%$\\
		&&&&&&&&&&$H^\pm \,H^\pm \rightarrow \gamma \gamma~~6\%$\\
		&&&&&&&&&&$H^\pm \,H_1\rightarrow \gamma W~~5\%$\\
		&&&&&&&&&&$H^\pm \,H^\pm\rightarrow \gamma Z~~4\%$\\
		&&&&&&&&&&$H_1 \,H_1\rightarrow W W~~4\%$\\
		&&&&&&&&&&$H_1 \,H_1\rightarrow Z Z~~4\%$\\
		&&&&&&&&&&$H_1 \,H_2\rightarrow WW~~3\%$\\
		&&&&&&&&&&$H_1 \,H_2\rightarrow ZZ~~2\%$\\
		&&&&&&&&&&$H_1 \,H_1\rightarrow l \overline{l}~~1\%$\\
		&&&&&&&&&&$H^\pm \,H^\pm\rightarrow l \overline{l}~~13\%$\\
		\hline
		&&&&&&&&&&$H^\pm \,H^\pm \rightarrow W W~~3\%$\\
			D-type&~0.9&~700.0&~701.61&~700.303&~701.605&$10^{-4}$&$10^{-4}$&0.1137&$2.2 \times 10^{-49}$& $A_1 \,A_1 \rightarrow W W~~14\%$\\
		 BP-8&&&&&&&&&&$H^\pm \,A_1\rightarrow \gamma W~~2\%$\\
		&&&&&&&&&&$A_1 \,A_1\rightarrow ZZ~~5\%$\\
		&&&&&&&&&&$H^\pm \,H_1\rightarrow \gamma W~~2\%$\\
		&&&&&&&&&&$H_1 \,H_1\rightarrow W W~~9\%$\\
		&&&&&&&&&&$H_2 \,H_2\rightarrow W W~~7\%$\\
		&&&&&&&&&&$H_2 \,H_1\rightarrow W W~~13\%$\\
		&&&&&&&&&&$H_1 \,H_1\rightarrow Z Z~~5\%$\\
		&&&&&&&&&&$H_1 \,H_2\rightarrow WW~~4\%$\\
		&&&&&&&&&&$H_1 \,H_2\rightarrow ZZ~~3\%$\\
		&&&&&&&&&&$H^\pm \,H^\pm \rightarrow \gamma \gamma~~1\%$\\
		&&&&&&&&&&$H^\pm \,H^\pm\rightarrow \gamma Z~~1\%$\\
		\hline
		\hline
	\end{tabular}}
\end{center}
	\caption{ Dominated (Co)annihilation processes. $H_1$ is mostly coming from CP-even component of $Z_2$-odd scalar doublet. The fermionic $t,u$-channel contributions almost zero.}
	\label{tabDM:2}
\end{table*}
%%%%%%%%%%%%%%%%%
%%%%%%%%%%%%%%%%%

\begin{figure}[h!]
	\begin{center}
		%\subfigure[]
		{
			\includegraphics[scale=0.4]{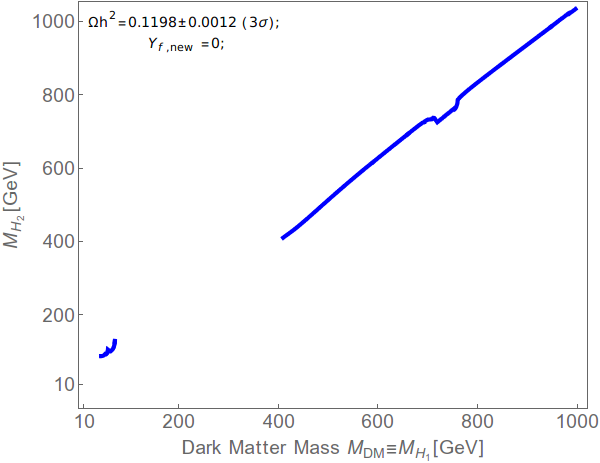}}
			
		\caption{  We varied DM (doublet type, i.e., mixing angle $\cos\beta=0.9$) mass and next-lightest $Z_2$ odd particle mass  $M_{H_2}$. The DM $Y_f=0$. Fermion masses $\sim1$ TeV, $M_{Z'}=1$ TeV. The blue line gives relic density $\Omega h^2=0.1198\pm 0.0012$ within $3\sigma$, its also allowed by the direct detection experimental data. The other region is ruled out either from direct detection and or invisible decay width data  or can not satisfy the exact relic density. }
		\label{fig:DarkPlotd1c}
	\end{center}
\end{figure}

For doublet type DM with $M_{H_1}  > 400$ GeV, the relic density can be precisely obtained while satisfying various experimental constraints when the mass of the DM particle and other $Z_2$-odd particles is approximately equal, which leads to an enhancement of the co-annihilation of DM with other dark sector particles.  We show a few benchmark points to highlight our results in Table~\ref{tabDM:2}.  Additionally, we have shown a plot of the DM relic density versus its mass in Fig.~\ref{fig:DarkPlotd1c}. Note that  the long break in the relic density curve for the mass range $ 72.5\, \rm{GeV} < M_{H_{1}} < 400\, \rm{GeV}$ corresponds to the case where the DM relic density cannot 
be satisfied due to the large annihilation rates of DM pair to SM gauge bosons, as discussed above. 
For $M_{H_1}  > 400$ GeV, the DM relic density is primarily satisfied due to the contributions coming from co-annihilation processes.
%%%%%%%%%%%%%%%%%
%%%%%%%%%%%%%%%%%
%%%%%%%%%%%%%%%%%
%%%%%%%%%%%%%%%%%
In this plot, the mixing angle is again fixed at $\cos\beta=0.9$, and the mass of the other $Z_2$ odd particles are chosen to be close to the DM mass. The heavy fermion masses are fixed as 
$(M_{N_1},M_{N_2},M_{N_3},M_{N_4}, M_{\psi^\pm})=(1.0,1.0,1.05,1.5,1.1)$ TeV in Fig.~\ref{fig:DarkPlotd1c}.
A small dip appears around DM mass $M_{DM}=750$ GeV. It is due to the 
resonance channel $H_i H_j \rightarrow WW, ZZ, hh~(i,j=1,2)$ via heavy CP-even Higgs $H$, with mass $M_H=1500$ GeV. This indicates that the presence of an extra channel for DM annihilation requires decreasing the co-annihilation contribution by increasing the mass difference between the DM and other dark sector particles.

So far, we have neglected any contributions from BSM fermions to the DM relic density by choosing relevant Yukawa couplings to be small. However, if we introduce these new fermionic interaction terms for doublet-type DM, we can achieve the correct DM relic density in the low DM mass region without relying on the SM Higgs resonance. This can be understood as follows: in the absence of new fermions, a significantly large Higgs portal coupling would be required to achieve the exact relic density. However, such a coupling is largely ruled out due to constraints from the invisible decay width of the Higgs boson and direct detection experiments. The terms $Y_{fd} \,\bar{\psi_L} \Phi_{2} e_R +Y_{f\nu} \,\bar{\ell_L} \tilde{\Phi}_{2} N_R + h.c.$ in the Lagrangian \ref{Eq:lagfermion} induce annihilation processes, $H_1 H_1 \rightarrow \ell \bar{\ell}$ through these new fermions in the $t$-channel. These interactions can greatly impact the overall DM relic density. Here, $l$ stands for SM leptons. Corresponding Feynman diagrams are shown in Fig. \ref{fig:DarkCoan}. We have shown a few corresponding benchmark points in the Table~\ref{tabDM:3} of this type. The term $Y_{fx}\,  \bar{\psi_R}\,S_{2} \,\ell_L + h.c.$ in the Lagrangian (eq.~\ref{Eq:lagfermion}) can contribute to DM relic density for the doublet type of DM scenario via the mixing of CP even scalar in the $Z_2$ odd sector through the above mentioned processes. We minimise the contribution of this term to the relic density by keeping $Y_{fx}$ small.
%%%%%%%%%%%%%%%%% 
\begin{table*}[h!]
\begin{center}\scalebox{0.8}{
	\begin{tabular}{|p{2.0cm}|p{1.25cm}|p{1.2cm}|p{1.2cm}|p{1.2cm}|p{1.4cm}|p{1.2cm}|p{1.2cm}|p{2.0cm}|p{4.4cm}|}
		\hline
		\hline
		Sl. No. &Mixing $\cos\beta$& $M_{DM}$ (GeV) & $M_{H_2}$ (GeV) & $M_{A_1}$ (GeV) & $Y_{fd,\nu}^{1,3}$ &~~$Y_{fx}^{1,3}$&$\Omega_{DM}h^2$&DD Cross (${\rm cm^3/s}$)&~~~~Contributions \\
		\hline
	    \hline
		&&&&&&&&&$H_1 \,H_1\rightarrow l \overline{l}~~96\%$\\
			D-type&~0.9&~10.0&~110.0&~310&$0.335$&$10^{-4}$&0.1198&$ 8.7 \times 10^{-47}$& $H_1 \,H_1\rightarrow \nu_l \overline{\nu}_l~~4\%$\\
			BP-9&&&&&&&&&\\
		\hline
		&&&&&&&&&$H_1 \,H_1\rightarrow l \overline{l}~~96\%$\\
			D-type&~0.9&~30.0&~130.0&~330&$0.3256$&$10^{-4}$&0.1198&$ 1.2 \times 10^{-47}$& $H_1 \,H_1\rightarrow \nu_l \overline{\nu}_l~~4\%$\\
			BP-10&&&&&&&&&\\
		\hline
		&&&&&&&&&$H_1 \,H_1\rightarrow l \overline{l}~~96\%$\\
			D-type&~0.9&~50.0&~150.0&350&$0.324$&$10^{-4}$&0.1198&$ 5.5 \times 10^{-48}$& $H_1 \,H_1\rightarrow \nu_l \overline{\nu}_l~~4\%$\\
			BP-11&&&&&&&&&\\
		\hline
		&&&&&&&&&$H_1 \,H_1\rightarrow l \overline{l}~~68\%$\\
			D-type&~0.9&~70.0&~170.0&~370&$0.288$&$10^{-4}$&0.1198&$ 2.5 \times 10^{-48}$& $H_1 \,H_1\rightarrow \nu_l \overline{\nu}_l~~3\%$\\
			BP-12&&&&&&&&&$H_1 \,H_1\rightarrow W  W^{*}~~34\%$\\
			&&&&&&&&&$H_1 \,H_1\rightarrow Z  Z^{*}~~4\%$\\
		\hline
		&&&&&&&&&$H_1 \,H_1\rightarrow l \overline{l}~~16\%$\\
			D-type&~0.9&~74.0&~173.8&~373.8&$0.1$&$10^{-4}$&0.1138&$ 1.2 \times 10^{-48}$& $H_1 \,H_1\rightarrow W  W^{*}~~77\%$\\
			BP-13&&&&&&&&&$H_1 \,H_1\rightarrow Z  Z^{*}~~6\%$\\
		\hline
		\hline
	\end{tabular}}
\end{center}
	\caption{ Dominated annihilation processes. $H_1$ is mostly coming from CP-even component of $Z_2$-odd scalar doublet. The fermionic $t,u$-channel contributions large. }
	\label{tabDM:3}
\end{table*}
%%%%%%%%%%%%%%%%%
%%%%%%%%%%%%%%%%%
\begin{figure}[h!]
	\begin{center}
		%\subfigure[]
		{
			\includegraphics[scale=0.4]{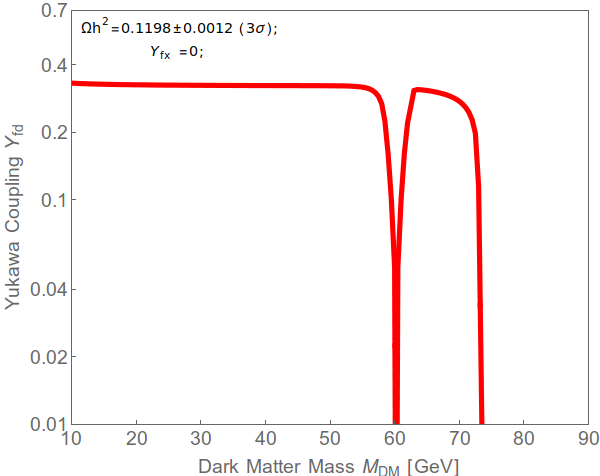}}
			
		\caption{  We varied DM (doublet type, i.e., mixing angle $0.9$ \textcolor{black}{(red)}mass and new Yukawa coupling $Y_{f_d}=Y_{f_\nu}$ with $Y_{f_x}=0$. Fermion masses $\sim1$ TeV, $M_{Z'}=1$ TeV. The red line gives relic density $\Omega h^2=0.1198\pm 0.0012$ within $3\sigma$, its also allowed by the invisible decay width and direct detection experimental data. }
		\label{fig:DarkPlotd1f}
	\end{center}
\end{figure}
%%%%%%%%%%%%%%%%%
%%%%%%%%%%%%%%%%%
One can see that the dominant contribution to the relic density in the low mass region mostly comes from the $t$ channels annihilation processes with the  BSM fermions in the propagator. For  $M_{\rm {DM}} > 40$ GeV,  the DM begins to annihilate into the SM gauge bosons via contact vertex. As these contributions become large, we need smaller Yukawa coupling to satisfy the DM relic density. This effect can be seen for the (D-type; BP-13) benchmark in the table. 

We have plotted the DM mass against the new Yukawa couplings $Y_{fd} \equiv Y_{f\nu}$ in Fig.~\ref{fig:DarkPlotd1f}. 
The red line gives the correct relic density ($\Omega h^2=0.1198\pm 0.0012$) within $3\sigma$ and is also allowed by the invisible decay width constraints coming from the Higgs observation. The points on the curve are also consistent with the direct detection experimental data.

\begin{figure}[h!]
	\begin{center}
		\subfigure[]{
			\includegraphics[scale=0.3]{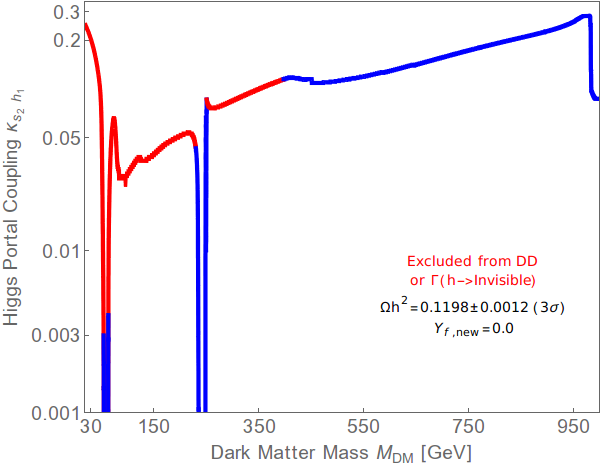}}	
		\caption{  We varied DM (mostly singlet type, i.e., mixing angle $\cos\beta=0.05$) mass and Higgs portal coupling $\kappa_{\phi_1 s_2}$ with $Y_{f,x,d,\nu}=0$, Fermion masses $\sim 1$ TeV, $M_{Z'}=1.5$ TeV, $M_H=500$ GeV. The red-blue line gives relic density $\Omega h^2=0.1198\pm 0.0012$ within $3\sigma$. The red line excluded from Higgs Invisible decay width and/or the direct detection cross-section. }
		\label{fig:DarkPlots1a}
	\end{center}
\end{figure}

The singlet-type DM scenario is realized for a small mixing angle, $\cos\beta \sim 0$. For $M_{\rm{DM}}<50$ GeV, the correct DM relic density can be again achieved through the Higgs portal coupling, where $\sin\beta \sim 1$:  
%%%%%
$$g_{h H_1 H_1} \sim \cos^2\beta (\lambda_3+\lambda_4)\, v_1 -\frac{2 \cos^2 \beta \, \sin^2 \beta \, (M_{H_2}^2-M_{H_1}^2)}{v_1}+\kappa _{1} \sin^2 \beta \, v_1 \sim \kappa _{1} \, v_1.$$ 
%%%%%
As before, there will be strong constraints on the strength of the Higgs portal coupling, coming from the invisible decay width of the SM Higgs boson as well as DD experiments. So we again rely on the contributions coming from the Higgs resonance region, i.e., at $M_{\rm{DM}} \approx M_h/2$, that leaves the possibility of a small Higgs portal coupling which can give $\langle\sigma v\rangle \sim 2\times 10^{-26}~{\rm cm^3/s}$. This can produce the correct relic density while remaining consistent with Higgs decay and DD constraints.
The annihilation could also occur via the additional heavy CP even scalar field $H$ which couples with the DM through  
%%%
$$g_{H H_1 H_1} \sim\frac{4 \mu_{\phi_{s2}}^2+2 \kappa_{\phi_1 s_2} v_1^2 - 3 M_{H_1}^2 - M_{A_2}^2 }{2 v_s}\,.$$
%%%
Due to this heavy Higgs, an additional resonance region can appear near $M_{H_1}= M_H/2$.
It is important to note that achieving relic density is possible across a spectrum of DM masses, specifically within the $65 < M_{H_1} < 650$ GeV~\footnote{This region is ruled out in pure singlet scenario from direct detection experiments \cite{Khan:2014kba}.}. This outcome is contingent upon the specific value of the heavy Higgs mass ($M_H$). 
We illustrated the relic density (dominated by $s$- channel and cross-channels only, see Fig.~\ref{fig:DarkAn}) for the low as well as the high-DM mass region in $\kappa_{\phi_1 s_2} - M_{H_1}$ plane in Fig.~\ref{fig:DarkPlots1a}. The solid red-blue band shows the relic density $\Omega h^2=0.1198\pm0.0012$ within $3\sigma$. The red band is, however, excluded from the Higgs invisible decay or DD constraints as a large Higgs portal coupling is needed to get the correct DM relic density. In this plot two funnel like region appear near $M_{DM}\approx \frac{M_h}{2}$ and $M_{DM}\approx \frac{M_H}{2}$ with $M_{H} = 500$ GeV, which corresponds to the DM annihilation through the $h$ and $H$ resonance respectively.
In these two funnel regions, the dominant channel is $H_1 \, H_1 \rightarrow b \overline{b}$ for the low-DM mass range, whereas the high-mass region is dominated by $H_1 \, H_1 \rightarrow \chi \chi$ channels. $\chi$ represents $W,Z,t$ and $h$ particles.
The process $H_1 H_1 \rightarrow HH$ also starts to contribute to the DM annihilation process near $M_{H_1} \approx M_H$. Hence a dip appears around $M_{\text{DM}}=500$ GeV.
The vicinity of the DM mass at $950$ GeV experiences a notable impact stemming from the coannihilation effect involving new heavy fermions, where the masses of these fermions are approximately $1$ TeV.
In this context, the DM annihilation through the $t$ and $u$-channel BSM fermion exchange is negligible due to the choice of small new Yukawa couplings $Y_{fd,\nu,x}^{1,3}\approx 10^{-4}$. 
We present corresponding benchmark points for all the region discussed above and the contributions of different annihilation channels in Table~\ref{tabDM:s1}.
%%%%%%%%%%%%%%%%%
%%%%%%%%%%%%%%%%%
%%%%%%%%%%%%%%%%%
%%%%%%%%%%%%%%%%%
%%%%%%%%%%%%%%%%%  
\begin{table*}[h!]
\begin{center}\scalebox{0.8}{
	\begin{tabular}{|p{2.0cm}|p{1.25cm}|p{1.2cm}|p{1.2cm}|p{1.2cm}|p{1.2cm}|p{1.4cm}|p{1.2cm}|p{1.2cm}|p{2.0cm}|p{4.6cm}|}
		\hline
		\hline
		Sl. No. &Mixing $\cos\beta$& $M_{DM}$ (GeV) & $M_{H_2}$ (GeV) & $M_{A_1}$ (GeV) & $M_{H^\pm}$ (GeV) & $\kappa_{\phi_1 s_2}$ &~~$Y_{fd,\nu,x}^{1,3}$&$\Omega_{DM}h^2$&DD Cross (${\rm cm^3/s}$)&~~~~Contributions \\
		\hline
		&&&&&&&&&&$H_1 \,H_1\rightarrow b \overline{b}~~77\%$\\
			S-type&~0.05&~56.0&~356.0&~ 356.0&~656.0&$0.00135$&$10^{-4}$&0.1196&$5.2 \times 10^{-47}$& $H_1 \,H_1\rightarrow WW^*\quad 13 \%$\\
		 BP-1&&&&&&&&&&$H_1 \,H_1\rightarrow c \overline{c}~~6\%$\\
		&&&&&&&&&&$H_1 \,H_1\rightarrow l \overline{l}~~4\%$\\
		&&&&&&&&&&$H_1 \,H_1\rightarrow ZZ^*~~1\%$\\
		\hline
		&&&&&&&&&&$H_1 \,H_1\rightarrow b \overline{b}~~71\%$\\
			S-type&~0.05&~62.9&~360.9&~ 360.9&~660.9&$0.001$&$10^{-4}$&0.1203&$3.1 \times 10^{-47}$& $H_1 \,H_1\rightarrow WW^*\quad 19 \%$\\
		 BP-2&&&&&&&&&&$H_1 \,H_1\rightarrow c \overline{c}~~5\%$\\
		&&&&&&&&&&$H_1 \,H_1\rightarrow l \overline{l}~~3\%$\\
		&&&&&&&&&&$H_1 \,H_1\rightarrow ZZ^*~~2\%$\\
		\hline
		\hline
		&&&&&&&&&&$H_1 \,H_1\rightarrow t \overline{t}~~7\%$\\
			S-type&~0.05&~230.0&~300.0&~300.0&~330.0&$0.044$&$10^{-4}$&0.1199&$3.4 \times 10^{-46}$& $H_1 \,H_1\rightarrow WW\quad 45 \%$\\
		 BP-3&&&&&&&&&&$H_1 \,H_1\rightarrow ZZ~~21\%$\\
		&&&&&&&&&&$H_1 \,H_1\rightarrow hh~~27\%$\\
		\hline
		&&&&&&&&&&$H_1 \,H_1\rightarrow t \overline{t}~~6\%$\\
			S-type&~0.05&~400.0&~470.0&~470.0&~500.0&$0.115$&$10^{-4}$&0.1211&$6.1 \times 10^{-46}$& $H_1 \,H_1\rightarrow WW\quad 46 \%$\\
		 BP-4&&&&&&&&&&$H_1 \,H_1\rightarrow ZZ~~23\%$\\
		&&&&&&&&&&$H_1 \,H_1\rightarrow hh~~24\%$\\
		\hline
		&&&&&&&&&&$H_1 \,H_1\rightarrow WW\quad 32 \%$\\
			S-type&~0.05&~950.0&~1020.0&~1020.0&~1050.0&$0.260$&$10^{-4}$&0.1211&$3.3 \times 10^{-46}$& $H_1 \,H_1\rightarrow ZZ\quad 16 \%$\\
		 BP-5&&&&&&&&&&$H_1 \,H_1\rightarrow hh~~16\%$\\
		&&&&&&&&&&$H_1 \,H_1\rightarrow HH~~6\%$\\
		&&&&&&&&&&$H_1 \,H^\pm\rightarrow hW~~4\%$\\
		&&&&&&&&&&$H_1 \,H^\pm\rightarrow HW~~3\%$\\
		\hline
		\hline
		&&&&&&&&&&$N_1 \,E_1^\pm \rightarrow ZW^\pm\quad 45 \%$\\
			S-type&~0.05&~995.0&~1060.0&~1060.0&~1095.0&$10^{-4}$&$10^{-4}$&0.1211&$5.1 \times 10^{-47}$& $N_1 \,E_1^\pm \rightarrow hW^\pm\quad 16 \%$\\
		 BP-6&&&&&&&&&&$N_1 \,N_1\rightarrow Zh~~9\%$\\
		&&&&&&&&&&$E_1 \,E_1\rightarrow EE~~13\%$\\
		&&&&&&&&&&$H_1 \,H_1\rightarrow WW~~2\%$\\
		&&&&&&&&&&$A_1 \,A_1\rightarrow WW~~1\%$\\
		\hline
		\hline
	\end{tabular}}
\end{center}
	\caption{ Dominated annihilation processes. $H_1$ is mostly coming from CP-even component of $Z_2$-odd scalar singlet. The fermionic $t,u$-channel contributions almost zero, co-annihilation present depending on mass. The Fermion masses $\sim 1$ TeV, $M_{Z'}=1.5$ TeV. }
	\label{tabDM:s1}
\end{table*}
%%%%%%%%%%%%%%%%%
%%%%%%%%%%%%%%%%%

Let us now examine how the BSM Yukawa couplings influence the computations of the relic density. The presence of the term $Y_{fx}\,  \bar{\psi}_R\,S_{2} \,\ell_L + h.c.$ in the Lagrangian (eq.\ref{Eq:lagfermion}) significantly enhance the annihilation rate of singlet-type DM as the DM can annihilate to SM leptons via t-channel diagram (see fig. \ref{fig:DarkCoan}).  
For this scenario, we keep $Y_{fx}^{1,3} = \mathcal{O}(0.1)$ while ensuring the second component of $Y_{fx}\approx\mathcal{O}(10^{-4})$ to address potential issues related to LFV, particularly the constraints from the branching ratio of $\mu \rightarrow e\gamma$.
%%%%%%%%%%%%%%%%%
%%%%%%%%%%%%%%%%%
\iffalse
\begin{figure}[h!]
	\begin{center}
		\subfigure[]{
			\includegraphics[scale=0.3]{./plot/Plot_SinglettypeCaseOnlyFermion0p1}}
	\hskip 1pt
			\subfigure[]{
			\includegraphics[scale=0.3]{./plot/Plot_SinglettypeCaseOnlyFermion0p26}}		
		\caption{  We varied DM (mostly singlet type, i.e., mixing angle $\cos\beta=0.05$) mass and Higgs portal coupling $\kappa_{\phi_1 s_2}$ with $Y_f=0.10$ and $0.26$, Fermion masses $\sim 1$ TeV, $M_{Z'}=1.5$ TeV, $M_{H}=500$ GeV. The red-blue line gives relic density $\Omega h^2=0.1198\pm 0.0012$ within $3\sigma$. The red line excluded from Higgs Invisible decay width and/or the direct detection cross-section. }
		\label{fig:DarkPlots1b}
	\end{center}
\end{figure}
\fi
%%%%%%%%%%%%%%%%%
%%%%%%%%%%%%%%%%%
Due to the presence of above mentioned Yukawa couplings ($Y_{fx}^{1,3}$) the observed relic can be achieved for a relatively low mass of 
DM ($\sim \mathcal{O}(10)$ GeV). These couplings compensate for the DM annihilation rate through the Higgs portal couplings by enabling new fermions-mediated channels. Therefore, for $M_{DM} < \frac{M_h}{2}$, the correct DM relic density can be achieved while  satisfying all relevant constraints, including the Higgs invisible decay width and direct detection constraints.
We show a few benchmark points for the singlet-type DM scenario with 
the corresponding contributions of different annihilation channels for them, in Table~\ref{tabDM:s2}.
From the table, we observe that achieving the exact relic density is feasible solely through t-channel DM annihilation to the SM leptons mediated by these BSM fermions. 
For $M_{H_1} > 100$ GeV, the annihilation processes of $H_1 H_1$ into $WW$, $ZZ$, and $hh$ arising due to the exchange of dark sector scalars or through contact vertex can contribute to the relic density as depicted in the table.
%%%%%%%%%%%%%%%%%  
%%%%%%%%%%%%%%%%%
%%%%%%%%%%%%%%%%%
\begin{table*}[h!]
\begin{center}\scalebox{0.8}{
	\begin{tabular}{|p{2.0cm}|p{1.25cm}|p{1.2cm}|p{1.2cm}|p{1.2cm}|p{1.2cm}|p{1.4cm}|p{1.2cm}|p{1.2cm}|p{2.0cm}|p{4.4cm}|}
		\hline
		\hline
		Sl. No. &Mixing $\cos\beta$& $M_{DM}$ (GeV) & $M_{H_2}$ (GeV) & $M_{A_1}$ (GeV) & $M_{H^\pm}$ (GeV) & $Y_{fd,\nu}^{1,3}$ &~~$Y_{fx}^{1,3}$&$\Omega_{DM}h^2$&DD Cross (${\rm cm^3/s}$)&~~~~Contributions \\
		\hline
	    \hline
		&&&&&&&&&&$H_1 \,H_1\rightarrow l \overline{l}~~67\%$\\
			S-type&~0.05&~10.0&~110.0&~109.81&~130.0&$10^{-4}$&$0.2696$&0.1198&$ 2.9 \times 10^{-45}$& $H_1 \,H_1\rightarrow \nu_l \overline{\nu}_l~~32\%$\\
			BP-7&&&&&&&&&&\\
		\hline
		&&&&&&&&&&$H_1 \,H_1\rightarrow l \overline{l}~~65\%$\\
			S-type&~0.05&~45.0&~145.0&~144.75&~155.0&$10^{-4}$&$0.2629$&0.1198&$ 1.4 \times 10^{-46}$& $H_1 \,H_1\rightarrow \nu_l \overline{\nu}_l~~34\%$\\
			BP-8&&&&&&&&&&\\
		\hline
		&&&&&&&&&&$H_1 \,H_1\rightarrow l \overline{l}~~64\%$\\
			S-type&~0.05&~100.0&~200.0&~199.70&~180.0&$10^{-4}$&$0.2631$&0.1196&$ 2.6 \times 10^{-47}$& $H_1 \,H_1\rightarrow \nu_l \overline{\nu}_l~~33\%$\\
			BP-9&&&&&&&&&&$H_1 \,H_1\rightarrow WW~~2\%$\\
		\hline
		&&&&&&&&&&$H_1 \,H_1\rightarrow l \overline{l}~~68\%$\\
			S-type&~0.05&~200.0&~300.0&~299.64&~280.0&$10^{-4}$&$0.2687$&0.1198&$ 5.7 \times 10^{-49}$& $H_1 \,H_1\rightarrow \nu_l \overline{\nu}_l~~30\%$\\
			BP-10&&&&&&&&&&$H_1 \,H_1\rightarrow WW~~1\%$\\
		\hline
		&&&&&&&&&&$H_1 \,H_1\rightarrow l \overline{l}~~56\%$\\
			S-type&~0.05&~600.0&~700.0&~699.56&~680.0&$10^{-4}$&$0.2924$&0.1198&$ 1.8 \times 10^{-47}$& $H_1 \,H_1\rightarrow \nu_l \overline{\nu}_l~~30\%$\\
			BP-11&&&&&&&&&&$H_1 \,H_1\rightarrow WW~~7\%$\\
			    &&&&&&&&&&$H_1 \,H_1\rightarrow ZZ~~4\%$\\
			    &&&&&&&&&&$H_1 \,H_1\rightarrow hh~~4\%$\\
		\hline
		\hline
	\end{tabular}}
\end{center}
	\caption{ Higgs portal coupling is tiny. $H_1$ is mostly coming from CP-even component of $Z_2$-odd scalar doublet. The fermionic $t,u$-channel contributions large. }
	\label{tabDM:s2}
\end{table*}
%%%%%%%%%%%%%%%%%
%%%%%%%%%%%%%%%%%
%%%%%%%%%%%%%%%%%
%%%%%%%%%%%%%%%%%
\begin{figure}[h!]
	\begin{center}
		\subfigure[]{
			\includegraphics[scale=0.3]{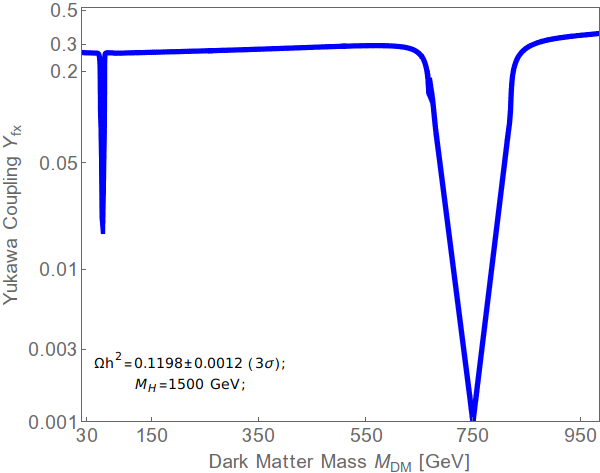}}
		\hskip 1pt
		\subfigure[]{
			\includegraphics[scale=0.3]{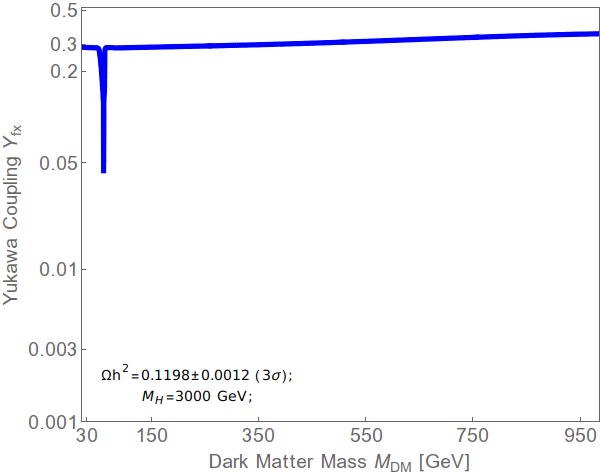}}
			
		\caption{  We varied DM (mostly singlet type, i.e., mixing angle $\cos\beta=0.05$) mass and new Yukawa coupling $Y_{f_x}$ with  $Y_{f,d,\nu}=0$, Fermion masses $\sim1$ TeV, $M_{H}=1.5$ TeV and  $M_{H}=3$ TeV . The blue line gives relic density $\Omega h^2=0.1198\pm 0.0012$ within $3\sigma$. }
		\label{fig:DarkPlots1c}
	\end{center}
\end{figure}
%%%%%%%%%%%%%%%%%

We illustrate a permissible range of relic density within the $Y_{fx} - M_{H_1}$ parameter space, as depicted in Figure~\ref{fig:DarkPlots1c}. The plots have been produced for two distinct heavy Higgs mass values, namely $M_{H} = 1.5$ TeV and $M_{H} = 3$ TeV while the mass of the newly introduced fermions is $1$ TeV. The funnel regions appear due to the CP even scalar resonance in the annihilation channel, which would mean a dip in the contributions coming from the fermionic channels and therefore smaller values (dips in the figure) for the Yukawa couplings.
%%%%%%%%%%%%%%%%%
\begin{figure}[h!]
	\begin{center}
%		\subfigure[]{
%			\includegraphics[scale=0.3]{./plot/PLot1NocoanniMDMkappaD.png}}
%		\hskip 1pt
		\subfigure[]{
			\includegraphics[scale=0.4]{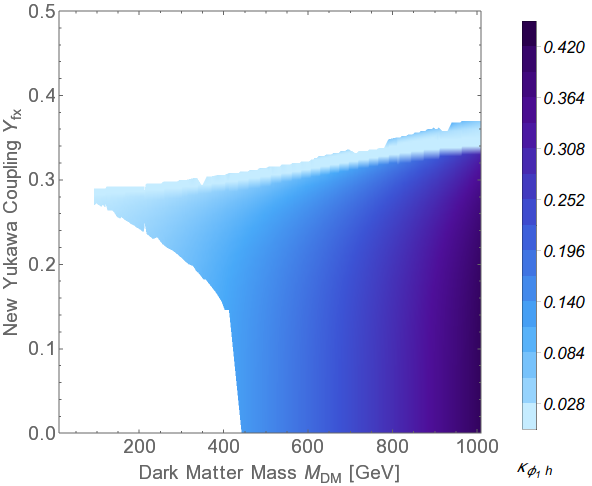}}			%
		\caption{  We varied DM mass, Higgs portal coupling $\kappa_{\phi_1 s_2}$ and Yukawa couplings $Y_{fx}$ here. All the blue shaded region provides relic density $\Omega h^2=0.1198\pm 0.0012$ within $3\sigma$, also satisfy all the theoretical and experimental constraints.}
		\label{fig:DarkPlot2}
	\end{center}
\end{figure}
%%%%%%%%%%%%%%%%%
%%%%%%%%%%%%%%%%%

We also conduct a parameter scan involving variations in the Higgs portal coupling, new Yukawa coupling $Y_{fx}$, and DM mass. The results of this scan are presented in Fig.~\ref{fig:DarkPlot2}. The plot highlights regions in the coupling strengths of both fermionic and scalar interactions with the DM that contribute together to give the correct relic density. The definitive breakup of the regions where each are dominant has been discussed earlier. This plot is aimed at showing the wide range of the parameter choices that can achieve the correct relic density while satisfying all experimental constraints in the model.  
%We have obtained a comprehensive view of the parameter space by exploring a range of values for the Higgs portal coupling, New Yukawa couplings, and DM mass. 
In our analysis, we have maintained certain parameters fixed, specifically setting $Y_{fd}=Y_{f\nu}\approx 0$, fermion masses at approximately 1 TeV, $M_H=3$ TeV, and $M_{Z'}=3$ TeV. The color points in the plot represent parameter combinations that satisfy the current relic density bound while adhering to all relevant theoretical and experimental constraints. Conversely, the empty region in the plot indicates parameter combinations that violate at least one of the imposed bounds. It could be due to the inconsistencies in DM relic density,  detection experiments, Higgs decay width when the DM mass is less than half of the Higgs mass or any other theoretical or experimental constraints considered in the analysis.

%%%%%%%%%%%%%%%%%
%%%%%%%%%%%%%%%%%
\begin{figure}[h!]
	\begin{center}
		\subfigure[]{
			\includegraphics[scale=0.42]{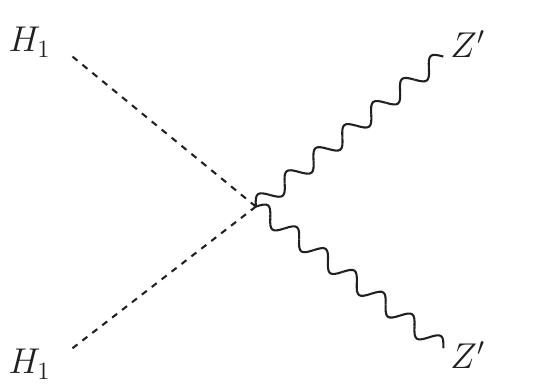}}
		\hskip 1pt
		\subfigure[]{
			\includegraphics[scale=0.42]{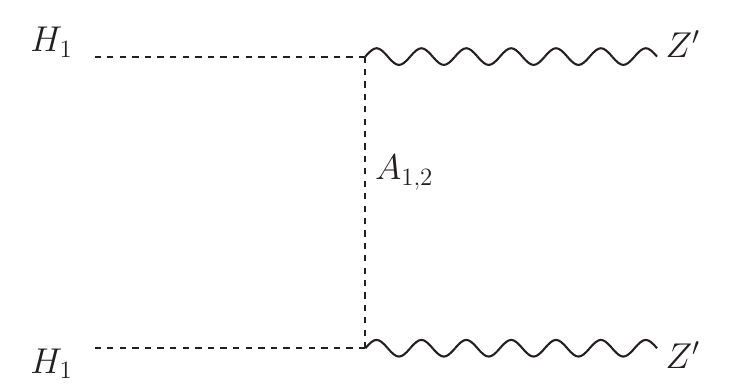}}
		\hskip 1pt	
		\subfigure[]{
			\includegraphics[scale=0.42]{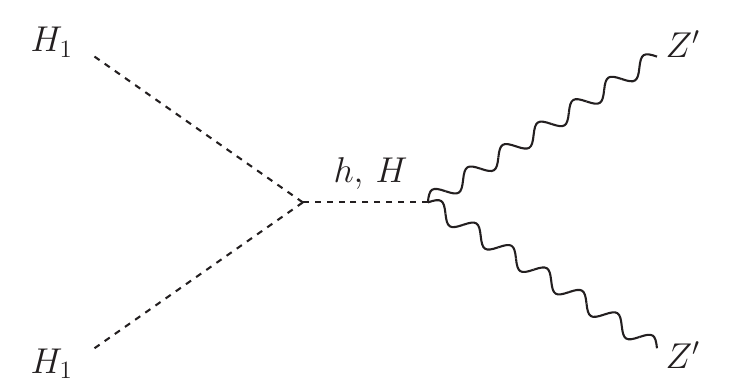}}
		\hskip 1pt	
		\subfigure[]{
			\includegraphics[scale=0.42]{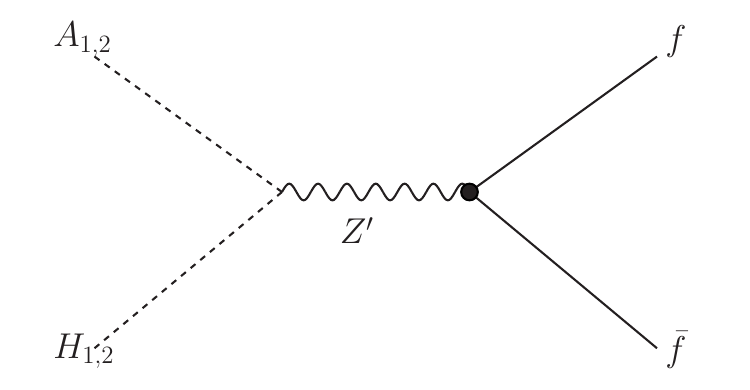}}		
		\caption{  Diagrams: DM annihilation into the new gauge bosons and fermions via gauge couplings.  $H_1$ is the DM candidate, can be doublet or singlet type or admixture of these depending on the mixing angle $\beta$.}
		\label{fig:Daigzpzp}
	\end{center}
\end{figure}
%%%%%%%%%%%%%%%%%
%%%%%%%%%%%%%%%%%

\begin{figure}[h!]
	\begin{center}
		\subfigure[]{
			\includegraphics[width=0.49\textwidth]{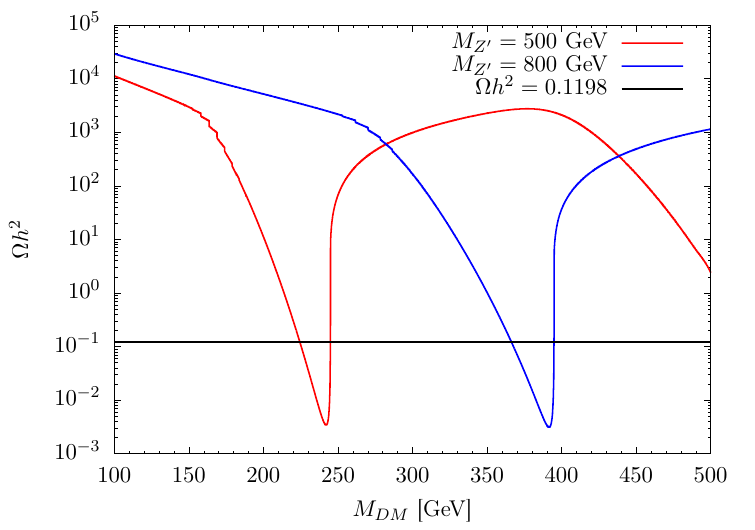}}	
		\caption{  We varied DM mass in the first plot for fixed $M_{ Z^\prime}=0.5$ and $0.8$ TeV. The kinetic mixing $g'_{x}=5\times10^{-3}$ and $ M_{A} - M_{H_1} = 10$ GeV. All the Higgs portal couplings are zero in this plot.}
		\label{fig:DarkPlotkinmix}
	\end{center}
\end{figure}

So far, we have discussed the role of scalar portal couplings and Yukawa couplings of the DM in achieving the observed DM relic density. When these couplings are very small, the DM annihilation is not efficient enough for obtaining the correct relic density. In such cases, we rely on the DM annihilation channels involving the new gauge boson $Z^\prime$ and the coupling $g_{x}$, as depicted in Fig.~\ref{fig:Daigzpzp}. The $Z^\prime$  directly couples to the DM particle, facilitating its production and annihilation processes in the early Universe and determining its relic density. Let's first consider the case where the DM mass ($M_{DM}$) is less than $M_{Z^\prime}$. In this scenario, DM can annihilate through the processes $H_1 H_1 \to Z^{\prime*}  Z^{\prime*}  \, (Z^{\prime*} \to f\bar{f})$ and/or $H_1 A_1 \to Z^\prime \to f\bar{f}$. Here, $A_1$ represents the singlet-type pseudo-scalar component, and $f$ denotes the SM fermions. 
The co-annihilation of $M_{H_1}$ and $M_{A_1}$ to SM fermions becomes significant near $M_{H_1} + M_{A_1} \sim M_{Z^\prime}$ due to a resonance effect in the channel $H_1 A_1 \to Z^\prime \to f\bar{f}$, that helps DM relic density to be satisfied . This effect is illustrated in Fig.~\ref{fig:DarkPlotkinmix} for $M_{Z^\prime}=500, 800$ GeV, where a non-zero value of gauge kinetic mixing $g'_x = 5\times 10^{-3}$ is considered. 
%%%%%%%%%%%%%%%%%
%%%%%%%%%%%%%%%%%

When the mass of DM exceeds the $M_{Z^\prime}$, the annihilation process of DM into $Z^\prime Z^\prime$ becomes possible kinematically, leading to an increase in the total annihilation cross-section. Therefore, in this region we rely on the DM annihilation process $H_1 H_1 \rightarrow Z^\prime Z^\prime$ to obtain the correct abundance of the DM in the Universe.  

We estimate the contributions of the diagrams for the above process in the limit where only the gauge interactions are considered. 
Hence, only diagrams (a) and (b) of Fig.~\ref{fig:Daigzpzp} contribute to the DM annihilation. 
We show some benchmark points for this scenario and the corresponding contributions of the annihilation channels in Table~\ref{tabDM:s1z}.
%%%%%%%%%%%%%%%%%  
\begin{table*}[h!]
\begin{center}\scalebox{0.8}{
	\begin{tabular}{|p{2.0cm}|p{1.25cm}|p{1.2cm}|p{1.2cm}|p{1.2cm}|p{1.7cm}|p{1.4cm}|p{1.2cm}|p{1.2cm}|p{2.0cm}|p{4.4cm}|}
		\hline
		\hline
		Sl. No. & $M_{DM}$ (GeV) & $M_{H_2}$ (GeV) & $M_{A_1}$ (GeV) & $M_{H^\pm}$ (GeV) &$M_{Z^\prime }$ (GeV)  & $g_{x}$ &~~$g_{k,mix}$&$\Omega_{DM}h^2$&DD Cross (${\rm cm^3/s}$)&~~~~Contributions \\
		\hline
	    \hline
		&&&&&&&&&&\\
			S-type&~300.0&~700.0&~600.0&500.0& $0.50\,M_{DM}$ &$0.1350$&$0.0$&0.1198&$ < 10^{-60}$& $H_1 \,H_1\rightarrow Z^\prime Z^\prime~~100\%$\\
			BP-12&&&&&&&&&&\\
	    \hline
		&&&&&&&&&&\\
			S-type&~300.0&~700.0&~600.0&500.0& $0.75\,M_{DM}$ &$0.2103$&$0.0$&0.1198&$ < 10^{-60}$& $H_1 \,H_1\rightarrow Z^\prime Z^\prime ~~100\%$\\
			BP-13&&&&&&&&&&\\
	    \hline
		&&&&&&&&&&\\
			S-type&~300.0&~700.0&~600.0&500.0& $0.90\,M_{DM}$ &$0.2626$&$0.0$&0.1198&$ < 10^{-60}$& $H_1 \,H_1\rightarrow Z^\prime Z^\prime ~~100\%$\\
			BP-14&&&&&&&&&&\\
		\hline
		\hline
		&&&&&&&&&&\\
			S-type&~800.0&~1000.0&~900.0&850.0& $0.50\,M_{DM}$ &$0.2794$&$0.0$&0.1198&$ < 10^{-60}$& $H_1 \,H_1\rightarrow Z^\prime Z^\prime ~~100\%$\\
			BP-15&&&&&&&&&&\\
	    \hline
		&&&&&&&&&&\\
			S-type&~800.0&~1000.0&~900.0&850.0& $0.75\,M_{DM}$ &$0.3863$&$0.0$&0.1198&$ < 10^{-60}$& $H_1 \,H_1\rightarrow Z^\prime Z^\prime ~~100\%$\\
			BP-16&&&&&&&&&&\\
	    \hline
		&&&&&&&&&&\\
			S-type&~800.0&~1000.0&~900.0&850.0& $0.90\,M_{DM}$ &$0.4497$&$0.0$&0.1198&$ < 10^{-60}$& $H_1 \,H_1\rightarrow Z^\prime Z^\prime ~~100\%$\\
			BP-17&&&&&&&&&&\\
		\hline
		\hline
	\end{tabular}}
\end{center}
	\caption{ Higgs portal coupling is tiny. $H_1$ is mostly coming from CP-even component of $Z_2$-odd scalar singlet, $\cos\beta=0.05$. The gauge kinetic mixing $g_{k,mix}$ is taken to be zero. The heavy scalar mass at $M_H=3.0$ TeV.}
	\label{tabDM:s1z}
\end{table*}
%%%%%%%%%%%%%%%%%
%%%%%%%%%%%%%%%%%
\begin{figure}[h!]
	\begin{center}
		\subfigure[]{
			\includegraphics[scale=0.350]{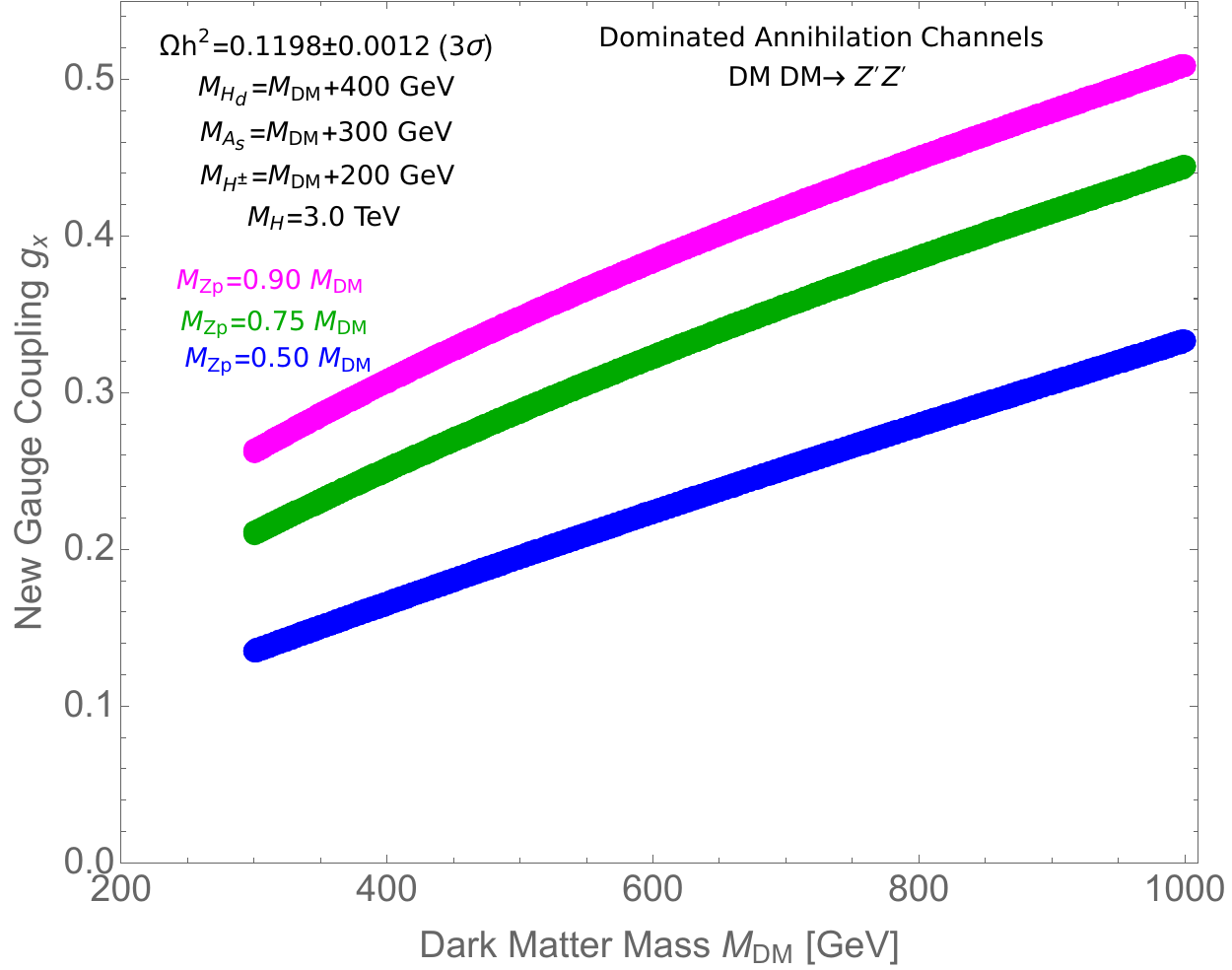}}	
		\subfigure[]{
			\includegraphics[scale=0.350]{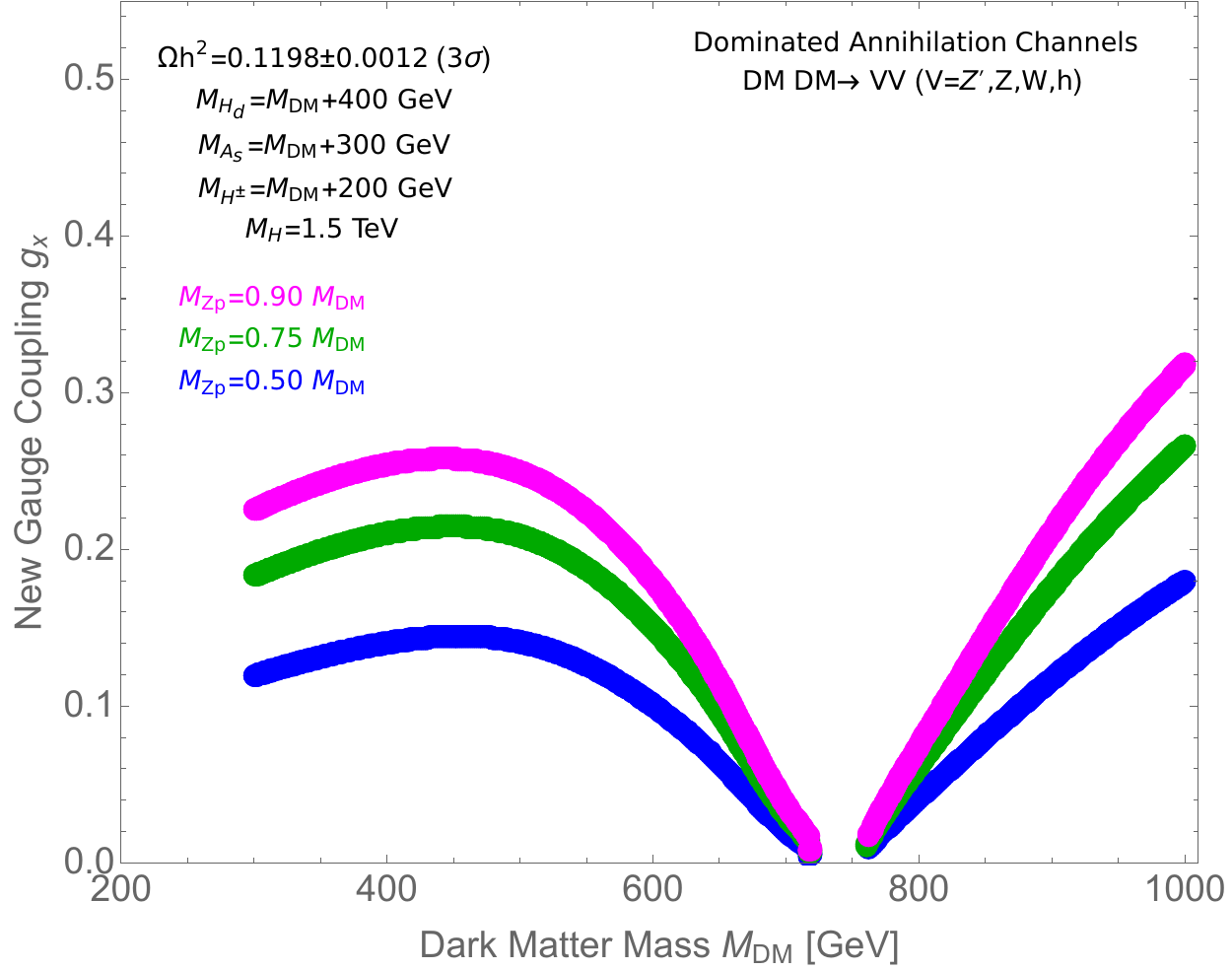}}
		\caption{ We varied DM mass, gauge coupling $g_{x}$. The line gives relic density $\Omega h^2=0.1198\pm 0.0012$ within $3\sigma$, also satisfy all the theoretical and experimental constraints.}
		\label{fig:DarkPlot2zz}
	\end{center}
\end{figure}
%%%%%%%%%%%%%%%%%
%%%%%%%%%%%%%%%%%
We present the observed $3\sigma$ band of the DM relic density in the 
$g_x$-$M_{\rm{DM}}$ plane for three different values of $n_f$, in Fig.~\ref{fig:DarkPlot2zz} (a), where $n_f=M_{Z^\prime}/M_{DM}$. The plot can be interpreted in the following manner: for a given value of $M_{DM}$, increasing the $Z'$ mass suppresses the DM annihilation cross section. A larger $g_{x}$ value will then be needed to enhance the $<\sigma v>$. Therefore the three curves representing the different ratios of the DM mass and $Z^\prime$ illustrate that heavier the DM mass, larger the gauge coupling $g_x$ helps satisfy the correct relic density.
%%%%%%%%%%%%%%%%%  
\begin{table*}[h!]
\begin{center}\scalebox{0.8}{
	\begin{tabular}{|p{2.0cm}|p{1.25cm}|p{1.2cm}|p{1.2cm}|p{1.2cm}|p{1.7cm}|p{1.4cm}|p{1.2cm}|p{1.2cm}|p{2.0cm}|p{4.4cm}|}
		\hline
		\hline
		Sl. No. & $M_{DM}$ (GeV) & $M_{H_2}$ (GeV) & $M_{A_1}$ (GeV) & $M_{H^\pm}$ (GeV) &$M_{Z^\prime }$ (GeV)  & $g_{x}$ &~~$g_{k,mix}$&$\Omega_{DM}h^2$&DD Cross (${\rm cm^3/s}$)&~~~~Contributions \\
		\hline
	    \hline
		&&&&&&&&&&$H_1 \,H_1\rightarrow Z^\prime Z^\prime ~~54\%$\\
			S-type&~300.0&~700.0&~600.0&500.0& $0.50\,M_{DM}$ &$0.1194$&$0.0$&0.1198&$ 3.0 \times 10^{-46}$& $H_1 \,H_1\rightarrow WW ~~21\%$\\
			BP-18&&&&&&&&&&$H_1 \,H_1\rightarrow hh ~~13\%$\\
			&&&&&&&&&&$H_1 \,H_1\rightarrow ZZ ~~9\%$\\
			&&&&&&&&&&$H_1 \,H_1\rightarrow t\bar{t} ~~4\%$\\
	    \hline
		&&&&&&&&&&$H_1 \,H_1\rightarrow Z^\prime Z^\prime ~~51\%$\\
			S-type&~300.0&~700.0&~600.0&500.0& $0.75\,M_{DM}$ &$0.1831$&$0.0$&0.1198&$ 2.8 \times 10^{-46}$& $H_1 \,H_1\rightarrow WW ~~22\%$\\
			BP-19&&&&&&&&&&$H_1 \,H_1\rightarrow hh ~~13\%$\\
			&&&&&&&&&&$H_1 \,H_1\rightarrow ZZ ~~10\%$\\
			&&&&&&&&&&$H_1 \,H_1\rightarrow t\bar{t} ~~4\%$\\
	    \hline
		&&&&&&&&&&$H_1 \,H_1\rightarrow Z^\prime Z^\prime ~~49\%$\\
			S-type&~300.0&~700.0&~600.0&500.0& $0.90\,M_{DM}$ &$0.2250$&$0.0$&0.1198&$ 2.7 \times 10^{-46}$& $H_1 \,H_1\rightarrow WW ~~23\%$\\
			BP-20&&&&&&&&&&$H_1 \,H_1\rightarrow hh ~~14\%$\\
			&&&&&&&&&&$H_1 \,H_1\rightarrow ZZ ~~10\%$\\
			&&&&&&&&&&$H_1 \,H_1\rightarrow t\bar{t} ~~4\%$\\
		\hline
		\hline
		%%%%%%%%%%%
		&&&&&&&&&&$H_1 \,H_1\rightarrow Z^\prime Z^\prime ~~5\%$\\
			S-type&~800.0&~1000.0&~900.0&800.0& $0.50\,M_{DM}$ &$0.0414$&$0.0$&0.1198&$ 3.0 \times 10^{-46}$& $H_1 \,H_1\rightarrow WW ~~46\%$\\
			BP-21&&&&&&&&&&$H_1 \,H_1\rightarrow hh ~~24\%$\\
			&&&&&&&&&&$H_1 \,H_1\rightarrow ZZ ~~20\%$\\
			&&&&&&&&&&$H_1 \,H_1\rightarrow t\bar{t} ~~4\%$\\
	    \hline
		&&&&&&&&&&$H_1 \,H_1\rightarrow Z^\prime Z^\prime ~~3\%$\\
			S-type&~800.0&~1000.0&~900.0&800.0& $0.75\,M_{DM}$ &$0.0621$&$0.0$&0.1198&$ 2.8 \times 10^{-46}$& $H_1 \,H_1\rightarrow WW ~~46\%$\\
			BP-22&&&&&&&&&&$H_1 \,H_1\rightarrow hh ~~25\%$\\
			&&&&&&&&&&$H_1 \,H_1\rightarrow ZZ ~~20\%$\\
			&&&&&&&&&&$H_1 \,H_1\rightarrow t\bar{t} ~~4\%$\\
	    \hline
		&&&&&&&&&&$H_1 \,H_1\rightarrow Z^\prime Z^\prime ~~2\%$\\
			S-type&~800.0&~1000.0&~900.0&800.0& $0.90\,M_{DM}$ &$0.0746$&$0.0$&0.1198&$ 2.7 \times 10^{-46}$& $H_1 \,H_1\rightarrow WW ~~47\%$\\
			BP-23&&&&&&&&&&$H_1 \,H_1\rightarrow hh ~~25\%$\\
			&&&&&&&&&&$H_1 \,H_1\rightarrow ZZ ~~20\%$\\
			&&&&&&&&&&$H_1 \,H_1\rightarrow t\bar{t} ~~4\%$\\
		\hline
		\hline
	\end{tabular}}
\end{center}
	\caption{ Higgs portal coupling is tiny. $H_1$ is mostly scalar singlet-type DM. The gauge kinetic mixing $g_{k,mix}$ is zero with $M_H=1.5$ TeV.}
	\label{tabDM:s2zp}
\end{table*}
%%%%%%%%%%%%%%%%%
%%%%%%%%%%%%%%%%%

We now examine how the presence of scalar portal coupling of the DM changes the DM relic density computation in the region $M_{\rm{DM}} > M_{Z'}$. We consider the scenario with the heavy Higgs-mediated channels alongside the above mentioned channels. This also allows DM to annihilate into two $Z'$ bosons via the heavy scalar $H$ in an s-channel process. Figure~\ref{fig:DarkPlot2zz}(b) illustrates the influence of these extra channels on DM relic density, with $M_{H} = 1.5$ TeV. The dip in $g_x$ around $M_{DM}\approx 750\, \rm{GeV}$ is a result of the heavy scalar resonance. Due to the presence of this additional annihilation channels for DM, a smaller $g_x$ is required compared to Figure~\ref{fig:DarkPlot2zz}(a) to achieve the correct DM relic density. We show some benchmark points of this type and corresponding contribution of the annihilation channels in Table~\ref{tabDM:s2zp}.

%\clearpage
%%%%%%%%%%%%%%%%%%%%%%%%%%%%%%%%%
%%%%%%%%%%%%%%%%%%%%%%%%%%%%%%%%%
\section{Collider Signature}
\label{sec:5}
%%%%%%%%%%%%%%%%%
%%%%%%%%%%%%%%%%%
We now briefly discuss some collider signatures of the model. Besides the usual signatures of a $Z^\prime$ at colliders \cite{Huitu:2008gf,Das:2017fjf,Grossmann:2010wm,Abdallah:2021npg,Abdallah:2021dul}, the presence of a new fermionic sector which contains an $SU(2)$ doublet and SM singlet in addition to being odd under a discrete $Z_2$ symmetry, can lead to many interesting signatures of the model at collider experiments. Because of the $Z_2$ symmetry, these fermions when produced will always lead to signatures with missing energy in the final state. We highlight only a 
few of the signals here that give the mono-lepton and di-lepton $+\slashed{E}_T$ with or without jets, multi-jet $+\slashed{E}_T$, etc. final states at the LHC. The production channels of these BSM scalars and fermions at LHC have been listed in table \ref{tab:lhc1} and \ref{tab:lhc2} along with their possible decay channels.
%%%%%%%%%%%%%%%%%%%%%%%%%%%%%%%%
%%%%%%%%%%%%%%%%%%%%%%%%%%%%%%%%%
\begin{table}[h]
	%\centering
	\begin{center}\scalebox{1.0}{
			\begin{tabular}{|c|c|c|}
				\hline Processes &  Mediators &  Decay   \\ \hline
				%\hline
				$p p \to H^+ H^-$  & $\gamma,Z,Z'$     & $H^\pm\to l^\pm N_{1},\psi^{\pm} \nu,W^\pm H_1$                   \\
               $p p \to H_1 H^\pm$  & $W^\pm$     & $H^\pm\to l^\pm N_{1},\psi^{\pm} \nu,W^\pm H_1$                   \\
				$p p \to H_1 A_1$  & $Z,Z'$      & $A_1\to \nu N_{1},Z H_1,Z'^{*} H_1, \psi^{\pm} l^\mp $                  \\
				$p p \to H_2 A_2$      & $Z'$  & $H_2 \to \psi^{\pm} l^\mp, \nu N_{1} $, $A_2 \to \psi^{\pm} l^\mp, \nu N_{1},Z'^{*} H_2$            \\ 
				 \hline
		\end{tabular}}
	\end{center}
	\caption{ The production channels of the dark sector scalars at LHC and their decay channels.}
	\label{tab:lhc1}
\end{table}
\begin{table}[h]
	%\centering
	\begin{center}\scalebox{1.0}{
			\begin{tabular}{|c|c|c|}
				\hline Processes &  Mediators &  Decay   \\ \hline
				%\hline
				$p p \to \psi^+ \psi^-$  & $\gamma,Z,Z'$     & $\psi^\pm\to l^\pm H_1, W^\pm N_{1}$                   \\
%				$p p \to N_i N_j$  & $Z,Z'$      & $N_i\to \nu H_{1} $                  \\
				$p p \to \psi^\pm N_{1}$      & $W^\pm$  & $\psi^\pm\to l^\pm H_1,W^\pm N_{1} $,$N_1\to \nu H_{1} $             \\ 
				 \hline
		\end{tabular}}
	\end{center}
	\caption{ The production channels of the dark sector fermions at LHC and their decay channels.}
	\label{tab:lhc2}
\end{table}

We now discuss how the above final states are realized at LHC. The process $p p \to \psi^+ \psi^-$, where $\psi^{\pm}$ represent the new charged fermions. These fermions can subsequently decay as $\psi^{\pm} \to W^\pm N_1$, with $N_1$ being the lightest BSM neutral fermion, followed by the decay $N_1 \to H_1 \, \nu_{i}$ which is a completely invisible decay channel since $H_1$ is the DM. Thus the pair production of $\psi^+ \psi^-$ will lead to $W^+ W^-$ and large missing energy in the final state. Therefore this channel with the above decay sequences gives rise to di-lepton$+\slashed{E}_T$, $mono-lepton +$dijet$+\slashed{E}_T$ or
$4$ jets$+\slashed{E}_T$ final states. The charged fermion $\psi^{\pm}$ 
 can also decay via $\psi^{\pm} \to H_{1} \, l$ for a given range of parameters. As $H_1$ is invisible, this decay sequence leads to the dilepton$+\slashed{E}_T$ signature. 
 
 Consider the process $p p \to \psi^\pm N_{1}$ which is mediated by the 
 $W$ boson. Following the sequence of decays discussed above, the 
 $\psi^{\pm} \to H_{1} \, l;  W^\pm N_1 $, both modes lead to mono-lepton$+\slashed{E}_T$ signature. Additionally, when the 
 $\psi^\pm$ decays to $W^\pm N_1$, and $W$ decays hadronically, we get the di-jet$+\slashed{E}_T$ final state as our signal. 
 
 The scalar production mode given by the process $p p \to H^+ H^-$, with the fermionic decay channels of $H^\pm$ as shown table~\ref{tab:lhc1}, also leads to the same final states as discussed above, with the only difference being the primary production channel of $H^+ \, H^-$ giving additional sources of missing energy in the cascades.  
 Similarly The process $p p \to H_1 H^\pm$ with $H^\pm \to W^\pm H_1$ can give rise to di-jet$+\slashed{E}_T$ and mono-lepton$+\slashed{E}_T$ final state. It is worth noting that these channels can exhibit a significant production cross-section at the LHC as these BSM particles have electroweak interactions. Moreover, as these particles are charged under the new gauge symmetry an additional diagram can contribute to the BSM scalar and fermion production at LHC via the $Z^\prime$ boson by introducing a non-zero gauge kinetic mixing. The $Z^\prime$ exchange in the $s$-channel can lead to a significant enhancement if the $Z^\prime$ is produced at resonance. In the process $p p \to A_{1,2} H_{1,2}$ the  pseudoscalar $A_{1,2}$ can subsequently decay as $A_{1,2} \to Z/Z' H_{1,2}$, where $Z/Z'$ denotes either the Standard Model $Z$ boson or a new neutral gauge boson. The $Z/Z'$ boson further decays in the visible channel as $Z/Z' \to ll$ or dijet, where $l$ represents the 
 SM charged lepton. Once again, this decay chain will result in the dilepton/dijet$+\slashed{E}_T$ signature. 

\begin{figure}[h!]
	\begin{center}
			\includegraphics[scale=0.350]{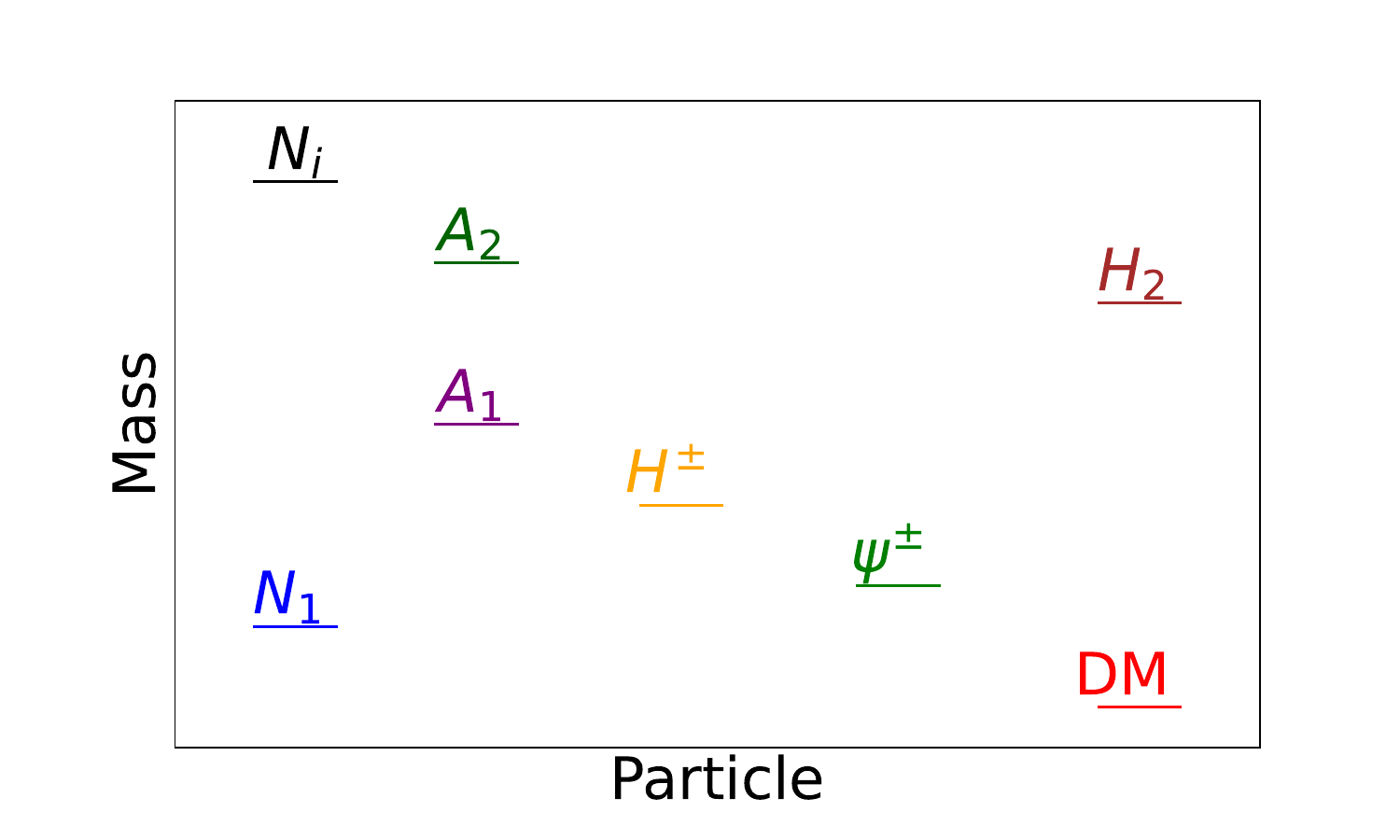}
		\caption{ The specific mass hierarchy for which collider signal has been discussed.}
		\label{collider}
	\end{center}
\end{figure}

It is worth noting that the discussion on the collider signal has been limited to a very specific mass hierarchy in the particle spectrum in our current discussion (see fig \ref{collider}). While there are indeed additional signals possible through various decay cascades of the produced particles in this model, a comprehensive analysis of these model particles and their properties is left for future work. These studies will aim to extract meaningful information about the properties of new particles, their interactions, and their role in the context of neutrino and DM including other constraints.
%%%%%%%%%%%%%%%%%
\section{Conclusions}
\label{sec:6}
%%%%%%%%%%%%%%%%%
%%%%%%%%%%%%%%%%%

In this study, we consider an additional $U(1)$ gauge symmetry along with the SM gauge symmetry. In addition to the standard model particles, our model incorporates vector-like doublet and singlet fermions, as well as two complex singlets and one doublet scalar field, which are charged under this new $U(1)$ gauge symmetry. All the SM particle contents are neutral under the new abelian gauge symmetry. We explore the neutrino masses, mixing, and scalar dark matter in this model. We successfully achieve neutrino oscillation parameters and DM density within the expected range by selecting this field content.
The neutrino mass and mixing angles are influenced significantly by the new Yukawa couplings and the mass differences between the $Z_2$-odd CP-even and CP-odd scalar particles. These parameters need to be extremely small to explain the observed neutrino properties. Interestingly, these same parameters are also directly linked to the DM abundance in the Universe. When the mass gap between these scalar particles is small, co-annihilation processes become more prominent, while smaller Yukawa couplings result in reduced annihilation rates. We have examined these parameters extensively through careful analysis and discovered a significant region within the parameter space that yields interesting results. 

To account for the observed relic density, we consider annihilation and co-annihilation processes involving other particles in the model. Moreover, the new Yukawa terms enable the DM to annihilate into standard model particles through cross-channel, $t$-channel, and $u$-channel interactions. The interplay between these channels, characterized by constructive or destructive interference, effectively modifies the annihilation cross-section. Consequently, our model obtains the correct relic density for the DM.

 The $Z^\prime$ boson is crucial in obtaining the correct DM relic density as the DM is charged under the new gauge symmetry. Even in the absence of other portal coupling, we find that the new gauge coupling provides correct DM density obeying all the experimental and theoretical bounds.

\subsection*{Acknowledgments}
The authors would like to acknowledge the support from the Department of Atomic Energy(DAE), India, for the Regional Centre for Accelerator-based Particle Physics (RECAPP), Harish Chandra Research Institute.

\section{Appendix}\textbf{}
\label{sec:app}
%%%%%%%%%%%%%%%%%%
\vspace*{-2cm}
\subsection{Stability constraints}
%%%%%%%%%%%%%%%%%%
The additional copositivity criteria~\cite{Chakrabortty:2013mha} to have scalar potential bounded from below all the way.
%%%%%%%%%%%%%%%%%%
%%%%%%%%%%%%%%%%%%
\bea
& \text{If} \,\, \left(\lambda _3+\lambda _4\right)<0 \,\&\, \lambda _{11}<0,\,\lambda _1 \kappa _{s_1 \phi _2}-\left(\lambda _3+\lambda _4\right) \lambda _{11}   \,\geq \,   \sqrt{\left( \,4 \, \lambda _1 \lambda _2-\left(\lambda _3+\lambda _4\right){}^2\right) \left( \,4 \, \lambda _1 \lambda _{s2}-\lambda _{11}^2\right)}, \nn\\
%%%%%%%%%%%%%%%%%
&  \text{If} \,\, \left(\lambda _3+\lambda _4\right)<0\, \&\,\lambda _{22}<0,\lambda _2 \kappa _{s_2 \phi _1}-\left(\lambda _3+\lambda _4\right) \lambda _{22}  \,\geq \,  \sqrt{\left( \,4 \, \lambda _1 \lambda _2-\left(\lambda _3+\lambda _4\right){}^2\right) \left( \,4 \, \lambda _2 \lambda _{s2}-\lambda _{22}^2\right)} ,\nn \\
%%%%%%%%%%%%%%%%%
&  \text{If} \,\, \left(\lambda _3+\lambda _4\right)<0\,\&\,\kappa _{s_2 \phi _1}<0,\lambda _1 \lambda _{22}-\left(\lambda _3+\lambda _4\right) \kappa _{s_2 \phi _1}  \,\geq \,  \sqrt{\left( \,4 \, \lambda _1 \lambda _2-\left(\lambda _3+\lambda _4\right){}^2\right) \left( \,4 \, \lambda _1 \lambda _{s2}-\kappa _{s_2 \phi _1}^2\right)} ,\nn\\
%%%%%%%%%%%%%%%%%
&  \text{If} \,\, \left(\lambda _3+\lambda _4\right)<0\,\&\,\kappa _{s_1 \phi _2}<0,\lambda _2 \lambda _{11}-\left(\lambda _3+\lambda _4\right) \kappa _{s_1 \phi _2}  \,\geq \,  \sqrt{\left( \,4 \, \lambda _1 \lambda _2-\left(\lambda _3+\lambda _4\right){}^2\right) \left( \,4 \, \lambda _2 \lambda _{s2}-\kappa _{s_1 \phi _2}^2\right)}, \nn\\
%%%%%%%%%%%%%%%%%
&  \text{If} \,\, \lambda _{11}<0\, \&\,\kappa _{s_2 \phi _1}<0,\lambda _1 \kappa _{3}-\lambda _{11} \kappa _{s_2 \phi _1}  \,\geq \,  \sqrt{\left( \,4 \, \lambda _1 \lambda _{s2}-\lambda _{11}^2\right) \left( \,4 \, \lambda _1 \lambda _{s2}-\kappa _{s_2 \phi _1}^2\right)},\nn \\
%%%%%%%%%%%%%%%%%
& \text{If} \,\, \kappa _{s_1 \phi _2}<0\,\&\,\lambda _{22}<0,\lambda _2 \kappa _{3}-\lambda _{22} \kappa _{s_1 \phi _2}  \,\geq \,  \sqrt{\left( \,4 \, \lambda _2 \lambda _{s2}-\lambda _{22}^2\right) \left( \,4 \, \lambda _2 \lambda _{s2}-\kappa _{s_1 \phi _2}^2\right)}, \nn \\
%%%%%%%%%%%%%%%%%
&  \text{If} \,\, \lambda _{22}<0\,\&\,\kappa _{s_1 \phi _2}<0,\lambda _2 \kappa _{3}-\lambda _{22} \kappa _{s_1 \phi _2}  \,\geq \,  \sqrt{\left(\lambda _2 \lambda _{s2}-\lambda _{22}^2\right) \left(\lambda _2 \lambda _{s2}-\kappa _{s_1 \phi _2}^2\right)},\nn \\
%%%%%%%%%%%%%%%%%
&  \text{If} \,\, \lambda _{11}<0\, \&\,\kappa _{s_1 \phi _2}<0,\left(\lambda _3+\lambda _4\right) \lambda _{s2}-\lambda _{11} \kappa _{s_1 \phi _2}  \,\geq \,  \sqrt{\left( \,4 \, \lambda _1 \lambda _{s2}-\lambda _{11}^2\right) \left( \,4 \, \lambda _2 \lambda _{s2}-\kappa _{s_1 \phi _2}^2\right)},\nn \\
%%%%%%%%%%%%%%%%%
&  \text{If} \,\, \lambda _{11}<0\,\&\,\kappa _{3}<0,\lambda _{s2} \kappa _{s_2 \phi _1}-\lambda _{11} \kappa _{3}  \,\geq \,  \sqrt{\left( \,4 \, \lambda _1 \lambda _{s2}-\lambda _{11}^2\right) \left( \,4 \, \lambda _{s2} \lambda _{s2}-\kappa _{3}^2\right)},\nn \\
%%%%%%%%%%%%%%%%%
&  \text{If} \,\, \kappa _{s_1 \phi _2}<0\,\&\,\kappa _{3}<0,\lambda _{22} \lambda _{s2}-\kappa _{3} \kappa _{s_1 \phi _2}  \,\geq \,  \sqrt{\left( \,4 \, \lambda _2 \lambda _{s2}-\kappa _{s_1 \phi _2}^2\right) \left( \,4 \, \lambda _{s2} \lambda _{s2}-\kappa _{3}^2\right)} \nn \\
%%%%%%%%%%%%%%%%%
& \text{If} \,\, \kappa _{3}<0\,\&\,\kappa _{s_1 \phi _2}<0,\lambda _{22} \lambda _{s2}-\kappa _{3} \kappa _{s_1 \phi _2}  \,\geq \,  \sqrt{\left( \,4 \, \lambda _2 \lambda _{s2}-\kappa _{s_1 \phi _2}^2\right) \left( \,4 \, \lambda _{s2} \lambda _{s2}-\kappa _{3}^2\right)}, \nn \\
%%%%%%%%%%%%%%%%%
& \text{If} \,\, \lambda _{22}<0\,\&\,\kappa _{s_2 \phi _1}<0,\left(\lambda _3+\lambda _4\right) \lambda _{s2}-\lambda _{22} \kappa _{s_2 \phi _1}  \,\geq \,  \sqrt{\left( \,4 \, \lambda _2 \lambda _{s2}-\lambda _{22}^2\right) \left( \,4 \, \lambda _1 \lambda _{s2}-\kappa _{s_2 \phi _1}^2\right)}, \nn \\
%%%%%%%%%%%%%%%%%
&  \text{If} \,\, \lambda _{22}<0\,\&\,\kappa _{3}<0,\lambda _{s2} \kappa _{s_1 \phi _2}-\lambda _{22} \kappa _{3}  \,\geq \,  \sqrt{\left( \,4 \, \lambda _2 \lambda _{s2}-\lambda _{22}^2\right) \left( \,4 \, \lambda _{s2} \lambda _{s2}-\kappa _{3}^2\right)}, \nn \\
%%%%%%%%%%%%%%%%%
& \text{If} \,\, \kappa _{3}<0\, \&\,\kappa _{s_2 \phi _1}<0,\lambda _{11} \lambda _{s2}-\kappa _{3} \kappa _{s_2 \phi _1}  \,\geq \,  \sqrt{\left( \,4 \, \lambda _{s2} \lambda _{s2}-\kappa _{3}^2\right) \left( \,4 \, \lambda _1 \lambda _{s2}-\kappa _{s_2 \phi _1}^2\right)} ,\nn \\
%%%%%%%%%%%%%%%%%
&  \text{If} \,\, \lambda _{11}<0\,\&\,\kappa _{s_1 \phi _2}<0,\left(\lambda _3+\lambda _4\right) \lambda _{s2}-\lambda _{11} \kappa _{s_1 \phi _2}  \,\geq \,  \sqrt{\left( \,4 \, \lambda _1 \lambda _{s2}-\lambda _{11}^2\right) \left( \,4 \, \lambda _2 \lambda _{s2}-\kappa _{s_1 \phi _2}^2\right)},\nn 
%%%%%%%%%%%%%%%%%
\label{eq:stabilityAbs2}
\eea
%%%%%%%%%%%%%%%%%
\vspace*{-2cm}
\subsection{Unitarity constraints}
%%%%%%%%%%%%%%%%%%
%%%%%%%%%%%%%%%%%%%%%%%%%%
Various quartic couplings in non-physical bases $\phi_{dr1},\, \phi_{dr2},\, \phi_{sr1},\, \phi_{sr2}, \, \phi_{di1}(\equiv=G^0),\, \phi_{di2},\, \phi_{si1},\, \phi_{si2}$, $\phi_{d1}^\pm(\equiv=G^\pm),\, \phi_{d2}^\pm,$.

%%%%%%%%%%%%%%%%%%%%%%%%%%
\begin{minipage}[c]{0.330\textwidth}
{\allowdisplaybreaks 
\begin{equation}
\resizebox{0.7\hsize}{!}{ $
\begin{split}
& \{ \phi_{dr1} \, \phi_{dr1} \, \phi_{dr1} \, \phi_{dr1} \}= 6 \lambda _1,\nn\\
& \{ \phi_{dr2} \, \phi_{dr2} \, \phi_{dr2} \, \phi_{dr2} \}= 6 \lambda _2,\nn\\
& \{ \phi_{sr1} \, \phi_{sr1} \, \phi_{sr1} \, \phi_{sr1} \}= 6 \lambda _{s1},\nn\\
& \{ \phi_{sr2} \, \phi_{sr2} \, \phi_{sr2} \, \phi_{sr2} \}= 6 \lambda _{s2},\nn\\
& \{ \phi_{di1} \, \phi_{di1} \, \phi_{di1} \, \phi_{di1} \}= 6 \lambda _1,\nn\\
& \{ \phi_{di2} \, \phi_{di2} \, \phi_{di2} \, \phi_{di2} \}= 6 \lambda _2,\nn\\
& \{ \phi_{si1} \, \phi_{si1} \, \phi_{si1} \, \phi_{si1} \}= 6 \lambda _{s1},\nn\\
& \{ \phi_{si2} \, \phi_{si2} \, \phi_{si2} \, \phi_{si2} \}= 6 \lambda _{s2},\nn\\
& \{ \phi_{dr1} \, \phi_{dr1} \, \phi_{dr2} \, \phi_{dr2} \}= \lambda _3+\lambda _4,\nn\\
& \{ \phi_{dr1} \, \phi_{dr1} \, \phi_{sr1} \, \phi_{sr1} \}= \lambda _{11},\nn\\
& \{ \phi_{dr1} \, \phi_{dr1} \, \phi_{sr2} \, \phi_{sr2} \}= \kappa _{s_2 \phi _1},\nn\\
& \{ \phi_{dr2} \, \phi_{dr2} \, \phi_{sr1} \, \phi_{sr1} \}= \kappa _{s_1 \phi _2},\nn\\
& \{ \phi_{dr2} \, \phi_{dr2} \, \phi_{sr2} \, \phi_{sr2} \}= \lambda _{22},\nn\\
& \{ \phi_{sr1} \, \phi_{sr1} \, \phi_{sr2} \, \phi_{sr2} \}= \kappa _{3},\nn\\
%%%
%%%%
& \{ \phi_{di1} \, \phi_{di1} \, \phi_{di2} \, \phi_{di2} \}= \lambda _3+\lambda _4,\nn\\
& \{ \phi_{di1} \, \phi_{di1} \, \phi_{si1} \, \phi_{si1} \}=  \lambda _{11},\nn\\
& \{ \phi_{di1} \, \phi_{di1} \, \phi_{si2} \, \phi_{si2} \}= \kappa _{s_2 \phi _1},\nn\\
& \{ \phi_{di2} \, \phi_{di2} \, \phi_{si1} \, \phi_{si1} \}= \kappa _{s_1 \phi _2},\nn\\
& \{ \phi_{di2} \, \phi_{di2} \, \phi_{si2} \, \phi_{si2} \}= \lambda _{22},\nn
\end{split}$}
\end{equation}}
\end{minipage}
%%%%%%%%%%%%%%%%%%%%
\hspace{-0.5cm}
\begin{minipage}[c]{0.330\textwidth}
{\allowdisplaybreaks 
\begin{equation*}
\resizebox{0.7\hsize}{!}{ $
\begin{split}
%%%%%%%%%%%
& \{ \phi_{si1} \, \phi_{si1} \, \phi_{si2} \, \phi_{si2} \}=  \kappa _{3},\nn\\
& \{ \phi_{dr1} \, \phi_{dr1} \, \phi_{di1} \, \phi_{di1} \}= 2 \lambda _1,\nn\\
& \{ \phi_{dr1} \, \phi_{dr1} \, \phi_{di2} \, \phi_{di2} \}= \lambda _3+\lambda _4,\nn\\
& \{ \phi_{dr1} \, \phi_{dr1} \, \phi_{si1} \, \phi_{si1} \}= \lambda _{11},\nn\\
& \{ \phi_{dr1} \, \phi_{dr1} \, \phi_{si2} \, \phi_{si2} \}= \kappa _{s_2 \phi _1},\nn\\
%%%%%%%%%%%%%%%%%%
& \{ \phi_{dr2} \, \phi_{dr2} \, \phi_{di1} \, \phi_{di1} \}=\lambda _3+\lambda _4,\nn\\
& \{ \phi_{dr2} \, \phi_{dr2} \, \phi_{di2} \, \phi_{di2} \}= 2 \lambda _2,\nn\\
& \{ \phi_{dr2} \, \phi_{dr2} \, \phi_{si1} \, \phi_{si1} \}= \kappa _{s_1 \phi _2},\nn\\
& \{ \phi_{dr2} \, \phi_{dr2} \, \phi_{si2} \, \phi_{si2} \}= \lambda _{22},\nn\\
%%%%%%%%%%%%
%%%%%%%%%%%%
& \{ \phi_{sr1} \, \phi_{sr1} \, \phi_{di1} \, \phi_{di1} \}= \lambda _{11},\nn\\
& \{ \phi_{sr1} \, \phi_{sr1} \, \phi_{di2} \, \phi_{di2} \}= \kappa _{s_1 \phi _2},\nn\\
& \{ \phi_{sr1} \, \phi_{sr1} \, \phi_{si1} \, \phi_{si1} \}= 2 \lambda _{\text{s1}},\nn\\
& \{ \phi_{sr1} \, \phi_{sr1} \, \phi_{si2} \, \phi_{si2} \}= \kappa _{3},\nn\\
%%%%%%%%%%%%
%%%%%%%%%%%%
& \{ \phi_{sr2} \, \phi_{sr2} \, \phi_{di1} \, \phi_{di1} \}= \kappa _{s_2 \phi _1},\nn\\
& \{ \phi_{sr2} \, \phi_{sr2} \, \phi_{di2} \, \phi_{di2} \}= \lambda _{22},\nn\\
& \{ \phi_{sr2} \, \phi_{sr2} \, \phi_{si1} \, \phi_{si1} \}= \kappa _{3},\nn\\
& \{ \phi_{sr2} \, \phi_{sr2} \, \phi_{si2} \, \phi_{si2} \}= 2 \lambda _{s2},\nn\\
%%%%%%%%%%%%%
& \{ \phi_{d1}^+ \, \phi_{d1}^- \, \phi_{dr1} \, \phi_{dr1} \}= 2 \lambda _1,\nn\\
& \{ \phi_{d1}^+ \, \phi_{d1}^- \, \phi_{dr2} \, \phi_{dr2} \}=  \lambda _{3},\nn
\end{split} $}
\end{equation*}}
\end{minipage}
%%%%%%%%%%%%%%%%%%%%
\hspace{-0.5cm}
\begin{minipage}[c]{0.330\textwidth}
{\allowdisplaybreaks 
\begin{equation*}
\resizebox{0.7\hsize}{!}{ $
\begin{split}
& \{ \phi_{d1}^+ \, \phi_{d1}^- \, \phi_{sr1} \, \phi_{sr1} \}= \lambda _{11},\nn\\
& \{ \phi_{d1}^+ \, \phi_{d1}^- \, \phi_{sr2} \, \phi_{sr2} \}= \kappa _{s_2 \phi _1},\nn\\
& \{ H^+ \, \phi_{d2}^- \, \phi_{dr1} \, \phi_{dr1} \}= \lambda _{22},\nn\\
& \{ H^+ \, \phi_{d2}^- \, \phi_{dr2} \, \phi_{dr2} \}=  \lambda _{3},\nn\\
& \{ H^+ \, \phi_{d2}^- \, \phi_{sr1} \, \phi_{sr1} \}= 2 \lambda _{2},\nn\\
& \{ H^+ \, \phi_{d2}^- \, \phi_{sr2} \, \phi_{sr2} \}= \kappa _{s_1 \phi _2},\nn\\
%
%%%%%%%%%%%%%%
%%%%%%%%%%%%
& \{ \phi_{d1}^+ \, \phi_{d1}^- \, \phi_{di1} \, \phi_{di1} \}= 2 \lambda _{1},\nn\\
& \{ \phi_{d1}^+ \, \phi_{d1}^- \, \phi_{di2} \, \phi_{di2} \}= \lambda _{3},\nn\\
& \{ \phi_{d1}^+ \, \phi_{d1}^- \, \phi_{si1} \, \phi_{si1} \}= 6 \lambda _{11},\nn\\
& \{ \phi_{d1}^+ \, \phi_{d1}^- \, \phi_{si2} \, \phi_{si2} \}= \kappa _{s_2 \phi _1},\nn\\
& \{ H^+ \, \phi_{d2}^- \, \phi_{di1} \, \phi_{di1} \}= \lambda _{3},\nn\\
& \{ H^+ \, \phi_{d2}^- \, \phi_{di2} \, \phi_{di2} \}= 2 \lambda _{2},\nn\\
& \{ H^+ \, \phi_{d2}^- \, \phi_{si1} \, \phi_{si1} \}= \kappa _{s_1 \phi _2},\nn\\
& \{ H^+ \, \phi_{d2}^- \, \phi_{si2} \, \phi_{si2} \}=  \lambda _{22},\nn\\
%
%%%%%%%%%%%%%%
%%%%%%%%%%%%
& \{ \phi_{d1}^+ \, \phi_{d1}^- \, \phi_{d1}^+ \, \phi_{d1}^- \}= 4 \lambda _{1},\nn\\
& \{ \phi_{d1}^+ \, \phi_{d1}^- \, H^+ \, \phi_{d2}^- \}= \lambda _3+\lambda _4,\nn\\
& \{ H^+ \, \phi_{d2}^- \, H^+ \, \phi_{d2}^- \}= 4 \lambda _{2},\nn
\end{split} $}
\end{equation*}}
\end{minipage}
%%%%%%%%%%%%%%%%%
%%%%%%%%%%%%%%%%%
The first $16 \times 16$ sub-matrix , corresponds to scattering processes whose initial and final states are one of
$\phi_{d1}^+\phi_{di1}  $,   $ \phi_{d1}^+\phi_{di2}  $,   $ \phi_{d1}^+\phi_{si1}  $,   $ \phi_{d1}^+\phi_{si2}  $,   $ \phi_{d1}^+\phi_{dr1}  $,   $ \phi_{d1}^+\phi_{dr2}  $,   $ \phi_{d1}^+\phi_{sr1}  $,   $ \phi_{d1}^+\phi_{sr2}  $,   $ H^+\phi_{di1}  $,   $ H^+\phi_{di2}  $,   $ H^+\phi_{si1}  $,   $ H^+\phi_{si2}  $,   $ H^+\phi_{dr1}  $,   $ H^+\phi_{dr2}  $,   $ H^+\phi_{sr1}  $ and  $ H^+\phi_{sr2}$ can be written as,
%%%%%%%%%%%%%%%%%%%%%%%%
%%%%%%%%%%%%%%%%%%%%
{\allowdisplaybreaks 
\begin{equation}
\resizebox{0.8\hsize}{!}{ $
\mathcal{M}_1=\left(
\begin{array}{cccccccccccccccc}
 2 \lambda _1 & 0 & 0 & 0 & 0 & 0 & 0 & 0 & 0 & \frac{\lambda _4}{2} & 0 & 0 & 0 & -\frac{1}{2} \left(i \lambda _4\right) & 0 & 0 \\
 0 & \lambda _3 & 0 & 0 & 0 & 0 & 0 & 0 & \frac{\lambda _4}{2} & 0 & 0 & 0 & \frac{i \lambda _4}{2} & 0 & 0 & 0 \\
 0 & 0 & \lambda _{11} & 0 & 0 & 0 & 0 & 0 & 0 & 0 & 0 & 0 & 0 & 0 & 0 & 0 \\
 0 & 0 & 0 & \kappa _{s_2 \phi _1} & 0 & 0 & 0 & 0 & 0 & 0 & 0 & 0 & 0 & 0 & 0 & 0 \\
 0 & 0 & 0 & 0 & 2 \lambda _1 & 0 & 0 & 0 & 0 & \frac{i \lambda _4}{2} & 0 & 0 & 0 & \frac{\lambda _4}{2} & 0 & 0 \\
 0 & 0 & 0 & 0 & 0 & \lambda _3 & 0 & 0 & -\frac{1}{2} \left(i \lambda _4\right) & 0 & 0 & 0 & \frac{\lambda _4}{2} & 0 & 0 & 0 \\
 0 & 0 & 0 & 0 & 0 & 0 & \lambda _{11} & 0 & 0 & 0 & 0 & 0 & 0 & 0 & 0 & 0 \\
 0 & 0 & 0 & 0 & 0 & 0 & 0 & \kappa _{s_2 \phi _1} & 0 & 0 & 0 & 0 & 0 & 0 & 0 & 0 \\
 0 & \frac{\lambda _4}{2} & 0 & 0 & 0 & \frac{i \lambda _4}{2} & 0 & 0 & \lambda _3 & 0 & 0 & 0 & 0 & 0 & 0 & 0 \\
 \frac{\lambda _4}{2} & 0 & 0 & 0 & -\frac{1}{2} \left(i \lambda _4\right) & 0 & 0 & 0 & 0 & 2 \lambda _2 & 0 & 0 & 0 & 0 & 0 & 0 \\
 0 & 0 & 0 & 0 & 0 & 0 & 0 & 0 & 0 & 0 & \kappa _{s_1 \phi _2} & 0 & 0 & 0 & 0 & 0 \\
 0 & 0 & 0 & 0 & 0 & 0 & 0 & 0 & 0 & 0 & 0 & \lambda _{22} & 0 & 0 & 0 & 0 \\
 0 & -\frac{1}{2} \left(i \lambda _4\right) & 0 & 0 & 0 & \frac{\lambda _4}{2} & 0 & 0 & 0 & 0 & 0 & 0 & \lambda _3 & 0 & 0 & 0 \\
 \frac{i \lambda _4}{2} & 0 & 0 & 0 & \frac{\lambda _4}{2} & 0 & 0 & 0 & 0 & 0 & 0 & 0 & 0 & 2 \lambda _2 & 0 & 0 \\
 0 & 0 & 0 & 0 & 0 & 0 & 0 & 0 & 0 & 0 & 0 & 0 & 0 & 0 & \kappa _{s_1 \phi _2} & 0 \\
 0 & 0 & 0 & 0 & 0 & 0 & 0 & 0 & 0 & 0 & 0 & 0 & 0 & 0 & 0 & \lambda _{22} \\
\end{array}
\right),
 $}
\label{eq:uniM1}
\end{equation}
%%%%%%%%%%%%%%%%%%%%
%%%%%%%%%%%%%%%%%%%%
with eigenvalues: $\kappa _{s_2 \phi _1}$,   $\kappa _{s_1 \phi _2}$,   $2 \lambda _1$, $\lambda _{11}$, $\lambda _{22}$,  $2 \lambda _2$, $\lambda _3$,   $\lambda _3\pm\lambda _4$,  $\lambda _1+\lambda _2 \pm \sqrt{\lambda _1^2-2 \lambda _2 \lambda _1+\lambda _2^2+\lambda _4^2}$.

The second $14 \times 14$ sub-matrix , corresponds to scattering processes whose initial and final states are one of $\phi _{d1}^+ \phi _{d2}^-$,   $\phi _{d1}^- \phi _{d2}^+$,   $\phi _{di1} \phi _{di2}$,   $\phi _{di1} \phi _{si1}$,   $\phi _{di1} \phi _{si2}$,   $\phi _{di2} \phi _{si1}$,   $\phi _{di2} \phi _{si2}$,   $\phi _{si1} \phi _{si2}$,   $\phi _{dr1} \phi _{dr2}$,   $\phi _{dr1} \phi _{sr1}$,   $\phi _{dr1} \phi _{sr2}$,   $\phi _{dr2} \phi _{sr1}$,   $\phi _{dr2} \phi _{sr2}$ and  $\phi _{sr1} \phi _{sr2}$ is given by,
%%%%%%%%%%%%%%%%%%%%%%%%
%%%%%%%%%%%%%%%%%%%%
{\allowdisplaybreaks 
\begin{equation}
\resizebox{0.8\hsize}{!}{ $
\mathcal{M}_2=\left(
\begin{array}{cccccccccccccc}
 \lambda _3+\lambda _4 & 0 & \frac{\lambda _4}{2} & 0 & 0 & 0 & 0 & 0 & \frac{\lambda _4}{2} & 0 & 0 & 0 & 0 & 0 \\
 0 & \lambda _3+\lambda _4 & \frac{\lambda _4}{2} & 0 & 0 & 0 & 0 & 0 & \frac{\lambda _4}{2} & 0 & 0 & 0 & 0 & 0 \\
 \frac{\lambda _4}{2} & \frac{\lambda _4}{2} & \lambda _3+\lambda _4 & 0 & 0 & 0 & 0 & 0 & 0 & 0 & 0 & 0 & 0 & 0 \\
 0 & 0 & 0 & \lambda _{11} & 0 & 0 & 0 & 0 & 0 & 0 & 0 & 0 & 0 & 0 \\
 0 & 0 & 0 & 0 & \kappa _{s_2 \phi _1} & 0 & 0 & 0 & 0 & 0 & 0 & 0 & 0 & 0 \\
 0 & 0 & 0 & 0 & 0 & \kappa _{s_1 \phi _2} & 0 & 0 & 0 & 0 & 0 & 0 & 0 & 0 \\
 0 & 0 & 0 & 0 & 0 & 0 & \lambda _{22} & 0 & 0 & 0 & 0 & 0 & 0 & 0 \\
 0 & 0 & 0 & 0 & 0 & 0 & 0 & \kappa _{3} & 0 & 0 & 0 & 0 & 0 & 0 \\
 \frac{\lambda _4}{2} & \frac{\lambda _4}{2} & 0 & 0 & 0 & 0 & 0 & 0 & \lambda _3+\lambda _4 & 0 & 0 & 0 & 0 & 0 \\
 0 & 0 & 0 & 0 & 0 & 0 & 0 & 0 & 0 & \lambda _{11} & 0 & 0 & 0 & 0 \\
 0 & 0 & 0 & 0 & 0 & 0 & 0 & 0 & 0 & 0 & \kappa _{s_2 \phi _1} & 0 & 0 & 0 \\
 0 & 0 & 0 & 0 & 0 & 0 & 0 & 0 & 0 & 0 & 0 & \kappa _{s_1 \phi _2} & 0 & 0 \\
 0 & 0 & 0 & 0 & 0 & 0 & 0 & 0 & 0 & 0 & 0 & 0 & \lambda _{22} & 0 \\
 0 & 0 & 0 & 0 & 0 & 0 & 0 & 0 & 0 & 0 & 0 & 0 & 0 & \kappa _{3} \\
\end{array}
\right),
 $}
\label{eq:uniM2}
\end{equation}
%%%%%%%%%%%%%%%%%%%%
%%%%%%%%%%%%%%%%%%%%
having eigenvalues: $\kappa _{s_2 \phi _1}$, $\kappa _{s_1 \phi _2}$,   $\kappa _{3}$, $\lambda _3$, $\lambda _3+\lambda _4$,   $\lambda _3+2 \lambda _4$,   $\lambda _{11}$, $\lambda _{22}$.

%%%$\kappa _{s_2 \phi _1}$,   $\kappa _{s_2 \phi _1}$,   $\kappa _{s_1 \phi _2}$,   $\kappa _{s_1 \phi _2}$,   $\kappa _{3}$,   $\kappa _{3}$,   $\lambda _3$,   $\lambda _3+\lambda _4$,   $\lambda _3+\lambda _4$,   $\lambda _3+2 \lambda _4$,   $\lambda _{11}$,   $\lambda _{11}$,   $\lambda _{22}$,   $\lambda _{22}$.

The third and most-stringent $10 \times 10$ sub-matrix $\mathcal{M}_3$, corresponds to scattering fields $\phi_{d1}^- \phi_{d1}^+$,   $\phi_{d2}^- H^+$,   $\frac{\phi_{di1} \phi_{di1}}{\sqrt{2}}$,   $\frac{\phi_{di2} \phi_{di2}}{\sqrt{2}}$,   $\frac{\phi_{si1} \phi_{si1}}{\sqrt{2}}$,   $\frac{\phi_{si2} \phi_{si2}}{\sqrt{2}}$,   $\frac{\phi_{dr1} \phi_{dr1}}{\sqrt{2}}$,   $\frac{\phi_{dr2} \phi_{dr2}}{\sqrt{2}}$,   $\frac{\phi_{sr1} \phi_{sr1}}{\sqrt{2}}$ and  $\frac{\phi_{sr2} \phi_{sr2}}{\sqrt{2}}$. The factor $\frac{1}{\sqrt{2}}$ has appeared due to the statistics of identical particles. $\mathcal{M}_3$ is,
%%%%%%%%%%%%%%%%%%%%%%%%
%%%%%%%%%%%%%%%%%%%%
{\allowdisplaybreaks 
\begin{equation}
\resizebox{0.8\hsize}{!}{ $
\mathcal{M}_3=\left(
\begin{array}{cccccccccc}
 4 \lambda _1 & \lambda _3+\lambda _4 & \sqrt{2} \lambda _1 & \frac{\lambda _3}{\sqrt{2}} & \frac{\lambda _{11}}{\sqrt{2}} & \frac{\kappa _{s_2 \phi _1}}{\sqrt{2}} & \sqrt{2} \lambda _1 & \frac{\lambda _3}{\sqrt{2}} & \frac{\lambda _{11}}{\sqrt{2}} & \frac{\kappa _{s_2 \phi _1}}{\sqrt{2}} \\
 \lambda _3+\lambda _4 & 4 \lambda _2 & \frac{\lambda _3}{\sqrt{2}} & \sqrt{2} \lambda _2 & \frac{\kappa _{s_1 \phi _2}}{\sqrt{2}} & \frac{\lambda _{22}}{\sqrt{2}} & \frac{\lambda _3}{\sqrt{2}} & \sqrt{2} \lambda _2 & \frac{\kappa _{s_1 \phi _2}}{\sqrt{2}} & \frac{\lambda _{22}}{\sqrt{2}} \\
 \sqrt{2} \lambda _1 & \frac{\lambda _3}{\sqrt{2}} & 3 \lambda _1 & \frac{1}{2} \left(\lambda _3+\lambda _4\right) & \frac{\lambda _{11}}{2} & \frac{1}{2} \kappa _{s_2 \phi _1} & \lambda _1 & \frac{1}{2} \left(\lambda _3+\lambda _4\right) & \frac{\lambda _{11}}{2} & \frac{1}{2} \kappa _{s_2 \phi _1} \\
 \frac{\lambda _3}{\sqrt{2}} & \sqrt{2} \lambda _2 & \frac{1}{2} \left(\lambda _3+\lambda _4\right) & 3 \lambda _2 & \frac{1}{2} \kappa _{s_1 \phi _2} & \frac{\lambda _{22}}{2} & \frac{1}{2} \left(\lambda _3+\lambda _4\right) & \lambda _2 & \frac{1}{2} \kappa _{s_1 \phi _2} & \frac{\lambda _{22}}{2} \\
 \frac{\lambda _{11}}{\sqrt{2}} & \frac{\kappa _{s_1 \phi _2}}{\sqrt{2}} & \frac{\lambda _{11}}{2} & \frac{1}{2} \kappa _{s_1 \phi _2} & 3 \lambda _{s1} & \frac{1}{2} \kappa _{3} & \frac{\lambda _{11}}{2} & \frac{1}{2} \kappa _{s_1 \phi _2} & \lambda _{s1} & \frac{1}{2} \kappa _{3} \\
 \frac{\kappa _{s_2 \phi _1}}{\sqrt{2}} & \frac{\lambda _{22}}{\sqrt{2}} & \frac{1}{2} \kappa _{s_2 \phi _1} & \frac{\lambda _{22}}{2} & \frac{1}{2} \kappa _{3} & 3 \lambda _{s2} & \frac{1}{2} \kappa _{s_2 \phi _1} & \frac{\lambda _{22}}{2} & \frac{1}{2} \kappa _{3} & \lambda _{s2} \\
 \sqrt{2} \lambda _1 & \frac{\lambda _3}{\sqrt{2}} & \lambda _1 & \frac{1}{2} \left(\lambda _3+\lambda _4\right) & \frac{\lambda _{11}}{2} & \frac{1}{2} \kappa _{s_2 \phi _1} & 3 \lambda _1 & \frac{1}{2} \left(\lambda _3+\lambda _4\right) & \frac{\lambda _{11}}{2} & \frac{1}{2} \kappa _{s_2 \phi _1} \\
 \frac{\lambda _3}{\sqrt{2}} & \sqrt{2} \lambda _2 & \frac{1}{2} \left(\lambda _3+\lambda _4\right) & \lambda _2 & \frac{1}{2} \kappa _{s_1 \phi _2} & \frac{\lambda _{22}}{2} & \frac{1}{2} \left(\lambda _3+\lambda _4\right) & 3 \lambda _2 & \frac{1}{2} \kappa _{s_1 \phi _2} & \frac{\lambda _{22}}{2} \\
 \frac{\lambda _{11}}{\sqrt{2}} & \frac{\kappa _{s_1 \phi _2}}{\sqrt{2}} & \frac{\lambda _{11}}{2} & \frac{1}{2} \kappa _{s_1 \phi _2} & \lambda _{s1} & \frac{1}{2} \kappa _{3} & \frac{\lambda _{11}}{2} & \frac{1}{2} \kappa _{s_1 \phi _2} & 3 \lambda _{s1} & \frac{1}{2} \kappa _{3} \\
 \frac{\kappa _{s_2 \phi _1}}{\sqrt{2}} & \frac{\lambda _{22}}{\sqrt{2}} & \frac{1}{2} \kappa _{s_2 \phi _1} & \frac{\lambda _{22}}{2} & \frac{1}{2} \kappa _{3} & \lambda _{s2} & \frac{1}{2} \kappa _{s_2 \phi _1} & \frac{\lambda _{22}}{2} & \frac{1}{2} \kappa _{3} & 3 \lambda _{s2} \\
\end{array}
\right),
 $}
\label{eq:uniM2}
\end{equation}
%%%%%%%%%%%%%%%%%%%%
%%%%%%%%%%%%%%%%%%%%
The eigenvalues can not be calculated analytically. We get the analytical values for the choice of $\lambda _{11}=\lambda _{22}=\kappa _{s_2 \phi _1}=\kappa _{s_1 \phi _2}=0$. The eigenvalues are now:
$2 \lambda_1$,   $2 \lambda_2$,   $2 \lambda_{s1}$,   $2 \lambda_{s2}$,   $\lambda_1+\lambda_2 \pm \sqrt{\left(\lambda_1-\lambda_2\right)^2+\lambda_4^2}$,   $3 \lambda_1+3 \lambda_2 \pm \sqrt{9 \left(\lambda_1-\lambda_2\right)^2+\left(2 \lambda_3+\lambda_4\right)^2}$ and   $ 2 \lambda_{s1}+2 \lambda_{s2}\pm \sqrt{\kappa_{s_1 s_2}^2+4 \left(\lambda_{s1}-\lambda_{s2}\right)^2}$.

%%%%%%%%%%%%%%%%%%
%%%%%%%%%%%%%%%%%%

%%%%%%%%%%%%%%%%%
%%%%%%%%%%%%%%%%%%%%%%%
%\newpage
%\bibliographystyle{plain}
%\bibliographystyle{unsrt}
%\bibliographystyle{apsrev4-1}
%\bibliographystyle{apsrev}
%\bibliographystyle{decsci}
%\bibliographystyle{utphys}
%\bibliography{tevportalnew}

\begin{thebibliography}{76}%
 	\makeatletter
 	\providecommand \@ifxundefined [1]{%
 		\@ifx{#1\undefined}
 	}%
 	\providecommand \@ifnum [1]{%
 		\ifnum #1\expandafter \@firstoftwo
 		\else \expandafter \@secondoftwo
 		\fi
 	}%
 	\providecommand \@ifx [1]{%
 		\ifx #1\expandafter \@firstoftwo
 		\else \expandafter \@secondoftwo
 		\fi
 	}%
 	\providecommand \natexlab [1]{#1}%
 	\providecommand \enquote  [1]{``#1''}%
 	\providecommand \bibnamefont  [1]{#1}%
 	\providecommand \bibfnamefont [1]{#1}%
 	\providecommand \citenamefont [1]{#1}%
 	\providecommand \href@noop [0]{\@secondoftwo}%
 	\providecommand \href [0]{\begingroup \@sanitize@url \@href}%
 	\providecommand \@href[1]{\@@startlink{#1}\@@href}%
 	\providecommand \@@href[1]{\endgroup#1\@@endlink}%
 	\providecommand \@sanitize@url [0]{\catcode `\\12\catcode `\$12\catcode
 		`\&12\catcode `\#12\catcode `\^12\catcode `\_12\catcode `\%12\relax}%
 	\providecommand \@@startlink[1]{}%
 	\providecommand \@@endlink[0]{}%
 	\providecommand \url  [0]{\begingroup\@sanitize@url \@url }%
 	\providecommand \@url [1]{\endgroup\@href {#1}{\urlprefix }}%
 	\providecommand \urlprefix  [0]{URL }%
 	\providecommand \Eprint [0]{\href }%
 	\providecommand \doibase [0]{http://dx.doi.org/}%
 	\providecommand \selectlanguage [0]{\@gobble}%
 	\providecommand \bibinfo  [0]{\@secondoftwo}%
 	\providecommand \bibfield  [0]{\@secondoftwo}%
 	\providecommand \translation [1]{[#1]}%
 	\providecommand \BibitemOpen [0]{}%
 	\providecommand \bibitemStop [0]{}%
 	\providecommand \bibitemNoStop [0]{.\EOS\space}%
 	\providecommand \EOS [0]{\spacefactor3000\relax}%
 	\providecommand \BibitemShut  [1]{\csname bibitem#1\endcsname}%
 	\let\auto@bib@innerbib\@empty
 	%</preamble>
 	\bibitem [{\citenamefont {Aad}\ \emph {et~al.}(2012)\citenamefont {Aad} \emph
 		{et~al.}}]{ATLAS:2012yve}%
 	\BibitemOpen
 	\bibfield  {author} {\bibinfo {author} {\bibfnamefont {G.}~\bibnamefont
 			{Aad}} \emph {et~al.} (\bibinfo {collaboration} {ATLAS}),\ }\href {\doibase
 		10.1016/j.physletb.2012.08.020} {\bibfield  {journal} {\bibinfo  {journal}
 			{Phys. Lett. B}\ }\textbf {\bibinfo {volume} {716}},\ \bibinfo {pages} {1}
 		(\bibinfo {year} {2012})},\ \Eprint {http://arxiv.org/abs/1207.7214}
 	{arXiv:1207.7214 [hep-ex]} \BibitemShut {NoStop}%
 	\bibitem [{\citenamefont {Chatrchyan}\ \emph {et~al.}(2012)\citenamefont
 		{Chatrchyan} \emph {et~al.}}]{CMS:2012qbp}%
 	\BibitemOpen
 	\bibfield  {author} {\bibinfo {author} {\bibfnamefont {S.}~\bibnamefont
 			{Chatrchyan}} \emph {et~al.} (\bibinfo {collaboration} {CMS}),\ }\href
 	{\doibase 10.1016/j.physletb.2012.08.021} {\bibfield  {journal} {\bibinfo
 			{journal} {Phys. Lett. B}\ }\textbf {\bibinfo {volume} {716}},\ \bibinfo
 		{pages} {30} (\bibinfo {year} {2012})},\ \Eprint
 	{http://arxiv.org/abs/1207.7235} {arXiv:1207.7235 [hep-ex]} \BibitemShut
 	{NoStop}%
 	\bibitem [{\citenamefont {Zwicky}(1933)}]{Zwicky:1933gu}%
 	\BibitemOpen
 	\bibfield  {author} {\bibinfo {author} {\bibfnamefont {F.}~\bibnamefont
 			{Zwicky}},\ }\href {\doibase 10.1007/s10714-008-0707-4} {\bibfield  {journal}
 		{\bibinfo  {journal} {Helv. Phys. Acta}\ }\textbf {\bibinfo {volume} {6}},\
 		\bibinfo {pages} {110} (\bibinfo {year} {1933})}\BibitemShut {NoStop}%
 	\bibitem [{\citenamefont {Clowe}\ \emph {et~al.}(2006)\citenamefont {Clowe},
 		\citenamefont {Bradac}, \citenamefont {Gonzalez}, \citenamefont {Markevitch},
 		\citenamefont {Randall}, \citenamefont {Jones},\ and\ \citenamefont
 		{Zaritsky}}]{Clowe:2006eq}%
 	\BibitemOpen
 	\bibfield  {author} {\bibinfo {author} {\bibfnamefont {D.}~\bibnamefont
 			{Clowe}}, \bibinfo {author} {\bibfnamefont {M.}~\bibnamefont {Bradac}},
 		\bibinfo {author} {\bibfnamefont {A.~H.}\ \bibnamefont {Gonzalez}}, \bibinfo
 		{author} {\bibfnamefont {M.}~\bibnamefont {Markevitch}}, \bibinfo {author}
 		{\bibfnamefont {S.~W.}\ \bibnamefont {Randall}}, \bibinfo {author}
 		{\bibfnamefont {C.}~\bibnamefont {Jones}}, \ and\ \bibinfo {author}
 		{\bibfnamefont {D.}~\bibnamefont {Zaritsky}},\ }\href {\doibase
 		10.1086/508162} {\bibfield  {journal} {\bibinfo  {journal} {Astrophys. J.
 				Lett.}\ }\textbf {\bibinfo {volume} {648}},\ \bibinfo {pages} {L109}
 		(\bibinfo {year} {2006})},\ \Eprint {http://arxiv.org/abs/astro-ph/0608407}
 	{arXiv:astro-ph/0608407} \BibitemShut {NoStop}%
 	\bibitem [{\citenamefont {Babcock}(1939)}]{1939LicOB..19...41B}%
 	\BibitemOpen
 	\bibfield  {author} {\bibinfo {author} {\bibfnamefont {H.~W.}\ \bibnamefont
 			{Babcock}},\ }\href {\doibase 10.5479/ADS/bib/1939LicOB.19.41B} {\bibfield
 		{journal} {\bibinfo  {journal} {Lick Observatory Bulletin}\ }\textbf
 		{\bibinfo {volume} {498}},\ \bibinfo {pages} {41} (\bibinfo {year}
 		{1939})}\BibitemShut {NoStop}%
 	\bibitem [{\citenamefont {Rubin}\ \emph {et~al.}(1980)\citenamefont {Rubin},
 		\citenamefont {Ford},\ and\ \citenamefont {Thonnard}}]{1980ApJ...238..471R}%
 	\BibitemOpen
 	\bibfield  {author} {\bibinfo {author} {\bibfnamefont {V.~C.}\ \bibnamefont
 			{Rubin}}, \bibinfo {author} {\bibfnamefont {J.}~\bibnamefont {Ford},
 			\bibfnamefont {W.~K.}}, \ and\ \bibinfo {author} {\bibfnamefont
 			{N.}~\bibnamefont {Thonnard}},\ }\href {\doibase 10.1086/158003} {\bibfield
 		{journal} {\bibinfo  {journal} {\apj}\ }\textbf {\bibinfo {volume} {238}},\
 		\bibinfo {pages} {471} (\bibinfo {year} {1980})}\BibitemShut {NoStop}%
 	\bibitem [{\citenamefont {Ahmad}\ \emph {et~al.}(2002)\citenamefont {Ahmad}
 		\emph {et~al.}}]{SNO:2002tuh}%
 	\BibitemOpen
 	\bibfield  {author} {\bibinfo {author} {\bibfnamefont {Q.~R.}\ \bibnamefont
 			{Ahmad}} \emph {et~al.} (\bibinfo {collaboration} {SNO}),\ }\href {\doibase
 		10.1103/PhysRevLett.89.011301} {\bibfield  {journal} {\bibinfo  {journal}
 			{Phys. Rev. Lett.}\ }\textbf {\bibinfo {volume} {89}},\ \bibinfo {pages}
 		{011301} (\bibinfo {year} {2002})},\ \Eprint
 	{http://arxiv.org/abs/nucl-ex/0204008} {arXiv:nucl-ex/0204008} \BibitemShut
 	{NoStop}%
 	\bibitem [{\citenamefont {Fukuda}\ \emph {et~al.}(1998)\citenamefont {Fukuda}
 		\emph {et~al.}}]{Super-Kamiokande:1998kpq}%
 	\BibitemOpen
 	\bibfield  {author} {\bibinfo {author} {\bibfnamefont {Y.}~\bibnamefont
 			{Fukuda}} \emph {et~al.} (\bibinfo {collaboration} {Super-Kamiokande}),\
 	}\href {\doibase 10.1103/PhysRevLett.81.1562} {\bibfield  {journal} {\bibinfo
 			{journal} {Phys. Rev. Lett.}\ }\textbf {\bibinfo {volume} {81}},\ \bibinfo
 		{pages} {1562} (\bibinfo {year} {1998})},\ \Eprint
 	{http://arxiv.org/abs/hep-ex/9807003} {arXiv:hep-ex/9807003} \BibitemShut
 	{NoStop}%
 	\bibitem [{\citenamefont {Langacker}(2009)}]{Langacker:2008yv}%
 	\BibitemOpen
 	\bibfield  {author} {\bibinfo {author} {\bibfnamefont {P.}~\bibnamefont
 			{Langacker}},\ }\href {\doibase 10.1103/RevModPhys.81.1199} {\bibfield
 		{journal} {\bibinfo  {journal} {Rev. Mod. Phys.}\ }\textbf {\bibinfo {volume}
 			{81}},\ \bibinfo {pages} {1199} (\bibinfo {year} {2009})},\ \Eprint
 	{http://arxiv.org/abs/0801.1345} {arXiv:0801.1345 [hep-ph]} \BibitemShut
 	{NoStop}%
 	\bibitem [{\citenamefont {Leike}(1999)}]{Leike:1998wr}%
 	\BibitemOpen
 	\bibfield  {author} {\bibinfo {author} {\bibfnamefont {A.}~\bibnamefont
 			{Leike}},\ }\href {\doibase 10.1016/S0370-1573(98)00133-1} {\bibfield
 		{journal} {\bibinfo  {journal} {Phys. Rept.}\ }\textbf {\bibinfo {volume}
 			{317}},\ \bibinfo {pages} {143} (\bibinfo {year} {1999})},\ \Eprint
 	{http://arxiv.org/abs/hep-ph/9805494} {arXiv:hep-ph/9805494} \BibitemShut
 	{NoStop}%
 	\bibitem [{\citenamefont {Rizzo}(2006)}]{Rizzo:2006nw}%
 	\BibitemOpen
 	\bibfield  {author} {\bibinfo {author} {\bibfnamefont {T.~G.}\ \bibnamefont
 			{Rizzo}},\ }in\ \href@noop {} {\emph {\bibinfo {booktitle} {{Theoretical
 					Advanced Study Institute in Elementary Particle Physics}: {Exploring New
 					Frontiers Using Colliders and Neutrinos}}}}\ (\bibinfo {year} {2006})\ pp.\
 	\bibinfo {pages} {537--575},\ \Eprint {http://arxiv.org/abs/hep-ph/0610104}
 	{arXiv:hep-ph/0610104} \BibitemShut {NoStop}%
 	\bibitem [{\citenamefont {Marshak}\ and\ \citenamefont
 		{Mohapatra}(1980)}]{Marshak:1979fm}%
 	\BibitemOpen
 	\bibfield  {author} {\bibinfo {author} {\bibfnamefont {R.~E.}\ \bibnamefont
 			{Marshak}}\ and\ \bibinfo {author} {\bibfnamefont {R.~N.}\ \bibnamefont
 			{Mohapatra}},\ }\href {\doibase 10.1016/0370-2693(80)90436-0} {\bibfield
 		{journal} {\bibinfo  {journal} {Phys. Lett. B}\ }\textbf {\bibinfo {volume}
 			{91}},\ \bibinfo {pages} {222} (\bibinfo {year} {1980})}\BibitemShut
 	{NoStop}%
 	\bibitem [{\citenamefont {Mohapatra}\ and\ \citenamefont
 		{Marshak}(1980)}]{Mohapatra:1980qe}%
 	\BibitemOpen
 	\bibfield  {author} {\bibinfo {author} {\bibfnamefont {R.~N.}\ \bibnamefont
 			{Mohapatra}}\ and\ \bibinfo {author} {\bibfnamefont {R.~E.}\ \bibnamefont
 			{Marshak}},\ }\href {\doibase 10.1103/PhysRevLett.44.1316} {\bibfield
 		{journal} {\bibinfo  {journal} {Phys. Rev. Lett.}\ }\textbf {\bibinfo
 			{volume} {44}},\ \bibinfo {pages} {1316} (\bibinfo {year} {1980})},\ \bibinfo
 	{note} {[Erratum: Phys.Rev.Lett. 44, 1643 (1980)]}\BibitemShut {NoStop}%
 	\bibitem [{\citenamefont {Khalil}(2008)}]{Khalil:2006yi}%
 	\BibitemOpen
 	\bibfield  {author} {\bibinfo {author} {\bibfnamefont {S.}~\bibnamefont
 			{Khalil}},\ }\href {\doibase 10.1088/0954-3899/35/5/055001} {\bibfield
 		{journal} {\bibinfo  {journal} {J. Phys. G}\ }\textbf {\bibinfo {volume}
 			{35}},\ \bibinfo {pages} {055001} (\bibinfo {year} {2008})},\ \Eprint
 	{http://arxiv.org/abs/hep-ph/0611205} {arXiv:hep-ph/0611205} \BibitemShut
 	{NoStop}%
 	\bibitem [{\citenamefont {Zyla}\ \emph {et~al.}(2020)\citenamefont {Zyla} \emph
 		{et~al.}}]{ParticleDataGroup:2020ssz}%
 	\BibitemOpen
 	\bibfield  {author} {\bibinfo {author} {\bibfnamefont {P.~A.}\ \bibnamefont
 			{Zyla}} \emph {et~al.} (\bibinfo {collaboration} {Particle Data Group}),\
 	}\href {\doibase 10.1093/ptep/ptaa104} {\bibfield  {journal} {\bibinfo
 			{journal} {PTEP}\ }\textbf {\bibinfo {volume} {2020}},\ \bibinfo {pages}
 		{083C01} (\bibinfo {year} {2020})}\BibitemShut {NoStop}%
 	\bibitem [{\citenamefont {Aaboud}\ \emph {et~al.}(2017)\citenamefont {Aaboud}
 		\emph {et~al.}}]{ATLAS:2017fih}%
 	\BibitemOpen
 	\bibfield  {author} {\bibinfo {author} {\bibfnamefont {M.}~\bibnamefont
 			{Aaboud}} \emph {et~al.} (\bibinfo {collaboration} {ATLAS}),\ }\href
 	{\doibase 10.1007/JHEP10(2017)182} {\bibfield  {journal} {\bibinfo  {journal}
 			{JHEP}\ }\textbf {\bibinfo {volume} {10}},\ \bibinfo {pages} {182} (\bibinfo
 		{year} {2017})},\ \Eprint {http://arxiv.org/abs/1707.02424} {arXiv:1707.02424
 		[hep-ex]} \BibitemShut {NoStop}%
 	\bibitem [{\citenamefont {Sirunyan}\ \emph {et~al.}(2020)\citenamefont
 		{Sirunyan} \emph {et~al.}}]{CMS:2019buh}%
 	\BibitemOpen
 	\bibfield  {author} {\bibinfo {author} {\bibfnamefont {A.~M.}\ \bibnamefont
 			{Sirunyan}} \emph {et~al.} (\bibinfo {collaboration} {CMS}),\ }\href
 	{\doibase 10.1103/PhysRevLett.124.131802} {\bibfield  {journal} {\bibinfo
 			{journal} {Phys. Rev. Lett.}\ }\textbf {\bibinfo {volume} {124}},\ \bibinfo
 		{pages} {131802} (\bibinfo {year} {2020})},\ \Eprint
 	{http://arxiv.org/abs/1912.04776} {arXiv:1912.04776 [hep-ex]} \BibitemShut
 	{NoStop}%
 	\bibitem [{\citenamefont {ACCOMANDO}\ \emph {et~al.}(2013)\citenamefont
 		{ACCOMANDO}, \citenamefont {Becciolini}, \citenamefont {Belyaev},
 		\citenamefont {De~Curtis}, \citenamefont {Dominici}, \citenamefont {King},
 		\citenamefont {Moretti},\ and\ \citenamefont
 		{Shepherd-Themistocleous}}]{ACCOMANDO:2013zz}%
 	\BibitemOpen
 	\bibfield  {author} {\bibinfo {author} {\bibfnamefont {E.}~\bibnamefont
 			{ACCOMANDO}}, \bibinfo {author} {\bibfnamefont {D.}~\bibnamefont
 			{Becciolini}}, \bibinfo {author} {\bibfnamefont {A.}~\bibnamefont {Belyaev}},
 		\bibinfo {author} {\bibfnamefont {S.}~\bibnamefont {De~Curtis}}, \bibinfo
 		{author} {\bibfnamefont {D.}~\bibnamefont {Dominici}}, \bibinfo {author}
 		{\bibfnamefont {S.~F.}\ \bibnamefont {King}}, \bibinfo {author}
 		{\bibfnamefont {S.}~\bibnamefont {Moretti}}, \ and\ \bibinfo {author}
 		{\bibfnamefont {C.~H.}\ \bibnamefont {Shepherd-Themistocleous}},\ }\href
 	{\doibase 10.22323/1.191.0125} {\bibfield  {journal} {\bibinfo  {journal}
 			{PoS}\ }\textbf {\bibinfo {volume} {DIS 2013}},\ \bibinfo {pages} {125}
 		(\bibinfo {year} {2013})}\BibitemShut {NoStop}%
 	\bibitem [{\citenamefont {Ma}(2006)}]{Ma:2006km}%
 	\BibitemOpen
 	\bibfield  {author} {\bibinfo {author} {\bibfnamefont {E.}~\bibnamefont
 			{Ma}},\ }\href {\doibase 10.1103/PhysRevD.73.077301} {\bibfield  {journal}
 		{\bibinfo  {journal} {Phys. Rev. D}\ }\textbf {\bibinfo {volume} {73}},\
 		\bibinfo {pages} {077301} (\bibinfo {year} {2006})},\ \Eprint
 	{http://arxiv.org/abs/hep-ph/0601225} {arXiv:hep-ph/0601225} \BibitemShut
 	{NoStop}%
 	\bibitem [{\citenamefont {del Aguila}\ \emph {et~al.}(1995)\citenamefont {del
 			Aguila}, \citenamefont {Masip},\ and\ \citenamefont
 		{Perez-Victoria}}]{delAguila:1995rb}%
 	\BibitemOpen
 	\bibfield  {author} {\bibinfo {author} {\bibfnamefont {F.}~\bibnamefont {del
 				Aguila}}, \bibinfo {author} {\bibfnamefont {M.}~\bibnamefont {Masip}}, \ and\
 		\bibinfo {author} {\bibfnamefont {M.}~\bibnamefont {Perez-Victoria}},\ }\href
 	{\doibase 10.1016/0550-3213(95)00511-6} {\bibfield  {journal} {\bibinfo
 			{journal} {Nucl. Phys. B}\ }\textbf {\bibinfo {volume} {456}},\ \bibinfo
 		{pages} {531} (\bibinfo {year} {1995})},\ \Eprint
 	{http://arxiv.org/abs/hep-ph/9507455} {arXiv:hep-ph/9507455} \BibitemShut
 	{NoStop}%
 	\bibitem [{\citenamefont {Chankowski}\ \emph {et~al.}(2006)\citenamefont
 		{Chankowski}, \citenamefont {Pokorski},\ and\ \citenamefont
 		{Wagner}}]{Chankowski:2006jk}%
 	\BibitemOpen
 	\bibfield  {author} {\bibinfo {author} {\bibfnamefont {P.~H.}\ \bibnamefont
 			{Chankowski}}, \bibinfo {author} {\bibfnamefont {S.}~\bibnamefont
 			{Pokorski}}, \ and\ \bibinfo {author} {\bibfnamefont {J.}~\bibnamefont
 			{Wagner}},\ }\href {\doibase 10.1140/epjc/s2006-02537-3} {\bibfield
 		{journal} {\bibinfo  {journal} {Eur. Phys. J. C}\ }\textbf {\bibinfo {volume}
 			{47}},\ \bibinfo {pages} {187} (\bibinfo {year} {2006})},\ \Eprint
 	{http://arxiv.org/abs/hep-ph/0601097} {arXiv:hep-ph/0601097} \BibitemShut
 	{NoStop}%
 	\bibitem [{\citenamefont {Chakrabortty}\ \emph {et~al.}(2014)\citenamefont
 		{Chakrabortty}, \citenamefont {Konar},\ and\ \citenamefont
 		{Mondal}}]{Chakrabortty:2013mha}%
 	\BibitemOpen
 	\bibfield  {author} {\bibinfo {author} {\bibfnamefont {J.}~\bibnamefont
 			{Chakrabortty}}, \bibinfo {author} {\bibfnamefont {P.}~\bibnamefont {Konar}},
 		\ and\ \bibinfo {author} {\bibfnamefont {T.}~\bibnamefont {Mondal}},\ }\href
 	{\doibase 10.1103/PhysRevD.89.095008} {\bibfield  {journal} {\bibinfo
 			{journal} {Phys. Rev. D}\ }\textbf {\bibinfo {volume} {89}},\ \bibinfo
 		{pages} {095008} (\bibinfo {year} {2014})},\ \Eprint
 	{http://arxiv.org/abs/1311.5666} {arXiv:1311.5666 [hep-ph]} \BibitemShut
 	{NoStop}%
 	\bibitem [{\citenamefont {Lee}\ \emph {et~al.}(1977{\natexlab{a}})\citenamefont
 		{Lee}, \citenamefont {Quigg},\ and\ \citenamefont {Thacker}}]{Lee:1977eg}%
 	\BibitemOpen
 	\bibfield  {author} {\bibinfo {author} {\bibfnamefont {B.~W.}\ \bibnamefont
 			{Lee}}, \bibinfo {author} {\bibfnamefont {C.}~\bibnamefont {Quigg}}, \ and\
 		\bibinfo {author} {\bibfnamefont {H.~B.}\ \bibnamefont {Thacker}},\ }\href
 	{\doibase 10.1103/PhysRevD.16.1519} {\bibfield  {journal} {\bibinfo
 			{journal} {Phys. Rev.}\ }\textbf {\bibinfo {volume} {D16}},\ \bibinfo {pages}
 		{1519} (\bibinfo {year} {1977}{\natexlab{a}})}\BibitemShut {NoStop}%
 	%%CITATION = PHRVA,D16,1519;%%
 	\bibitem [{\citenamefont {Lee}\ \emph {et~al.}(1977{\natexlab{b}})\citenamefont
 		{Lee}, \citenamefont {Quigg},\ and\ \citenamefont {Thacker}}]{Lee:1977yc}%
 	\BibitemOpen
 	\bibfield  {author} {\bibinfo {author} {\bibfnamefont {B.~W.}\ \bibnamefont
 			{Lee}}, \bibinfo {author} {\bibfnamefont {C.}~\bibnamefont {Quigg}}, \ and\
 		\bibinfo {author} {\bibfnamefont {H.~B.}\ \bibnamefont {Thacker}},\ }\href
 	{\doibase 10.1103/PhysRevLett.38.883} {\bibfield  {journal} {\bibinfo
 			{journal} {Phys. Rev. Lett.}\ }\textbf {\bibinfo {volume} {38}},\ \bibinfo
 		{pages} {883} (\bibinfo {year} {1977}{\natexlab{b}})}\BibitemShut {NoStop}%
 	\bibitem [{\citenamefont {Peskin}\ and\ \citenamefont
 		{Takeuchi}(1992)}]{Peskin:1991sw}%
 	\BibitemOpen
 	\bibfield  {author} {\bibinfo {author} {\bibfnamefont {M.~E.}\ \bibnamefont
 			{Peskin}}\ and\ \bibinfo {author} {\bibfnamefont {T.}~\bibnamefont
 			{Takeuchi}},\ }\href {\doibase 10.1103/PhysRevD.46.381} {\bibfield  {journal}
 		{\bibinfo  {journal} {Phys. Rev. D}\ }\textbf {\bibinfo {volume} {46}},\
 		\bibinfo {pages} {381} (\bibinfo {year} {1992})}\BibitemShut {NoStop}%
 	\bibitem [{\citenamefont {Altarelli}\ \emph {et~al.}(1993)\citenamefont
 		{Altarelli}, \citenamefont {Barbieri},\ and\ \citenamefont
 		{Caravaglios}}]{Altarelli:1993bh}%
 	\BibitemOpen
 	\bibfield  {author} {\bibinfo {author} {\bibfnamefont {G.}~\bibnamefont
 			{Altarelli}}, \bibinfo {author} {\bibfnamefont {R.}~\bibnamefont {Barbieri}},
 		\ and\ \bibinfo {author} {\bibfnamefont {F.}~\bibnamefont {Caravaglios}},\
 	}\href {\doibase 10.1016/0370-2693(93)91249-M} {\bibfield  {journal}
 		{\bibinfo  {journal} {Phys. Lett. B}\ }\textbf {\bibinfo {volume} {314}},\
 		\bibinfo {pages} {357} (\bibinfo {year} {1993})}\BibitemShut {NoStop}%
 	\bibitem [{\citenamefont {Baak}\ \emph {et~al.}(2014)\citenamefont {Baak},
 		\citenamefont {Cúth}, \citenamefont {Haller}, \citenamefont {Hoecker},
 		\citenamefont {Kogler}, \citenamefont {Mönig}, \citenamefont {Schott},\ and\
 		\citenamefont {Stelzer}}]{Baak:2014ora}%
 	\BibitemOpen
 	\bibfield  {author} {\bibinfo {author} {\bibfnamefont {M.}~\bibnamefont
 			{Baak}}, \bibinfo {author} {\bibfnamefont {J.}~\bibnamefont {Cúth}},
 		\bibinfo {author} {\bibfnamefont {J.}~\bibnamefont {Haller}}, \bibinfo
 		{author} {\bibfnamefont {A.}~\bibnamefont {Hoecker}}, \bibinfo {author}
 		{\bibfnamefont {R.}~\bibnamefont {Kogler}}, \bibinfo {author} {\bibfnamefont
 			{K.}~\bibnamefont {Mönig}}, \bibinfo {author} {\bibfnamefont
 			{M.}~\bibnamefont {Schott}}, \ and\ \bibinfo {author} {\bibfnamefont
 			{J.}~\bibnamefont {Stelzer}} (\bibinfo {collaboration} {Gfitter Group}),\
 	}\href {\doibase 10.1140/epjc/s10052-014-3046-5} {\bibfield  {journal}
 		{\bibinfo  {journal} {Eur. Phys. J.}\ }\textbf {\bibinfo {volume} {C74}},\
 		\bibinfo {pages} {3046} (\bibinfo {year} {2014})},\ \Eprint
 	{http://arxiv.org/abs/1407.3792} {arXiv:1407.3792 [hep-ph]} \BibitemShut
 	{NoStop}%
 	%%CITATION = ARXIV:1407.3792;%%
 	\bibitem [{\citenamefont {Arhrib}\ \emph {et~al.}(2012)\citenamefont {Arhrib},
 		\citenamefont {Benbrik},\ and\ \citenamefont {Gaur}}]{Arhrib:2012ia}%
 	\BibitemOpen
 	\bibfield  {author} {\bibinfo {author} {\bibfnamefont {A.}~\bibnamefont
 			{Arhrib}}, \bibinfo {author} {\bibfnamefont {R.}~\bibnamefont {Benbrik}}, \
 		and\ \bibinfo {author} {\bibfnamefont {N.}~\bibnamefont {Gaur}},\ }\href
 	{\doibase 10.1103/PhysRevD.85.095021} {\bibfield  {journal} {\bibinfo
 			{journal} {Phys. Rev.}\ }\textbf {\bibinfo {volume} {D85}},\ \bibinfo {pages}
 		{095021} (\bibinfo {year} {2012})},\ \Eprint {http://arxiv.org/abs/1201.2644}
 	{arXiv:1201.2644 [hep-ph]} \BibitemShut {NoStop}%
 	%%CITATION = ARXIV:1201.2644;%%
 	\bibitem [{\citenamefont {Barbieri}\ \emph {et~al.}(2006)\citenamefont
 		{Barbieri}, \citenamefont {Hall},\ and\ \citenamefont
 		{Rychkov}}]{Barbieri:2006dq}%
 	\BibitemOpen
 	\bibfield  {author} {\bibinfo {author} {\bibfnamefont {R.}~\bibnamefont
 			{Barbieri}}, \bibinfo {author} {\bibfnamefont {L.~J.}\ \bibnamefont {Hall}},
 		\ and\ \bibinfo {author} {\bibfnamefont {V.~S.}\ \bibnamefont {Rychkov}},\
 	}\href {\doibase 10.1103/PhysRevD.74.015007} {\bibfield  {journal} {\bibinfo
 			{journal} {Phys. Rev. D}\ }\textbf {\bibinfo {volume} {74}},\ \bibinfo
 		{pages} {015007} (\bibinfo {year} {2006})},\ \Eprint
 	{http://arxiv.org/abs/hep-ph/0603188} {arXiv:hep-ph/0603188} \BibitemShut
 	{NoStop}%
 	\bibitem [{\citenamefont {Joglekar}\ \emph {et~al.}(2012)\citenamefont
 		{Joglekar}, \citenamefont {Schwaller},\ and\ \citenamefont
 		{Wagner}}]{Joglekar:2012vc}%
 	\BibitemOpen
 	\bibfield  {author} {\bibinfo {author} {\bibfnamefont {A.}~\bibnamefont
 			{Joglekar}}, \bibinfo {author} {\bibfnamefont {P.}~\bibnamefont {Schwaller}},
 		\ and\ \bibinfo {author} {\bibfnamefont {C.~E.~M.}\ \bibnamefont {Wagner}},\
 	}\href {\doibase 10.1007/JHEP12(2012)064} {\bibfield  {journal} {\bibinfo
 			{journal} {JHEP}\ }\textbf {\bibinfo {volume} {12}},\ \bibinfo {pages} {064}
 		(\bibinfo {year} {2012})},\ \Eprint {http://arxiv.org/abs/1207.4235}
 	{arXiv:1207.4235 [hep-ph]} \BibitemShut {NoStop}%
 	\bibitem [{\citenamefont {Haller}\ \emph {et~al.}(2018)\citenamefont {Haller},
 		\citenamefont {Hoecker}, \citenamefont {Kogler}, \citenamefont {M\"onig},
 		\citenamefont {Peiffer},\ and\ \citenamefont {Stelzer}}]{Haller:2018nnx}%
 	\BibitemOpen
 	\bibfield  {author} {\bibinfo {author} {\bibfnamefont {J.}~\bibnamefont
 			{Haller}}, \bibinfo {author} {\bibfnamefont {A.}~\bibnamefont {Hoecker}},
 		\bibinfo {author} {\bibfnamefont {R.}~\bibnamefont {Kogler}}, \bibinfo
 		{author} {\bibfnamefont {K.}~\bibnamefont {M\"onig}}, \bibinfo {author}
 		{\bibfnamefont {T.}~\bibnamefont {Peiffer}}, \ and\ \bibinfo {author}
 		{\bibfnamefont {J.}~\bibnamefont {Stelzer}},\ }\href {\doibase
 		10.1140/epjc/s10052-018-6131-3} {\bibfield  {journal} {\bibinfo  {journal}
 			{Eur. Phys. J. C}\ }\textbf {\bibinfo {volume} {78}},\ \bibinfo {pages} {675}
 		(\bibinfo {year} {2018})},\ \Eprint {http://arxiv.org/abs/1803.01853}
 	{arXiv:1803.01853 [hep-ph]} \BibitemShut {NoStop}%
 	\bibitem [{\citenamefont {Arroyo-Ure\~na}\ \emph {et~al.}(2023)\citenamefont
 		{Arroyo-Ure\~na}, \citenamefont {Fern\'andez-T\'ellez},\ and\ \citenamefont
 		{Tavares-Velasco}}]{Arroyo-Urena:2019fyd}%
 	\BibitemOpen
 	\bibfield  {author} {\bibinfo {author} {\bibfnamefont {M.~A.}\ \bibnamefont
 			{Arroyo-Ure\~na}}, \bibinfo {author} {\bibfnamefont {A.}~\bibnamefont
 			{Fern\'andez-T\'ellez}}, \ and\ \bibinfo {author} {\bibfnamefont
 			{G.}~\bibnamefont {Tavares-Velasco}},\ }\href {\doibase
 		10.31349/RevMexFis.69.020803} {\bibfield  {journal} {\bibinfo  {journal}
 			{Rev. Mex. Fis.}\ }\textbf {\bibinfo {volume} {69}},\ \bibinfo {pages}
 		{020803} (\bibinfo {year} {2023})},\ \Eprint
 	{http://arxiv.org/abs/1906.07821} {arXiv:1906.07821 [hep-ph]} \BibitemShut
 	{NoStop}%
 	\bibitem [{\citenamefont {Arroyo-Ure\~na}\ \emph {et~al.}(2022)\citenamefont
 		{Arroyo-Ure\~na}, \citenamefont {Chakraborty}, \citenamefont
 		{D\'\i{}az-Cruz}, \citenamefont {Ghosh}, \citenamefont {Khan},\ and\
 		\citenamefont {Moretti}}]{Arroyo-Urena:2022oft}%
 	\BibitemOpen
 	\bibfield  {author} {\bibinfo {author} {\bibfnamefont {M.~A.}\ \bibnamefont
 			{Arroyo-Ure\~na}}, \bibinfo {author} {\bibfnamefont {A.}~\bibnamefont
 			{Chakraborty}}, \bibinfo {author} {\bibfnamefont {J.~L.}\ \bibnamefont
 			{D\'\i{}az-Cruz}}, \bibinfo {author} {\bibfnamefont {D.~K.}\ \bibnamefont
 			{Ghosh}}, \bibinfo {author} {\bibfnamefont {N.}~\bibnamefont {Khan}}, \ and\
 		\bibinfo {author} {\bibfnamefont {S.}~\bibnamefont {Moretti}},\ }\href@noop
 	{} {\  (\bibinfo {year} {2022})},\ \Eprint {http://arxiv.org/abs/2205.12641}
 	{arXiv:2205.12641 [hep-ph]} \BibitemShut {NoStop}%
 	\bibitem [{\citenamefont {Tumasyan}\ \emph {et~al.}(2022)\citenamefont
 		{Tumasyan} \emph {et~al.}}]{CMS:2022qva}%
 	\BibitemOpen
 	\bibfield  {author} {\bibinfo {author} {\bibfnamefont {A.}~\bibnamefont
 			{Tumasyan}} \emph {et~al.} (\bibinfo {collaboration} {CMS}),\ }\href
 	{\doibase 10.1103/PhysRevD.105.092007} {\bibfield  {journal} {\bibinfo
 			{journal} {Phys. Rev. D}\ }\textbf {\bibinfo {volume} {105}},\ \bibinfo
 		{pages} {092007} (\bibinfo {year} {2022})},\ \Eprint
 	{http://arxiv.org/abs/2201.11585} {arXiv:2201.11585 [hep-ex]} \BibitemShut
 	{NoStop}%
 	\bibitem [{\citenamefont {Aad}\ \emph {et~al.}(2023)\citenamefont {Aad} \emph
 		{et~al.}}]{ATLAS:2022tnm}%
 	\BibitemOpen
 	\bibfield  {author} {\bibinfo {author} {\bibfnamefont {G.}~\bibnamefont
 			{Aad}} \emph {et~al.} (\bibinfo {collaboration} {ATLAS}),\ }\href {\doibase
 		10.1007/JHEP07(2023)088} {\bibfield  {journal} {\bibinfo  {journal} {JHEP}\
 		}\textbf {\bibinfo {volume} {07}},\ \bibinfo {pages} {088} (\bibinfo {year}
 		{2023})},\ \Eprint {http://arxiv.org/abs/2207.00348} {arXiv:2207.00348
 		[hep-ex]} \BibitemShut {NoStop}%
 	\bibitem [{\citenamefont {Djouadi}(2008)}]{Djouadi:2005gj}%
 	\BibitemOpen
 	\bibfield  {author} {\bibinfo {author} {\bibfnamefont {A.}~\bibnamefont
 			{Djouadi}},\ }\href {\doibase 10.1016/j.physrep.2007.10.005} {\bibfield
 		{journal} {\bibinfo  {journal} {Phys. Rept.}\ }\textbf {\bibinfo {volume}
 			{459}},\ \bibinfo {pages} {1} (\bibinfo {year} {2008})},\ \Eprint
 	{http://arxiv.org/abs/hep-ph/0503173} {arXiv:hep-ph/0503173} \BibitemShut
 	{NoStop}%
 	\bibitem [{\citenamefont {Baldini}\ \emph {et~al.}(2018)\citenamefont {Baldini}
 		\emph {et~al.}}]{Baldini:2018nnn}%
 	\BibitemOpen
 	\bibfield  {author} {\bibinfo {author} {\bibfnamefont {A.~M.}\ \bibnamefont
 			{Baldini}} \emph {et~al.} (\bibinfo {collaboration} {MEG II}),\ }\href
 	{\doibase 10.1140/epjc/s10052-018-5845-6} {\bibfield  {journal} {\bibinfo
 			{journal} {Eur. Phys. J.}\ }\textbf {\bibinfo {volume} {C78}},\ \bibinfo
 		{pages} {380} (\bibinfo {year} {2018})},\ \Eprint
 	{http://arxiv.org/abs/1801.04688} {arXiv:1801.04688 [physics.ins-det]}
 	\BibitemShut {NoStop}%
 	%%CITATION = ARXIV:1801.04688;%%
 	\bibitem [{\citenamefont {Aubert}\ \emph {et~al.}(2010)\citenamefont {Aubert}
 		\emph {et~al.}}]{BaBar:2009hkt}%
 	\BibitemOpen
 	\bibfield  {author} {\bibinfo {author} {\bibfnamefont {B.}~\bibnamefont
 			{Aubert}} \emph {et~al.} (\bibinfo {collaboration} {BaBar}),\ }\href
 	{\doibase 10.1103/PhysRevLett.104.021802} {\bibfield  {journal} {\bibinfo
 			{journal} {Phys. Rev. Lett.}\ }\textbf {\bibinfo {volume} {104}},\ \bibinfo
 		{pages} {021802} (\bibinfo {year} {2010})},\ \Eprint
 	{http://arxiv.org/abs/0908.2381} {arXiv:0908.2381 [hep-ex]} \BibitemShut
 	{NoStop}%
 	\bibitem [{\citenamefont {Aguillard}\ \emph {et~al.}(2023)\citenamefont
 		{Aguillard} \emph {et~al.}}]{Muong-2:2023cdq}%
 	\BibitemOpen
 	\bibfield  {author} {\bibinfo {author} {\bibfnamefont {D.~P.}\ \bibnamefont
 			{Aguillard}} \emph {et~al.} (\bibinfo {collaboration} {Muon g-2}),\
 	}\href@noop {} {\  (\bibinfo {year} {2023})},\ \Eprint
 	{http://arxiv.org/abs/2308.06230} {arXiv:2308.06230 [hep-ex]} \BibitemShut
 	{NoStop}%
 	\bibitem [{\citenamefont {Bennett}\ \emph {et~al.}(2006)\citenamefont {Bennett}
 		\emph {et~al.}}]{Muong-2:2006rrc}%
 	\BibitemOpen
 	\bibfield  {author} {\bibinfo {author} {\bibfnamefont {G.~W.}\ \bibnamefont
 			{Bennett}} \emph {et~al.} (\bibinfo {collaboration} {Muon g-2}),\ }\href
 	{\doibase 10.1103/PhysRevD.73.072003} {\bibfield  {journal} {\bibinfo
 			{journal} {Phys. Rev. D}\ }\textbf {\bibinfo {volume} {73}},\ \bibinfo
 		{pages} {072003} (\bibinfo {year} {2006})},\ \Eprint
 	{http://arxiv.org/abs/hep-ex/0602035} {arXiv:hep-ex/0602035} \BibitemShut
 	{NoStop}%
 	\bibitem [{\citenamefont {Abi}\ \emph {et~al.}(2021)\citenamefont {Abi} \emph
 		{et~al.}}]{Muong-2:2021ojo}%
 	\BibitemOpen
 	\bibfield  {author} {\bibinfo {author} {\bibfnamefont {B.}~\bibnamefont
 			{Abi}} \emph {et~al.} (\bibinfo {collaboration} {Muon g-2}),\ }\href
 	{\doibase 10.1103/PhysRevLett.126.141801} {\bibfield  {journal} {\bibinfo
 			{journal} {Phys. Rev. Lett.}\ }\textbf {\bibinfo {volume} {126}},\ \bibinfo
 		{pages} {141801} (\bibinfo {year} {2021})},\ \Eprint
 	{http://arxiv.org/abs/2104.03281} {arXiv:2104.03281 [hep-ex]} \BibitemShut
 	{NoStop}%
 	\bibitem [{\citenamefont {Venanzoni}(2024)}]{Venanzoni:2023mbe}%
 	\BibitemOpen
 	\bibfield  {author} {\bibinfo {author} {\bibfnamefont {G.}~\bibnamefont
 			{Venanzoni}} (\bibinfo {collaboration} {Muon g-2}),\ }\href {\doibase
 		10.22323/1.449.0037} {\bibfield  {journal} {\bibinfo  {journal} {PoS}\
 		}\textbf {\bibinfo {volume} {EPS-HEP2023}},\ \bibinfo {pages} {037} (\bibinfo
 		{year} {2024})},\ \Eprint {http://arxiv.org/abs/2311.08282} {arXiv:2311.08282
 		[hep-ex]} \BibitemShut {NoStop}%
 	\bibitem [{\citenamefont {Aoyama}\ \emph {et~al.}(2020)\citenamefont {Aoyama}
 		\emph {et~al.}}]{Aoyama:2020ynm}%
 	\BibitemOpen
 	\bibfield  {author} {\bibinfo {author} {\bibfnamefont {T.}~\bibnamefont
 			{Aoyama}} \emph {et~al.},\ }\href {\doibase 10.1016/j.physrep.2020.07.006}
 	{\bibfield  {journal} {\bibinfo  {journal} {Phys. Rept.}\ }\textbf {\bibinfo
 			{volume} {887}},\ \bibinfo {pages} {1} (\bibinfo {year} {2020})},\ \Eprint
 	{http://arxiv.org/abs/2006.04822} {arXiv:2006.04822 [hep-ph]} \BibitemShut
 	{NoStop}%
 	\bibitem [{\citenamefont {Kurz}\ \emph {et~al.}(2014)\citenamefont {Kurz},
 		\citenamefont {Liu}, \citenamefont {Marquard},\ and\ \citenamefont
 		{Steinhauser}}]{Kurz:2014wya}%
 	\BibitemOpen
 	\bibfield  {author} {\bibinfo {author} {\bibfnamefont {A.}~\bibnamefont
 			{Kurz}}, \bibinfo {author} {\bibfnamefont {T.}~\bibnamefont {Liu}}, \bibinfo
 		{author} {\bibfnamefont {P.}~\bibnamefont {Marquard}}, \ and\ \bibinfo
 		{author} {\bibfnamefont {M.}~\bibnamefont {Steinhauser}},\ }\href {\doibase
 		10.1016/j.physletb.2014.05.043} {\bibfield  {journal} {\bibinfo  {journal}
 			{Phys. Lett. B}\ }\textbf {\bibinfo {volume} {734}},\ \bibinfo {pages} {144}
 		(\bibinfo {year} {2014})},\ \Eprint {http://arxiv.org/abs/1403.6400}
 	{arXiv:1403.6400 [hep-ph]} \BibitemShut {NoStop}%
 	\bibitem [{\citenamefont {Davier}\ \emph {et~al.}(2020)\citenamefont {Davier},
 		\citenamefont {Hoecker}, \citenamefont {Malaescu},\ and\ \citenamefont
 		{Zhang}}]{Davier:2019can}%
 	\BibitemOpen
 	\bibfield  {author} {\bibinfo {author} {\bibfnamefont {M.}~\bibnamefont
 			{Davier}}, \bibinfo {author} {\bibfnamefont {A.}~\bibnamefont {Hoecker}},
 		\bibinfo {author} {\bibfnamefont {B.}~\bibnamefont {Malaescu}}, \ and\
 		\bibinfo {author} {\bibfnamefont {Z.}~\bibnamefont {Zhang}},\ }\href
 	{\doibase 10.1140/epjc/s10052-020-7792-2} {\bibfield  {journal} {\bibinfo
 			{journal} {Eur. Phys. J. C}\ }\textbf {\bibinfo {volume} {80}},\ \bibinfo
 		{pages} {241} (\bibinfo {year} {2020})},\ \bibinfo {note} {[Erratum:
 		Eur.Phys.J.C 80, 410 (2020)]},\ \Eprint {http://arxiv.org/abs/1908.00921}
 	{arXiv:1908.00921 [hep-ph]} \BibitemShut {NoStop}%
 	\bibitem [{\citenamefont {Davier}\ \emph {et~al.}(2017)\citenamefont {Davier},
 		\citenamefont {Hoecker}, \citenamefont {Malaescu},\ and\ \citenamefont
 		{Zhang}}]{Davier:2017zfy}%
 	\BibitemOpen
 	\bibfield  {author} {\bibinfo {author} {\bibfnamefont {M.}~\bibnamefont
 			{Davier}}, \bibinfo {author} {\bibfnamefont {A.}~\bibnamefont {Hoecker}},
 		\bibinfo {author} {\bibfnamefont {B.}~\bibnamefont {Malaescu}}, \ and\
 		\bibinfo {author} {\bibfnamefont {Z.}~\bibnamefont {Zhang}},\ }\href
 	{\doibase 10.1140/epjc/s10052-017-5161-6} {\bibfield  {journal} {\bibinfo
 			{journal} {Eur. Phys. J. C}\ }\textbf {\bibinfo {volume} {77}},\ \bibinfo
 		{pages} {827} (\bibinfo {year} {2017})},\ \Eprint
 	{http://arxiv.org/abs/1706.09436} {arXiv:1706.09436 [hep-ph]} \BibitemShut
 	{NoStop}%
 	\bibitem [{\citenamefont {Keshavarzi}\ \emph {et~al.}(2018)\citenamefont
 		{Keshavarzi}, \citenamefont {Nomura},\ and\ \citenamefont
 		{Teubner}}]{Keshavarzi:2018mgv}%
 	\BibitemOpen
 	\bibfield  {author} {\bibinfo {author} {\bibfnamefont {A.}~\bibnamefont
 			{Keshavarzi}}, \bibinfo {author} {\bibfnamefont {D.}~\bibnamefont {Nomura}},
 		\ and\ \bibinfo {author} {\bibfnamefont {T.}~\bibnamefont {Teubner}},\ }\href
 	{\doibase 10.1103/PhysRevD.97.114025} {\bibfield  {journal} {\bibinfo
 			{journal} {Phys. Rev. D}\ }\textbf {\bibinfo {volume} {97}},\ \bibinfo
 		{pages} {114025} (\bibinfo {year} {2018})},\ \Eprint
 	{http://arxiv.org/abs/1802.02995} {arXiv:1802.02995 [hep-ph]} \BibitemShut
 	{NoStop}%
 	\bibitem [{\citenamefont {Colangelo}\ \emph {et~al.}(2019)\citenamefont
 		{Colangelo}, \citenamefont {Hoferichter},\ and\ \citenamefont
 		{Stoffer}}]{Colangelo:2018mtw}%
 	\BibitemOpen
 	\bibfield  {author} {\bibinfo {author} {\bibfnamefont {G.}~\bibnamefont
 			{Colangelo}}, \bibinfo {author} {\bibfnamefont {M.}~\bibnamefont
 			{Hoferichter}}, \ and\ \bibinfo {author} {\bibfnamefont {P.}~\bibnamefont
 			{Stoffer}},\ }\href {\doibase 10.1007/JHEP02(2019)006} {\bibfield  {journal}
 		{\bibinfo  {journal} {JHEP}\ }\textbf {\bibinfo {volume} {02}},\ \bibinfo
 		{pages} {006} (\bibinfo {year} {2019})},\ \Eprint
 	{http://arxiv.org/abs/1810.00007} {arXiv:1810.00007 [hep-ph]} \BibitemShut
 	{NoStop}%
 	\bibitem [{\citenamefont {Hoferichter}\ \emph {et~al.}(2019)\citenamefont
 		{Hoferichter}, \citenamefont {Hoid},\ and\ \citenamefont
 		{Kubis}}]{Hoferichter:2019mqg}%
 	\BibitemOpen
 	\bibfield  {author} {\bibinfo {author} {\bibfnamefont {M.}~\bibnamefont
 			{Hoferichter}}, \bibinfo {author} {\bibfnamefont {B.-L.}\ \bibnamefont
 			{Hoid}}, \ and\ \bibinfo {author} {\bibfnamefont {B.}~\bibnamefont {Kubis}},\
 	}\href {\doibase 10.1007/JHEP08(2019)137} {\bibfield  {journal} {\bibinfo
 			{journal} {JHEP}\ }\textbf {\bibinfo {volume} {08}},\ \bibinfo {pages} {137}
 		(\bibinfo {year} {2019})},\ \Eprint {http://arxiv.org/abs/1907.01556}
 	{arXiv:1907.01556 [hep-ph]} \BibitemShut {NoStop}%
 	\bibitem [{\citenamefont {Keshavarzi}\ \emph {et~al.}(2020)\citenamefont
 		{Keshavarzi}, \citenamefont {Nomura},\ and\ \citenamefont
 		{Teubner}}]{Keshavarzi:2019abf}%
 	\BibitemOpen
 	\bibfield  {author} {\bibinfo {author} {\bibfnamefont {A.}~\bibnamefont
 			{Keshavarzi}}, \bibinfo {author} {\bibfnamefont {D.}~\bibnamefont {Nomura}},
 		\ and\ \bibinfo {author} {\bibfnamefont {T.}~\bibnamefont {Teubner}},\ }\href
 	{\doibase 10.1103/PhysRevD.101.014029} {\bibfield  {journal} {\bibinfo
 			{journal} {Phys. Rev. D}\ }\textbf {\bibinfo {volume} {101}},\ \bibinfo
 		{pages} {014029} (\bibinfo {year} {2020})},\ \Eprint
 	{http://arxiv.org/abs/1911.00367} {arXiv:1911.00367 [hep-ph]} \BibitemShut
 	{NoStop}%
 	\bibitem [{\citenamefont {Blum}\ \emph {et~al.}(2020)\citenamefont {Blum},
 		\citenamefont {Christ}, \citenamefont {Hayakawa}, \citenamefont {Izubuchi},
 		\citenamefont {Jin}, \citenamefont {Jung},\ and\ \citenamefont
 		{Lehner}}]{Blum:2019ugy}%
 	\BibitemOpen
 	\bibfield  {author} {\bibinfo {author} {\bibfnamefont {T.}~\bibnamefont
 			{Blum}}, \bibinfo {author} {\bibfnamefont {N.}~\bibnamefont {Christ}},
 		\bibinfo {author} {\bibfnamefont {M.}~\bibnamefont {Hayakawa}}, \bibinfo
 		{author} {\bibfnamefont {T.}~\bibnamefont {Izubuchi}}, \bibinfo {author}
 		{\bibfnamefont {L.}~\bibnamefont {Jin}}, \bibinfo {author} {\bibfnamefont
 			{C.}~\bibnamefont {Jung}}, \ and\ \bibinfo {author} {\bibfnamefont
 			{C.}~\bibnamefont {Lehner}},\ }\href {\doibase
 		10.1103/PhysRevLett.124.132002} {\bibfield  {journal} {\bibinfo  {journal}
 			{Phys. Rev. Lett.}\ }\textbf {\bibinfo {volume} {124}},\ \bibinfo {pages}
 		{132002} (\bibinfo {year} {2020})},\ \Eprint
 	{http://arxiv.org/abs/1911.08123} {arXiv:1911.08123 [hep-lat]} \BibitemShut
 	{NoStop}%
 	\bibitem [{\citenamefont {Borsanyi}\ \emph {et~al.}(2021)\citenamefont
 		{Borsanyi} \emph {et~al.}}]{Borsanyi:2020mff}%
 	\BibitemOpen
 	\bibfield  {author} {\bibinfo {author} {\bibfnamefont {S.}~\bibnamefont
 			{Borsanyi}} \emph {et~al.},\ }\href {\doibase 10.1038/s41586-021-03418-1}
 	{\bibfield  {journal} {\bibinfo  {journal} {Nature}\ }\textbf {\bibinfo
 			{volume} {593}},\ \bibinfo {pages} {51} (\bibinfo {year} {2021})},\ \Eprint
 	{http://arxiv.org/abs/2002.12347} {arXiv:2002.12347 [hep-lat]} \BibitemShut
 	{NoStop}%
 	\bibitem [{\citenamefont {C\`e}\ \emph {et~al.}(2022)\citenamefont {C\`e} \emph
 		{et~al.}}]{Ce:2022kxy}%
 	\BibitemOpen
 	\bibfield  {author} {\bibinfo {author} {\bibfnamefont {M.}~\bibnamefont
 			{C\`e}} \emph {et~al.},\ }\href {\doibase 10.1103/PhysRevD.106.114502}
 	{\bibfield  {journal} {\bibinfo  {journal} {Phys. Rev. D}\ }\textbf {\bibinfo
 			{volume} {106}},\ \bibinfo {pages} {114502} (\bibinfo {year} {2022})},\
 	\Eprint {http://arxiv.org/abs/2206.06582} {arXiv:2206.06582 [hep-lat]}
 	\BibitemShut {NoStop}%
 	\bibitem [{\citenamefont {Alexandrou}\ \emph {et~al.}(2023)\citenamefont
 		{Alexandrou} \emph {et~al.}}]{ExtendedTwistedMass:2022jpw}%
 	\BibitemOpen
 	\bibfield  {author} {\bibinfo {author} {\bibfnamefont {C.}~\bibnamefont
 			{Alexandrou}} \emph {et~al.} (\bibinfo {collaboration} {Extended Twisted
 			Mass}),\ }\href {\doibase 10.1103/PhysRevD.107.074506} {\bibfield  {journal}
 		{\bibinfo  {journal} {Phys. Rev. D}\ }\textbf {\bibinfo {volume} {107}},\
 		\bibinfo {pages} {074506} (\bibinfo {year} {2023})},\ \Eprint
 	{http://arxiv.org/abs/2206.15084} {arXiv:2206.15084 [hep-lat]} \BibitemShut
 	{NoStop}%
 	\bibitem [{\citenamefont {Chao}\ \emph {et~al.}(2023)\citenamefont {Chao},
 		\citenamefont {Meyer},\ and\ \citenamefont {Parrino}}]{Chao:2022ycy}%
 	\BibitemOpen
 	\bibfield  {author} {\bibinfo {author} {\bibfnamefont {E.-H.}\ \bibnamefont
 			{Chao}}, \bibinfo {author} {\bibfnamefont {H.~B.}\ \bibnamefont {Meyer}}, \
 		and\ \bibinfo {author} {\bibfnamefont {J.}~\bibnamefont {Parrino}},\ }\href
 	{\doibase 10.1103/PhysRevD.107.054505} {\bibfield  {journal} {\bibinfo
 			{journal} {Phys. Rev. D}\ }\textbf {\bibinfo {volume} {107}},\ \bibinfo
 		{pages} {054505} (\bibinfo {year} {2023})},\ \Eprint
 	{http://arxiv.org/abs/2211.15581} {arXiv:2211.15581 [hep-lat]} \BibitemShut
 	{NoStop}%
 	\bibitem [{\citenamefont {Ignatov}\ \emph {et~al.}(2023)\citenamefont {Ignatov}
 		\emph {et~al.}}]{CMD-3:2023alj}%
 	\BibitemOpen
 	\bibfield  {author} {\bibinfo {author} {\bibfnamefont {F.~V.}\ \bibnamefont
 			{Ignatov}} \emph {et~al.} (\bibinfo {collaboration} {CMD-3}),\ }\href@noop {}
 	{\  (\bibinfo {year} {2023})},\ \Eprint {http://arxiv.org/abs/2302.08834}
 	{arXiv:2302.08834 [hep-ex]} \BibitemShut {NoStop}%
 	\bibitem [{\citenamefont {Esteban}\ \emph {et~al.}(2020)\citenamefont
 		{Esteban}, \citenamefont {Gonzalez-Garcia}, \citenamefont {Maltoni},
 		\citenamefont {Schwetz},\ and\ \citenamefont {Zhou}}]{Esteban:2020cvm}%
 	\BibitemOpen
 	\bibfield  {author} {\bibinfo {author} {\bibfnamefont {I.}~\bibnamefont
 			{Esteban}}, \bibinfo {author} {\bibfnamefont {M.~C.}\ \bibnamefont
 			{Gonzalez-Garcia}}, \bibinfo {author} {\bibfnamefont {M.}~\bibnamefont
 			{Maltoni}}, \bibinfo {author} {\bibfnamefont {T.}~\bibnamefont {Schwetz}}, \
 		and\ \bibinfo {author} {\bibfnamefont {A.}~\bibnamefont {Zhou}},\ }\href
 	{\doibase 10.1007/JHEP09(2020)178} {\bibfield  {journal} {\bibinfo  {journal}
 			{JHEP}\ }\textbf {\bibinfo {volume} {09}},\ \bibinfo {pages} {178} (\bibinfo
 		{year} {2020})},\ \Eprint {http://arxiv.org/abs/2007.14792} {arXiv:2007.14792
 		[hep-ph]} \BibitemShut {NoStop}%
 	\bibitem [{\citenamefont {Gonzalez-Garcia}\ \emph {et~al.}(2021)\citenamefont
 		{Gonzalez-Garcia}, \citenamefont {Maltoni},\ and\ \citenamefont
 		{Schwetz}}]{Gonzalez-Garcia:2021dve}%
 	\BibitemOpen
 	\bibfield  {author} {\bibinfo {author} {\bibfnamefont {M.~C.}\ \bibnamefont
 			{Gonzalez-Garcia}}, \bibinfo {author} {\bibfnamefont {M.}~\bibnamefont
 			{Maltoni}}, \ and\ \bibinfo {author} {\bibfnamefont {T.}~\bibnamefont
 			{Schwetz}},\ }\href {\doibase 10.3390/universe7120459} {\bibfield  {journal}
 		{\bibinfo  {journal} {Universe}\ }\textbf {\bibinfo {volume} {7}},\ \bibinfo
 		{pages} {459} (\bibinfo {year} {2021})},\ \Eprint
 	{http://arxiv.org/abs/2111.03086} {arXiv:2111.03086 [hep-ph]} \BibitemShut
 	{NoStop}%
 	\bibitem [{\citenamefont {Aghanim}\ \emph {et~al.}(2018)\citenamefont {Aghanim}
 		\emph {et~al.}}]{Aghanim:2018eyx}%
 	\BibitemOpen
 	\bibfield  {author} {\bibinfo {author} {\bibfnamefont {N.}~\bibnamefont
 			{Aghanim}} \emph {et~al.} (\bibinfo {collaboration} {Planck}),\ }\href@noop
 	{} {\  (\bibinfo {year} {2018})},\ \Eprint {http://arxiv.org/abs/1807.06209}
 	{arXiv:1807.06209 [astro-ph.CO]} \BibitemShut {NoStop}%
 	\bibitem [{\citenamefont {Aprile}\ \emph {et~al.}(2018)\citenamefont {Aprile}
 		\emph {et~al.}}]{Aprile:2018dbl}%
 	\BibitemOpen
 	\bibfield  {author} {\bibinfo {author} {\bibfnamefont {E.}~\bibnamefont
 			{Aprile}} \emph {et~al.} (\bibinfo {collaboration} {XENON}),\ }\href
 	{\doibase 10.1103/PhysRevLett.121.111302} {\bibfield  {journal} {\bibinfo
 			{journal} {Phys. Rev. Lett.}\ }\textbf {\bibinfo {volume} {121}},\ \bibinfo
 		{pages} {111302} (\bibinfo {year} {2018})},\ \Eprint
 	{http://arxiv.org/abs/1805.12562} {arXiv:1805.12562 [astro-ph.CO]}
 	\BibitemShut {NoStop}%
 	%%CITATION = ARXIV:1805.12562;%%
 	\bibitem [{\citenamefont {Aalbers}\ \emph {et~al.}(2022)\citenamefont {Aalbers}
 		\emph {et~al.}}]{LZ:2022ufs}%
 	\BibitemOpen
 	\bibfield  {author} {\bibinfo {author} {\bibfnamefont {J.}~\bibnamefont
 			{Aalbers}} \emph {et~al.} (\bibinfo {collaboration} {LZ}),\ }\href@noop {} {\
 		(\bibinfo {year} {2022})},\ \Eprint {http://arxiv.org/abs/2207.03764}
 	{arXiv:2207.03764 [hep-ex]} \BibitemShut {NoStop}%
 	\bibitem [{\citenamefont {Tan}\ \emph {et~al.}(2016)\citenamefont {Tan} \emph
 		{et~al.}}]{PandaX:2016pdl}%
 	\BibitemOpen
 	\bibfield  {author} {\bibinfo {author} {\bibfnamefont {A.}~\bibnamefont
 			{Tan}} \emph {et~al.} (\bibinfo {collaboration} {PandaX}),\ }\href {\doibase
 		10.1103/PhysRevD.93.122009} {\bibfield  {journal} {\bibinfo  {journal} {Phys.
 				Rev. D}\ }\textbf {\bibinfo {volume} {93}},\ \bibinfo {pages} {122009}
 		(\bibinfo {year} {2016})},\ \Eprint {http://arxiv.org/abs/1602.06563}
 	{arXiv:1602.06563 [hep-ex]} \BibitemShut {NoStop}%
 	\bibitem [{\citenamefont {Alloul}\ \emph {et~al.}(2014)\citenamefont {Alloul},
 		\citenamefont {Christensen}, \citenamefont {Degrande}, \citenamefont {Duhr},\
 		and\ \citenamefont {Fuks}}]{Alloul:2013bka}%
 	\BibitemOpen
 	\bibfield  {author} {\bibinfo {author} {\bibfnamefont {A.}~\bibnamefont
 			{Alloul}}, \bibinfo {author} {\bibfnamefont {N.~D.}\ \bibnamefont
 			{Christensen}}, \bibinfo {author} {\bibfnamefont {C.}~\bibnamefont
 			{Degrande}}, \bibinfo {author} {\bibfnamefont {C.}~\bibnamefont {Duhr}}, \
 		and\ \bibinfo {author} {\bibfnamefont {B.}~\bibnamefont {Fuks}},\ }\href
 	{\doibase 10.1016/j.cpc.2014.04.012} {\bibfield  {journal} {\bibinfo
 			{journal} {Comput. Phys. Commun.}\ }\textbf {\bibinfo {volume} {185}},\
 		\bibinfo {pages} {2250} (\bibinfo {year} {2014})},\ \Eprint
 	{http://arxiv.org/abs/1310.1921} {arXiv:1310.1921 [hep-ph]} \BibitemShut
 	{NoStop}%
 	\bibitem [{\citenamefont {Bélanger}\ \emph {et~al.}(2018)\citenamefont
 		{Bélanger}, \citenamefont {Boudjema}, \citenamefont {Goudelis},
 		\citenamefont {Pukhov},\ and\ \citenamefont {Zaldivar}}]{Belanger:2018mqt}%
 	\BibitemOpen
 	\bibfield  {author} {\bibinfo {author} {\bibfnamefont {G.}~\bibnamefont
 			{Bélanger}}, \bibinfo {author} {\bibfnamefont {F.}~\bibnamefont {Boudjema}},
 		\bibinfo {author} {\bibfnamefont {A.}~\bibnamefont {Goudelis}}, \bibinfo
 		{author} {\bibfnamefont {A.}~\bibnamefont {Pukhov}}, \ and\ \bibinfo {author}
 		{\bibfnamefont {B.}~\bibnamefont {Zaldivar}},\ }\href {\doibase
 		10.1016/j.cpc.2018.04.027} {\bibfield  {journal} {\bibinfo  {journal}
 			{Comput. Phys. Commun.}\ }\textbf {\bibinfo {volume} {231}},\ \bibinfo
 		{pages} {173} (\bibinfo {year} {2018})},\ \Eprint
 	{http://arxiv.org/abs/1801.03509} {arXiv:1801.03509 [hep-ph]} \BibitemShut
 	{NoStop}%
 	%%CITATION = ARXIV:1801.03509;%%
 	\bibitem [{\citenamefont {Staub}(2014)}]{Staub:2013tta}%
 	\BibitemOpen
 	\bibfield  {author} {\bibinfo {author} {\bibfnamefont {F.}~\bibnamefont
 			{Staub}},\ }\href {\doibase 10.1016/j.cpc.2014.02.018} {\bibfield  {journal}
 		{\bibinfo  {journal} {Comput. Phys. Commun.}\ }\textbf {\bibinfo {volume}
 			{185}},\ \bibinfo {pages} {1773} (\bibinfo {year} {2014})},\ \Eprint
 	{http://arxiv.org/abs/1309.7223} {arXiv:1309.7223 [hep-ph]} \BibitemShut
 	{NoStop}%
 	\bibitem [{\citenamefont {Staub}(2015)}]{Staub:2015kfa}%
 	\BibitemOpen
 	\bibfield  {author} {\bibinfo {author} {\bibfnamefont {F.}~\bibnamefont
 			{Staub}},\ }\href {\doibase 10.1155/2015/840780} {\bibfield  {journal}
 		{\bibinfo  {journal} {Adv. High Energy Phys.}\ }\textbf {\bibinfo {volume}
 			{2015}},\ \bibinfo {pages} {840780} (\bibinfo {year} {2015})},\ \Eprint
 	{http://arxiv.org/abs/1503.04200} {arXiv:1503.04200 [hep-ph]} \BibitemShut
 	{NoStop}%
 	\bibitem [{\citenamefont {Porod}\ and\ \citenamefont
 		{Staub}(2012)}]{Porod:2011nf}%
 	\BibitemOpen
 	\bibfield  {author} {\bibinfo {author} {\bibfnamefont {W.}~\bibnamefont
 			{Porod}}\ and\ \bibinfo {author} {\bibfnamefont {F.}~\bibnamefont {Staub}},\
 	}\href {\doibase 10.1016/j.cpc.2012.05.021} {\bibfield  {journal} {\bibinfo
 			{journal} {Comput. Phys. Commun.}\ }\textbf {\bibinfo {volume} {183}},\
 		\bibinfo {pages} {2458} (\bibinfo {year} {2012})},\ \Eprint
 	{http://arxiv.org/abs/1104.1573} {arXiv:1104.1573 [hep-ph]} \BibitemShut
 	{NoStop}%
 	\bibitem [{\citenamefont {Hall}\ \emph {et~al.}(2010)\citenamefont {Hall},
 		\citenamefont {Jedamzik}, \citenamefont {March-Russell},\ and\ \citenamefont
 		{West}}]{Hall:2009bx}%
 	\BibitemOpen
 	\bibfield  {author} {\bibinfo {author} {\bibfnamefont {L.~J.}\ \bibnamefont
 			{Hall}}, \bibinfo {author} {\bibfnamefont {K.}~\bibnamefont {Jedamzik}},
 		\bibinfo {author} {\bibfnamefont {J.}~\bibnamefont {March-Russell}}, \ and\
 		\bibinfo {author} {\bibfnamefont {S.~M.}\ \bibnamefont {West}},\ }\href
 	{\doibase 10.1007/JHEP03(2010)080} {\bibfield  {journal} {\bibinfo  {journal}
 			{JHEP}\ }\textbf {\bibinfo {volume} {03}},\ \bibinfo {pages} {080} (\bibinfo
 		{year} {2010})},\ \Eprint {http://arxiv.org/abs/0911.1120} {arXiv:0911.1120
 		[hep-ph]} \BibitemShut {NoStop}%
 	\bibitem [{\citenamefont {Bauer}\ and\ \citenamefont
 		{Plehn}(2019)}]{Bauer:2017qwy}%
 	\BibitemOpen
 	\bibfield  {author} {\bibinfo {author} {\bibfnamefont {M.}~\bibnamefont
 			{Bauer}}\ and\ \bibinfo {author} {\bibfnamefont {T.}~\bibnamefont {Plehn}},\
 	}\href {\doibase 10.1007/978-3-030-16234-4} {\emph {\bibinfo {title} {{Yet
 					Another Introduction to Dark Matter}: {The Particle Physics Approach}}}},\
 	\bibinfo {series} {Lecture Notes in Physics}, Vol.\ \bibinfo {volume} {959}\
 	(\bibinfo  {publisher} {Springer},\ \bibinfo {year} {2019})\ \Eprint
 	{http://arxiv.org/abs/1705.01987} {arXiv:1705.01987 [hep-ph]} \BibitemShut
 	{NoStop}%
 	\bibitem [{\citenamefont {Khan}\ and\ \citenamefont
 		{Rakshit}(2015)}]{Khan:2015ipa}%
 	\BibitemOpen
 	\bibfield  {author} {\bibinfo {author} {\bibfnamefont {N.}~\bibnamefont
 			{Khan}}\ and\ \bibinfo {author} {\bibfnamefont {S.}~\bibnamefont {Rakshit}},\
 	}\href {\doibase 10.1103/PhysRevD.92.055006} {\bibfield  {journal} {\bibinfo
 			{journal} {Phys. Rev. D}\ }\textbf {\bibinfo {volume} {92}},\ \bibinfo
 		{pages} {055006} (\bibinfo {year} {2015})},\ \Eprint
 	{http://arxiv.org/abs/1503.03085} {arXiv:1503.03085 [hep-ph]} \BibitemShut
 	{NoStop}%
 	\bibitem [{\citenamefont {Khan}\ and\ \citenamefont
 		{Rakshit}(2014)}]{Khan:2014kba}%
 	\BibitemOpen
 	\bibfield  {author} {\bibinfo {author} {\bibfnamefont {N.}~\bibnamefont
 			{Khan}}\ and\ \bibinfo {author} {\bibfnamefont {S.}~\bibnamefont {Rakshit}},\
 	}\href {\doibase 10.1103/PhysRevD.90.113008} {\bibfield  {journal} {\bibinfo
 			{journal} {Phys. Rev. D}\ }\textbf {\bibinfo {volume} {90}},\ \bibinfo
 		{pages} {113008} (\bibinfo {year} {2014})},\ \Eprint
 	{http://arxiv.org/abs/1407.6015} {arXiv:1407.6015 [hep-ph]} \BibitemShut
 	{NoStop}%
 	\bibitem [{\citenamefont {Huitu}\ \emph {et~al.}(2008)\citenamefont {Huitu},
 		\citenamefont {Khalil}, \citenamefont {Okada},\ and\ \citenamefont
 		{Rai}}]{Huitu:2008gf}%
 	\BibitemOpen
 	\bibfield  {author} {\bibinfo {author} {\bibfnamefont {K.}~\bibnamefont
 			{Huitu}}, \bibinfo {author} {\bibfnamefont {S.}~\bibnamefont {Khalil}},
 		\bibinfo {author} {\bibfnamefont {H.}~\bibnamefont {Okada}}, \ and\ \bibinfo
 		{author} {\bibfnamefont {S.~K.}\ \bibnamefont {Rai}},\ }\href {\doibase
 		10.1103/PhysRevLett.101.181802} {\bibfield  {journal} {\bibinfo  {journal}
 			{Phys. Rev. Lett.}\ }\textbf {\bibinfo {volume} {101}},\ \bibinfo {pages}
 		{181802} (\bibinfo {year} {2008})},\ \Eprint {http://arxiv.org/abs/0803.2799}
 	{arXiv:0803.2799 [hep-ph]} \BibitemShut {NoStop}%
 	\bibitem [{\citenamefont {Das}\ \emph {et~al.}(2018)\citenamefont {Das},
 		\citenamefont {Li}, \citenamefont {Nandi},\ and\ \citenamefont
 		{Rai}}]{Das:2017fjf}%
 	\BibitemOpen
 	\bibfield  {author} {\bibinfo {author} {\bibfnamefont {K.}~\bibnamefont
 			{Das}}, \bibinfo {author} {\bibfnamefont {T.}~\bibnamefont {Li}}, \bibinfo
 		{author} {\bibfnamefont {S.}~\bibnamefont {Nandi}}, \ and\ \bibinfo {author}
 		{\bibfnamefont {S.~K.}\ \bibnamefont {Rai}},\ }\href {\doibase
 		10.1140/epjc/s10052-017-5495-0} {\bibfield  {journal} {\bibinfo  {journal}
 			{Eur. Phys. J. C}\ }\textbf {\bibinfo {volume} {78}},\ \bibinfo {pages} {35}
 		(\bibinfo {year} {2018})},\ \Eprint {http://arxiv.org/abs/1708.00328}
 	{arXiv:1708.00328 [hep-ph]} \BibitemShut {NoStop}%
 	\bibitem [{\citenamefont {Grossmann}\ \emph {et~al.}(2010)\citenamefont
 		{Grossmann}, \citenamefont {McElrath}, \citenamefont {Nandi},\ and\
 		\citenamefont {Rai}}]{Grossmann:2010wm}%
 	\BibitemOpen
 	\bibfield  {author} {\bibinfo {author} {\bibfnamefont {B.~N.}\ \bibnamefont
 			{Grossmann}}, \bibinfo {author} {\bibfnamefont {B.}~\bibnamefont {McElrath}},
 		\bibinfo {author} {\bibfnamefont {S.}~\bibnamefont {Nandi}}, \ and\ \bibinfo
 		{author} {\bibfnamefont {S.~K.}\ \bibnamefont {Rai}},\ }\href {\doibase
 		10.1103/PhysRevD.82.055021} {\bibfield  {journal} {\bibinfo  {journal} {Phys.
 				Rev. D}\ }\textbf {\bibinfo {volume} {82}},\ \bibinfo {pages} {055021}
 		(\bibinfo {year} {2010})},\ \Eprint {http://arxiv.org/abs/1006.5019}
 	{arXiv:1006.5019 [hep-ph]} \BibitemShut {NoStop}%
 	\bibitem [{\citenamefont {Abdallah}\ \emph {et~al.}(2021)\citenamefont
 		{Abdallah}, \citenamefont {Barik}, \citenamefont {Rai},\ and\ \citenamefont
 		{Samui}}]{Abdallah:2021npg}%
 	\BibitemOpen
 	\bibfield  {author} {\bibinfo {author} {\bibfnamefont {W.}~\bibnamefont
 			{Abdallah}}, \bibinfo {author} {\bibfnamefont {A.~K.}\ \bibnamefont {Barik}},
 		\bibinfo {author} {\bibfnamefont {S.~K.}\ \bibnamefont {Rai}}, \ and\
 		\bibinfo {author} {\bibfnamefont {T.}~\bibnamefont {Samui}},\ }\href
 	{\doibase 10.1103/PhysRevD.104.095031} {\bibfield  {journal} {\bibinfo
 			{journal} {Phys. Rev. D}\ }\textbf {\bibinfo {volume} {104}},\ \bibinfo
 		{pages} {095031} (\bibinfo {year} {2021})},\ \Eprint
 	{http://arxiv.org/abs/2106.01362} {arXiv:2106.01362 [hep-ph]} \BibitemShut
 	{NoStop}%
 	\bibitem [{\citenamefont {Abdallah}\ \emph {et~al.}(2023)\citenamefont
 		{Abdallah}, \citenamefont {Barik}, \citenamefont {Rai},\ and\ \citenamefont
 		{Samui}}]{Abdallah:2021dul}%
 	\BibitemOpen
 	\bibfield  {author} {\bibinfo {author} {\bibfnamefont {W.}~\bibnamefont
 			{Abdallah}}, \bibinfo {author} {\bibfnamefont {A.~K.}\ \bibnamefont {Barik}},
 		\bibinfo {author} {\bibfnamefont {S.~K.}\ \bibnamefont {Rai}}, \ and\
 		\bibinfo {author} {\bibfnamefont {T.}~\bibnamefont {Samui}},\ }\href
 	{\doibase 10.1103/PhysRevD.107.015026} {\bibfield  {journal} {\bibinfo
 			{journal} {Phys. Rev. D}\ }\textbf {\bibinfo {volume} {107}},\ \bibinfo
 		{pages} {015026} (\bibinfo {year} {2023})},\ \Eprint
 	{http://arxiv.org/abs/2109.07980} {arXiv:2109.07980 [hep-ph]} \BibitemShut
 	{NoStop}%
 \end{thebibliography}
%\begin{thebibliography}{70}%

%\end{thebibliography}%
\end{document}